\documentclass[11pt]{article}
 \UseRawInputEncoding
\usepackage[a4paper,includeheadfoot,top=1cm,bottom=2cm,left=2cm,right=2cm]{geometry}
\usepackage{here}
\usepackage{sectsty}
\usepackage{graphicx}
\usepackage{gnuplottex}
\usepackage[colorlinks=true,urlcolor=blue]{hyperref}
\usepackage{authblk}
\usepackage{caption}
\usepackage{pgfgantt}
\usepackage{tcolorbox}
\usepackage{mathtools}
\usepackage{amsmath}
\usepackage{subcaption}
\usepackage{graphicx} 
\usepackage{arydshln}
\usepackage{gensymb}

\usepackage{authblk}

\begin{document}
\title{Machine learning of twin/matrix interfaces from local stress field}

\author[1]{Javier F. Troncoso}
\author[2]{Yang Hu}
\author[3]{Nicolo M. della Ventura}
\author[3]{Amit Sharma}
\author[3]{Xavier Maeder}
\author[1]{Vladyslav Turlo}
\affil[1]{Laboratory for Advanced Materials Processing, Empa - Swiss Federal Laboratories for Materials Science and Technology, Switzerland}
\affil[2]{Department of Mechanical and Process Engineering, ETH Zürich, Switzerland}
\affil[3]{Laboratory for Mechanics of Materials and Nanostructures, Empa - Swiss Federal Laboratories for Materials Science and Technology, Switzerland}

\maketitle

\begin{abstract}
Twinning is an important deformation mode in plastically deformed hexagonal close-packed materials. The extremely high twin growth rates at the nanoscale make atomistic simulations an attractive method for investigating the role of individual twin/matrix interfaces such as twin boundaries and basal-prismatic interfaces in twin growth kinetics. Unfortunately, there is no single framework that allows researchers to differentiate such interfaces automatically, neither in experimental in-situ transmission electron microscopy analysis images nor in atomistic simulations. Moreover, the presence of alloying elements introduces substantial noise to local atomic environments, making it nearly impossible to identify which atoms belong to which interface. Here, with the help of advanced machine learning methods, we provide a proof-of-concept way of using the local stress field distribution as an indicator for the presence of interfaces and for determining their types. We apply such an analysis to the growth of twin embryos in Mg-10 at.\% Al alloys under constant stress and constant strain conditions, corresponding to two extremes of high and low strain rates, respectively. We discover that the kinetics of such growth is driven by high-energy basal-prismatic interfaces, in line with our experimental observations for pure Mg. 
\end{abstract}

\section{Introduction}\label{sec1}

As a promising structural material in the aerospace and automobile industries, magnesium (Mg) has drawn attention due to its low density and high strength/weight ratio. The wide application of Mg and its alloys in the transportation industry can result in reduced fuel consumption and, therefore, reduced environmental impact. However, due to the hexagonal close-packed (hcp) lattice structure under ambient conditions that leads to complex plastic accommodation mechanisms, Mg does not possess the high plastic deformability and ductility observed in face-centered cubic (fcc) or body-centered cubic (bcc) metals  \cite{Hutchinson2010,Sanchez-Martin2014}. In particular, Mg has a limited number of easily activated slip systems, often insufficient to accommodate homogeneous plastic deformations under various loading conditions. Alloying is used as a common strategy to activate/promote non-basal slip, which should lead to a higher ductility of Mg alloys \cite{Stricker2020c,Sandlobes2013b,Ahmad2020c,Wu2015}. Alternatively, twinning deformation can be used to improve the ductility of Mg alloys, in particular, by designing twin meshes \cite{Barnett2007,Barnett2007a,Wang2020b}.
Among all twin types, \{10$\bar{1}$2\} tension twins are profusely observed in Mg \cite{ElKadiri2015,Wang2010,DellaVentura2021}, and are characterized by a rotation of 86.3$\degree$ around the \textit{a}-axis of the hcp crystal structure. 
Such tension twins can grow to large sizes and consume the entire parent structure \cite{ElKadiri2013b}, allowing the material to release the global and locally stored stresses that build as a consequence of an increasing applied load, and facilitating the activation of other deformation modes due to the lattice reorientation of the twinned regions. Together with the effect of grain size and texture as well as pre-strained conditions and the presence of alloying elements, the orientation of the hcp structure with respect to the loading direction selectively activates different deformation modes (slip and twinning), each of which has a distinct dependence on changes in temperature and strain rate \cite{Yu2010,DellaVentura2021,DellaVentura2022b}. The competition of these factors results in changes in the crystallographic characteristics of the twin domains (e.g. interfaces, misorientations) as well as in their evolution under different testing conditions, which, in turn, affect the mechanical response of the material. To understand the mechanism underlying deformation twinning, the precise atomic movements that occur during twin growth need to be characterized and tracked, which is hard to be achieved experimentally, especially in the early stages of nucleation and twin growth, as their direct experimental observation is technically challenging.

Numerical and theoretical methods, such as the fast Fourier transformation (FFT)-based formalism \cite{ArulKumar2015}, the phase field method \cite{Spearot2020}, or the dislocation theory \cite{Capolungo2008}, have been employed to study twin growth in Mg and Mg alloys. However, these methods focus on the interface/dislocation kinetics without accounting for the atomistic mechanisms of the interface motion. Molecular dynamics (MD) simulations have been shown to be an efficient tool to investigate twinning in Mg and its alloys at the atomic level \cite{Xu2013,Spearot2020,Hu2020a,Hu2020,Gong2021,Huang2021}, covering also interactions of twins with other crystal defects such as grain boundaries and dislocations \cite{Dang2021,ArulKumar2022}. At the atomic scale, the motion of the twin/matrix interfaces is governed by the nucleation and glide of disconnections, defined as interfacial dislocations with a step character \cite{Howe2009}. For example, on the \{10$\bar{1}$2\} twin boundary, twinning disconnections with a step height of two interplanar spacings of the twin plane have been found \cite{Pond2016}, one-layer (step height equal to c/2) and two-layer disconnections (step height equal to c) have also been observed at the basal-prismatic/prismatic-basal interfaces \cite{Zu2017}. Previous works show that disconnection velocity and twin facet velocity measured from atomistic simulations are important parameters that assist in developing phenomenological models which manage to describe twin growth at a much larger length and time scales and therefore bridge the gap between atomistic simulations and experiments \cite{Hu2020a}. The generic formulation of the disconnection-mediated interface migration in relation to different driving forces is still in its infancy, especially for anisotropic crystal structures such as hcp; however, substantial progress in this direction has recently been made for cubic materials \cite{Han2022,Salvalaglio2022}. Another level of complexity is added by alloying elements/solute atoms, which can affect both nucleation and propagation of disconnections in rather stochastic ways. For example, MD simulations of twin embryo growth in heavily doped Mg-Al alloys demonstrate that disconnection nucleation and propagation is affected by solutes in the way that the net effect leads to different twins growing in different directions, i.e. activating the random selection of twin variant at larger length scales \cite{Hu2020}. Although such findings demonstrate that solute additions facilitate the formation of twin meshes in Mg-Al alloys, in line with experimental observations \cite{Drozdenko2019}, reproducing such effects in continuum models would require a complete revision of current deterministic models with the integration of stochastic terms describing solute-disconnection interactions. The derivation of such terms and their parameterization would be another challenge which could again be addressed with the help of high-throughput MD simulations.

However, to bolster progress in this direction, MD simulations need to be further supplemented with structure analysis to differentiate and track the time evolution of different twin facets. At the atomic scale, common topological analysis tools include the common neighbor analysis method \cite{Honeycutt1987}, the bond angle analysis method \cite{Ackland2006}, and the polyhedral template matching (PTM) method \cite{Larsen2016a}, all of which classify atoms according to the bulk crystalline structure, such as bcc, fcc, hcp, etc. More information about non-crystalline arrangements can be extracted by using Voronoi decomposition \cite{Voronoi1908}, in which an atom is enclosed by the Voronoi polyhedron (Wigner-Seitz cell) defined by the nearest neighbors. However, all of these traditional methods are highly sensitive to perturbations of atomic positions resulting from local thermal fluctuations, the presence of solutes with contrasting atomic sizes, and crystal defects such as vacancies and dislocations. An alternative solution is provided by newly emerged machine learning algorithms, developed to extract patterns from experimental and modeling data that cannot be derived with traditional methods \cite{Carleo2019}. Such algorithms include artificial neural networks (NN) \cite{Behler2016}, Gaussian processes \cite{Rasmussen2006}, and support vector machines \cite{Cristianini2000}, with proven success in bioinformatics, physics, engineering, and economics \cite{Rabunal2005}. Machine learning algorithms are generally classified as supervised learning, unsupervised learning, and reinforcement learning \cite{Carleo2019}. In supervised learning, algorithms learn the mapping function that relates inputs to outputs. For example, machine learning interatomic potentials recently adopted by the atomistic modeling community relate local atomic environment descriptors to atomic energies and forces, providing near \textit{ab initio} precision for problems of interest \cite{Mishin2021a}. The most frequently used descriptors of local atomic environments are Coulomb matrices \cite{Rupp2012}, sine matrices \cite{Faber2015}, atomic-centered symmetry functions \cite{Behler2011}, and smooth overlap of atomic orbitals (SOAP) \cite{Bartok2013}. These descriptors are invariant to rotations, translations, and permutations of atoms, and can be used in combination with unsupervised learning algorithms to identify structural building blocks of grain boundaries \cite{Rosenbrock2017,Homer2019,Troncoso2021}. 

Although advanced descriptors have already been developed for atomistic simulations, they cannot be applied to both experimental characterization data and simulation data. On the other hand, the local volume-averaged orientation and stresses/strains are universally accessible, and can be used as input features in machine learning-based analysis tools, promoting their transferability to experimental data while reducing noise due to various atomic-scale heterogeneities in atomistic simulations. As a proof-of-concept, we analyze large-scale MD simulations of the twin growth process in Mg-10 at.\% Al alloy under constant strain and constant stress conditions, representing two extremes of low and high strain rates, respectively. Supervised learning is applied to detect twin/matrix interfaces, and unsupervised learning is applied to distinguish their types, based on local spatial distribution of stress. As a result, we determine that at high strain rates, twin growth is driven by high-energy basal/prismatic interfaces, whereas at low strain rates, it is driven by low-energy twin boundaries, which is consistent with our experiments for pure magnesium. 

\section{Methods} \label{sec:methods}

\subsection{Atomistic simulations and machine learning}
 The atomistic simulation data analyzed in this work was gathered from our previous work \cite{Hu2020} on twin embryo growth in Mg-Al alloys under constant shear strains. The simulation box is set with the X axis along the [1$\bar{2}$10]-direction, the Y axis along the twinning direction ([10$\bar{1}$1]-direction), and the Z-axis perpendicular to the \{10$\bar{1}$2\} plane. Periodic boundary conditions are applied in all directions. The dimensions of the simulation box are about 5.1 × 77.2 × 77.0 $nm^{3}$ in the X, Y, and Z directions, respectively, and the sample contains approximately 1.3 million atoms. Additional simulations were performed to analyze twin embryo growth under constant shear stress conditions, in contrast to those under constant strain conditions. Under constant stress conditions, the twin embryo is growing faster (see Supplementary Figure S1), therefore, requiring a larger simulation box, which is ~1.8 times bigger along the Y and Z axes, and contains around 4.4 million atoms. A twin embryo with a length of 7.6 nm and a thickness of 4.1 nm was inserted at the center of each simulation box, using the Eshelby method reported by Xu et al. \cite{Xu2013}. To create Mg-Al solid solutions, 10 at.\% of Mg atoms were randomly replaced by Al solutes.
 
Molecular dynamics simulations were performed in LAMMPS \cite{lammpsWeb,Plimpton1995} with an embedded atom method (EAM) potential proposed by Liu et al. \cite{Liu1998}. LAMMPS, the acronym of Large-scale Atomic / Molecular Massively Parallel Simulator, is a free, parallelized and open-source software for MD simulations and energy calculations developed by Sandia National Laboratories. To activate twin growth, constant shear strains/stresses were applied to the simulation box. In simulations with constant shear strains, a global shear stress of around 1.05 GPa was first applied using an NPT (isothermal-isobaric) ensemble at 1 K. The simulation box was held for 1 ps to reach an equilibrated shear stress. After this time, an NVT (isothermal-isochoric) ensemble was used to fix the shear strain value that was responsible for the $\sim$1.05 GPa stress and relax the simulation cell. In simulations with constant shear stresses, atoms were relaxed under an NPT (isothermal-isobaric) ensemble at 1.0 GPa and 1 K, with the temperature adjusted every 100 time steps with one integration step of 0.1 fs. While a constant shear strain condition allows us to study decaying dynamics with twin growth via reducing the growth driving force, the constant shear stress condition allows us to explore mobile interfaces under constant growth driving force. The atomic stresses were also computed at each time step from the atomic kinetic energy and the virial contribution due to interatomic interactions \cite{Thompson2009}. Post-simulation analysis is performed with the Open Visualization Tool (OVITO) \cite{Stukowski2010}, and the Polyhedral Template Matching (PTM) method \cite{Larsen2016a} is utilized to differentiate the twinned region from the matrix, as well as atoms at the interfaces (hcp vs. non-hcp). 

The per-particle stress field  $S$ is calculated in LAMMPS at each output time step in MD simulations. This particle property is mapped into the bin grid and the representative value of the stress field $\sigma$ at each grid point is obtained by summing the stresses  $S$ of all particles in the bin volume and then being divided by the volume of the respective bin. Several bin sized have been tested (see Supplementary Figure S3). The spatial binning tool in OVITO is used to calculate the average local stress in each bin, whereas the presence of non-hcp atoms in the bin was used to separate interface bins from the other ones. For supervised learning, the number of input features varied between $n=4-18$ (see Supplementary Table S1), while the number of output nodes is $n=1$, with a binary output value corresponding to 0 for the interface bins/pixels, and 1 for the rest. A hidden layer with $h=10$ nodes is used between the input and output nodes. The sigmoid function is used as a neuron activation function \cite{FausettLaurene1994}, and the backpropagation algorithm is used to minimize the binary cross-entropy loss \cite{Larose2004}. For unsupervised learning, \textit{k}-means analysis and SOMs are used with three assumed cluster types, applied only to interface bins using 18 input features (Supplementary Table S1, set \#8). In the training, snapshots of the MD trajectory separated by 2,000 time steps were employed, and 70\% of the configurations were used as the training set. For this study, only bins containing grain boundaries according to the PTM method are used for the classification.
The errors in the training of the k-means algorithm and SOM are minimized by excluding highly isolated bin points or by averaging position sizes for consecutive snapshots of MD simulations.

\subsection{Experimental validation}
Two micropillar structures compressed in the work described in Ref. \cite{DellaVentura2022b} were further investigated for the purpose of this article to provide experimental support of the different character of the parent/twin interfaces formed under quasi-static and shock deformations, i.e. $10^{-4}$ and 500 $s^{-1}$. The micro-mechanical tests were conducted in displacement-controlled mode using a dedicated \textit{in situ} Alemnis AG nanoindenter setup for quasi-static and high strain rate deformation conditions mounted inside a scanning electron microscope (SEM) (Philips XL30). In particular, by setting the input voltage amplitude to a piezoelectric actuator, a predefined displacement profile (or velocity profile) has been induced in the tested structures. The piezoelectric actuator are specially designed for detecting these high-speed displacements with a resolution of ~15 nm. The displacement voltage signals are detected using a data acquisition board capable of fast sampling, up to 50k data point per second, providing a reliable feedback of the actual applied strain rate. To investigate the microstructure of both deformed test-pieces, EBSD patterns were collected inside a field emission gun (FEG) SEM (Tescan, Lyra3) and recorded for off-line analyses (electron beam conditions of 20 kV and 10 nA) with a Symmetry detector and Aztec 4.2 software (Oxford Instruments, UK), using a 2×2 binning (78 x 53 $px^{2}$) and 100 nm step size. The HR-EBSD cross-correlation was performed using the  CrossCourt V.4.3 (BLG Vantage, UK) software, with elastic constants of Mg (in GPa): \textit{$C_{11}$} = 59.7, \textit{$C_{33}$} = 61.7, \textit{$C_{44}$} = 16.4, \textit{$C_{12}$} = 26.2, \textit{$C_{13}$} = 21.7. The details about the HR-EBSD cross correlation technique can be found in Refs. \cite{Angus2006,Angus2010}. 
To analyze the atomic structure of the different twin boundaries, both the deformed pillars were lifted out and further thinned down to 120 nm. The high atomic resolution images were acquired using a ThermoFischer Themis 200 G3 spherical aberration (probe) corrected transmission electron microscope (TEM) operating at 200 kV. The $[2\bar{1}\bar{1}0]$ direction (the \textit{a}-axis), being the rotation axis of the $\{10\bar{1}2\} $ twin, was selected to be the zone axis.

\section{Results}

\begin{figure}[h!]
\centering
  \begin{center}

    \includegraphics[width=0.8\linewidth]{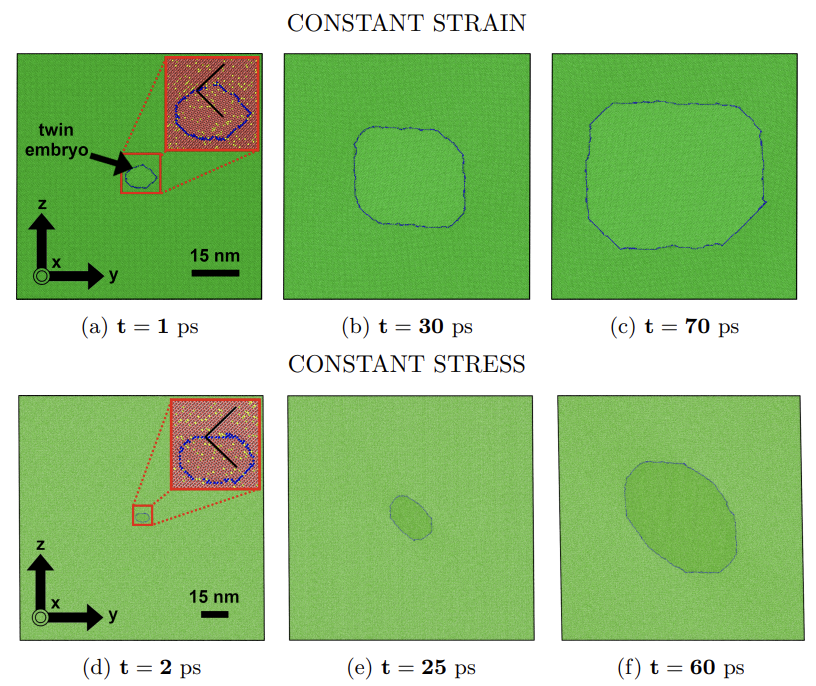}
  \end{center}
  \hspace*{\fill}   
\caption{\label{fig:simulationboxraw} (a-c) Twin embryo growth in Mg-10 at.\% Al, in a 5$\times$76$\times$76 nm$^3$ simulation box under constant shear strain. Green atoms are hcp atoms, while blue atoms correspond to twin/matrix interfaces.  (d-f) Growth of twin embryos in Mg-10 at\% Al, in a 5$\times$141$\times$138 nm$^3$ simulation box under constant shear stress. In the insets, the chemical information is provided, with red Mg atoms, yellow Al atoms, and black lines that highlight the basal plane orientation. }
\end{figure}

Quasi-2D MD simulations of twin embryo growth in Mg-10 at.\%Al are compared under constant shear strain and constant shear stress conditions, representative of low- and high-strain-rate deformations, respectively (further details are provided in the Methods section). The resulting time evolution of the twin configuration is shown in Fig. \ref{fig:simulationboxraw}, with blue atoms belonging to twin/matrix interfaces (identified as non-hcp by the PTM analysis). Fig. \ref{fig:simulationboxraw} shows that twin growth is significantly dependent on loading conditions, altering the direction and rate of growth, the shape of the embryo, and the curvature of twin/matrix interfaces. Although changes in twin morphology induced by varying loading conditions have also been revealed elsewhere \cite{Dudamell2013,Zhou2021a,DellaVentura2021,Zhang2021}, the thorough analysis of the effects of different loading states on the character of individual interfaces is still under debate. Figure \ref{fig:simulationboxraw} shows that the twin grows primarily in the horizontal direction under constant shear strain, while growing toward the diagonal direction under constant shear stress. At 60 ps, the twin volume reaches 10,533 nm$^3$ at constant applied strains, while it reaches 14,825 nm$^3$ at constant applied stress. The initial twin embryo are 171.6 nm$^3$ and 157.8 nm$^3$, respectively; the higher twin growth rate in the constant stress condition is due to the strain energies constantly added to the system, maintaining the average global stress constant, as one can see in Supplementary Fig. S1. During the twin growth process, the formation of disconnections (interface dislocations with a step character) is also observed in different twin facets. The pile-up of disconnections can be noted, for example, at constant stress, leading to more curved interfaces than at constant strain. In general, the interface curvature is the result of the interplay between disconnection nucleation and disconnection propagation/pinning, which is discussed in detail in our previous works \cite{Hu2020a,Hu2020}. Here, we focus on the role of individual twin/matrix interfaces in accommodating crystal deformation at different loading conditions.  

\begin{figure}[h!]
\centering
    \includegraphics[width=0.8\linewidth]{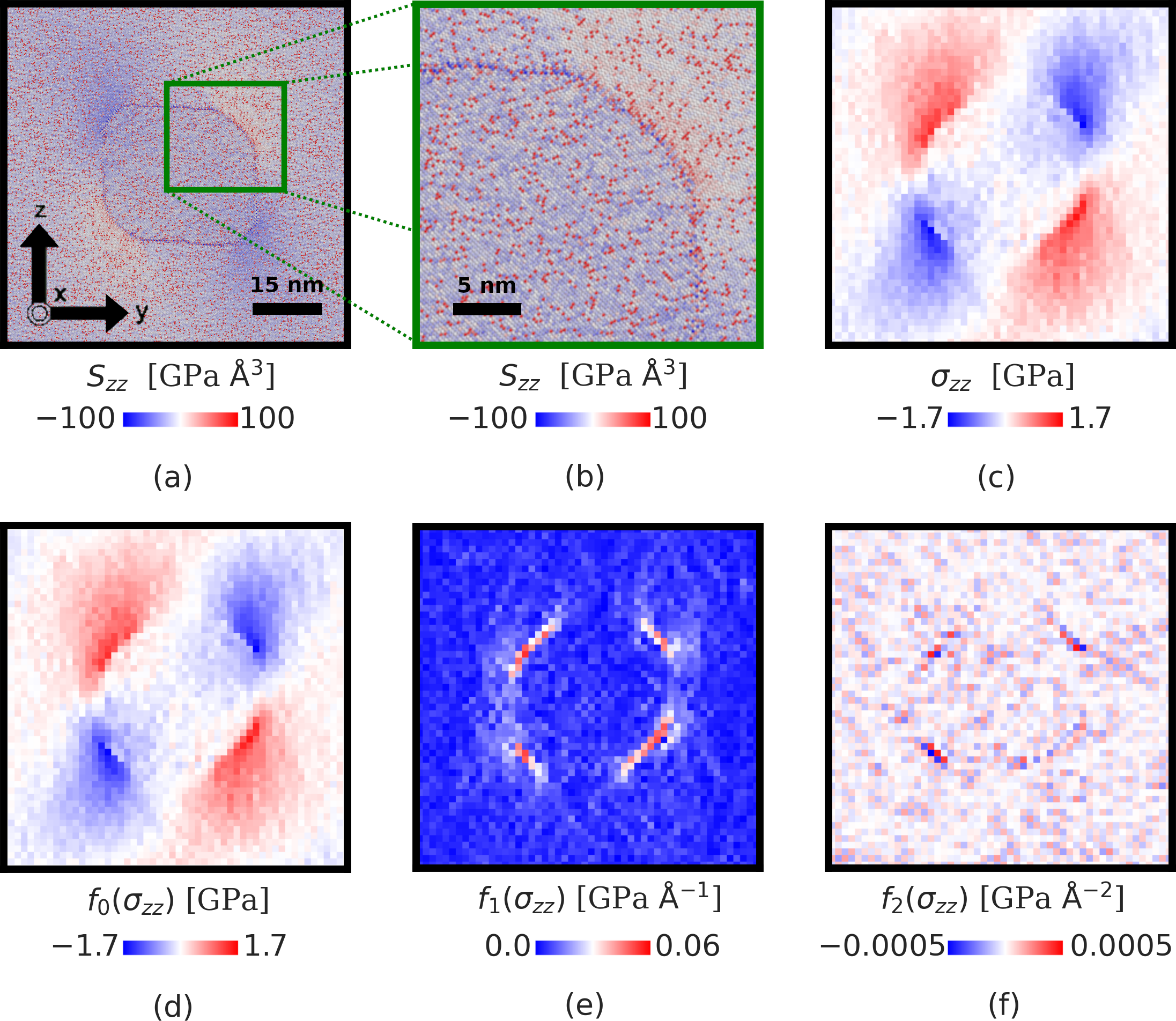}
\caption{\label{fig:allstresses} Twin embryo in Mg-10 at.\% Al in a 5×76×76 nm3 simulation box.   (a) Values of the  per-particle stress field tensor, $S^v$, of the configuration shown in Fig. \hyperref[fig:simulationboxraw]{1b}. (b) Zoom of (a) showing more details about the atomic configuration at the grain boundary. Values of the (c) stress field tensor $\sigma$ of the configuration shown in Fig. \hyperref[fig:simulationboxraw]{1b} in a 1$\times$52$\times$52 bin grid with a bin volume of 5$\times$1.5$\times$1.5 nm$^{3}$, and of the operators (d) $f_0$, (e) $f_1$ and (f) $f_2$ evaluated. Green boxes in figures c-d highlight the region zommed out in figure b. }
\end{figure}

The atomic structures of the twin/matrix interfaces depend on the lattice misorientation, resulting in a distinct stress field around the twin. Using the twin configuration in Fig. \hyperref[fig:simulationboxraw]{1b} as an example, the third component of the per-atom stress tensor, $S_{zz}$ is shown in Fig. \hyperref[fig:allstresses]{2a}, with units of pressure multiplied by volume \cite{Thompson2009}. Other per-atom stress tensor components for both loading conditions can be found in Supplementary Fig. S2. The main feature of Mg-Al alloys is the local stress heterogeneity and noise introduced by the solute atoms, which is better visible in the amplified view in Fig. \hyperref[fig:allstresses]{2b}. Al atoms are characterized by a smaller atomic volume than Mg, thus leading to strong tensile (i.e. positive) atomic stresses for diagonal (xx, yy, zz) components of the stress tensor, interfering with macroscopic interface stresses as one can see in Supplementary Fig. S2. In addition, thermal fluctuations of atoms vibrating around their equilibrium positions introduce additional noise into the atomic stress field, which can be seen in Fig. \hyperref[fig:allstresses]{2b}, even though we choose the extremely low temperature of 1 K for our simulations to minimize the effect of thermal noise on twin embryo growth. To reduce such noises and derive proper local macroscopic stresses $\sigma$, atomic stresses must be summed and averaged over reasonable volume. On the larger scale, this allows us, for example, to track the evolution of not only the average shear stress in the entire sample, but also the average shear stresses associated with the matrix and the twinned regions that are shown in Supplementary Fig. S1. The magnitude of twin stress is significantly lower than the matrix stress, thus driving twin embryo growth. For the constant stress conditions, such difference in stresses stays constant while for the constant strain conditions, such difference is getting smaller, slowing down the twin growth at the later stages. 

Based on such twin/matrix stress difference, we hypothesize in this work that the position and type of twin matrix interfaces can be identified just by analyzing local distribution of macroscopic stresses $\sigma$, which can be achieved through the spatial binning and volume averaging of atomic stresses $S$. The choice of the bin size is decisive here. As demonstrated in Supplementary Fig. S3, the small bin size leads to noisy stress field caused by random distribution of solutes and thermal fluctuations, while the larger bin size leads to the loss of information about the individual interfaces or even about the presence of the twin. Based on such assessment, the optimal bin size of 1.5 nm in both $y$ and $z$ directions was identified and used for further analysis. Figure \hyperref[fig:allstresses]{2c} shows the component $\sigma_{zz}$ after binning, whilst other stress components can be found in Supplementary Fig. S4 for both loading conditions. 

Next, we can test if position of twin/matrix interfaces and their types can be linked to the local stress field by applying advanced data analysis techniques based on supervised and unsupervised machine learning techniques. However, as both local and global stresses are evolving during twin growth, the raw $\sigma$ fields are not suitable for such analysis. In order to make more reliable time-independent prediction and include information about the local environment, three operators were applied to each component of the stress tensor, as defined on page 6 in Supplementary Material. The operator $f_0$ represents the deviation from the mean stress value at any given time step, $f_1$ corresponds to the modulus of the gradient vector in 2D with the compact stencil, and $f_2$ is the bi-harmonic operator, which includes information about the first and second neighboring bins \cite{Hawick2010}. These operators were applied to the components of the stress field, $\sigma$, obtained from MD trajectories. The example of applying such operators to $\sigma_{zz}$ are shown in Fig. \hyperref[fig:allstresses]{2d-f}, and the other components can be found in Supplementary Figs. S5-7 for both loading conditions. Due to the quasi-2D atomic structure used in atomistic simulations, not all operators and stress tensor components provide visible distinction for twin/matrix interfaces, so several sets of input features were chosen for supervised machine learning and listed in Supplementary Table S1. The NN with one hidden layer of 10 nodes and one binary output feature is used in this work to prove the direct connection between the local stress field and the presence of twin/matrix interfaces. The binary output feature takes value 0 for each bin that contains at least one non-hcp atom (i.e., belonging to twin/matrix interface due to the absence of any other structural defects), and takes value 1 for all other bins.

Three different snapshots from the constant strain simulation with the twin at the initial, intermediate and final stages of its growth shown as grey vertical lines in Fig. \hyperref[fig:accuracyvstime]{3a} were used in the training of NNs. Then, the data is separated into training, validation, and testing datasets, in a 70\%-15\%-15\% ratio, respectively. The last column in Supplementary Table S1 shows that the accuracy in the test datasets for different sets of input features is above 96\%. Remarkably, this points out that the stress field, even by considering a limited number of its components and transformation operators, can capture all the relevant information about the presence of interfaces. Nevertheless, the highest accuracy is achieved when all components of the stress field, $\sigma$, and all operators ($f_0$,$f_1$, and $f_2$) are used in the training (dataset \#8), which might indicate about hidden features in the noisy data that can be captured by machine learning algorithms. To prove the transferability of our NN analysis tool, we applied it to other time steps and other loading conditions with the results given in Supplementary Figs. S8 and S9. One major finding from such results is that the accuracy of the prediction decreases with time (see, for example, Fig. \hyperref[fig:accuracyvstime]{3a}), which might be the result of twin-twin interactions with itself through periodic boundaries. The twin-twin interactions inevitably alter the stress/strain fields around the facets. Figure \hyperref[fig:accuracyvstime]{3b} represents the summary of results from Supplementary Figs. S8 and S9, as probability distributions for different input sets and loading conditions, with an example of the NN prediction (Fig. \hyperref[fig:accuracyvstime]{3c}) versus ground truth (Fig. \hyperref[fig:accuracyvstime]{3d}) for the snapshot shown in Figure \hyperref[fig:allstresses]{2}. From Fig. \hyperref[fig:accuracyvstime]{3b}, one may notice that the accuracy is larger than 90\% even for the constant stress conditions that was not used in the fitting procedure and is characterized by larger simulations box and different shape of the twin embryo (see Fig. \ref{fig:simulationboxraw}). This proves the reliability and transferability of our method for locating interfaces with distinct local stress fields in large-scale atomistic simulations.

\begin{figure}[h!]
\includegraphics[width=0.8\linewidth]{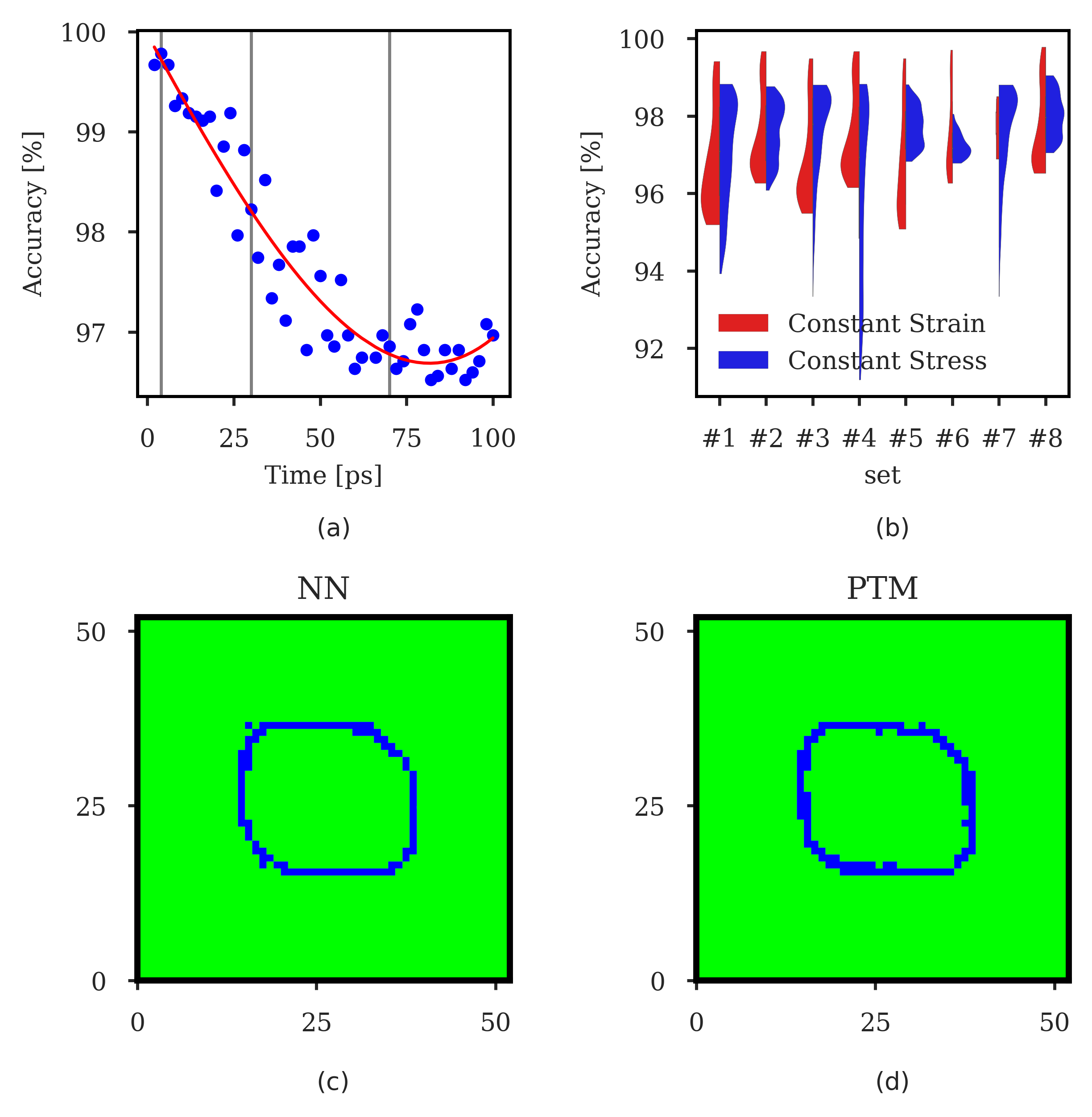}  
\centering
\caption{\label{fig:accuracyvstime} (a) Accuracy of the NN trained with the input set \#8 listed in Supplementary Table S1 in the detection of the grain boundary at different time steps in the MD simulation at constant strain. Vertical grey lines correspond to the time steps used in the training. (b) Comparison in the accuracy of the NNs trained with the different input datasets shown in Supplementary Table S1 and under different conditions: constant strain (red) and constant stress (blue). Classification into bulk (green) and interfacial (blue) bins given by (c) the NN (set \#8) and compared with the ground truth given by (d) the PTM analysis.}
\end{figure}

As one may fairly notice, atomistic structural analysis tools such as PTM, which was used as ground truth in our supervised learning, can locate interfaces and other crystal defects with atomic precision. However, such tools are susceptible to thermal and chemical noise, and cannot identify interface types automatically. More sophisticated algorithms were built over the last decade, which are able to not only identify atoms located at interfaces, but also relate the atomic structures of interfaces with interface crystallography. Yet even though these algorithms have been proved to be powerful, it is costly to track the position and size of individual interfaces regarding the processing time, especially for large systems with millions of atoms \cite{panzarino2015quantitative,priedeman2018quantifying}. Here, as a proof-of-concept, we propose to sacrifice some of the atomistic details of each individual facet bounding twin embryos, but focus on the interface identification based on the local stress signatures. In our quasi-2D case, the twin embryo is surrounded by twin boundaries (TB), forward TBs (f-TB), basal-prismatic (BP), and prismatic-basal (PB) interfaces \cite{Hu2020}. As the maximum number of possible boundary types is known, we hypothesize that unsupervised learning algorithms such as k-means and Self-Organized Maps (SOMs) can be used to divide interfacial (i.e., containing non-hcp atoms according to PTM) bins into several classes based on provided input features. For the latter, we use an all-inclusive set \#8 as defined in Supplementary Table S1. Though f-TBs and TBs differ in plane orientation, both generate similar weak stress fields, we choose three distinguishable facet types such as TB/f-TB, PB, and BP. As both k-means and SOMs were leading to the same/similar results, we further present only the results obtained by the k-means algorithm. More details about the training procedure are given in the Methods section. 


\begin{figure}[h!]
\centering
    \includegraphics[width=0.8\linewidth]{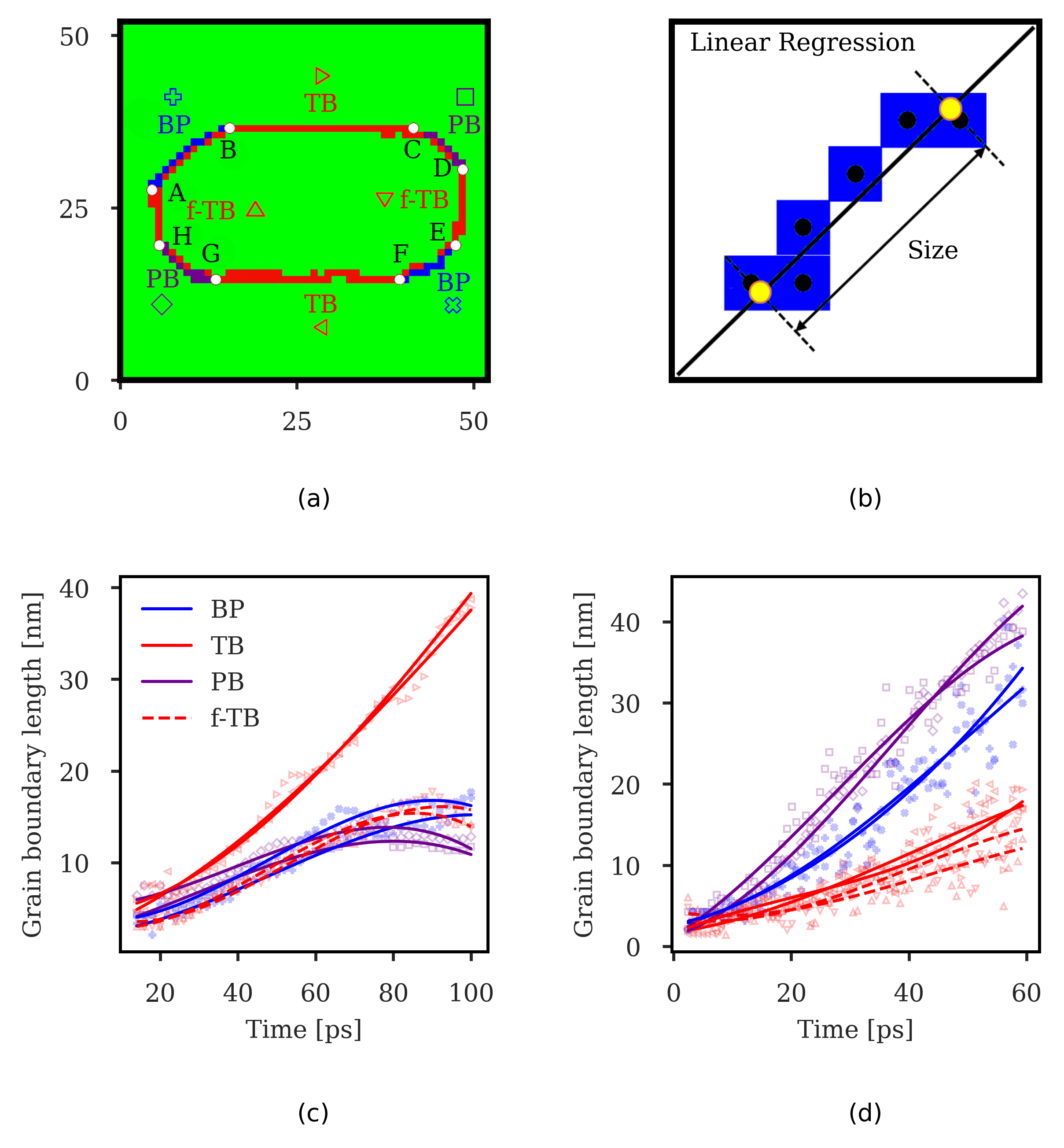}
\caption{\label{fig:supervisedresults} (a) Classification into non-defected hcp structure (green), twin boundaries (red), basal prismatic facets (blue) and prismatic basal facets (violet). Letters (next to white points) represent the limits of each boundary type. (b) Construction used in the present work to determine the size of the facets. Evolution of the size of each facet within the simulation at (c) constant strain and (d) constant stress conditions. }
\end{figure}

Figure  \hyperref[fig:supervisedresults]{4a} shows the interfacial bins belonging to different boundary types as predicted by the k-means algorithm, with PB and BP interfaces correctly highlighted by violet and blue bins, respectively. The segmentation of twin/matrix interfaces for all time steps and both loading conditions can be found in Supplementary Movies.  The areas marked in red correspond to twin boundaries, but also highlight the inner (twin side) parts in the PB and BP interfaces, which can be the result of weak stress fields in this area (see Figure 2, for example). However, the boundary points (A,B,...) of individual PB and BP facets can be well differentiated by using the construction shown in Fig. \hyperref[fig:supervisedresults]{4b}. In particular, the facet plane is obtained from the linear regression of the position of the bin centers of all bins forming the corresponding facet, with the boundary points of the facet obtained from the projection of the extreme points onto the linear fit. After all such points (A,B,...) are defined, all eight individual facets (AB,BC,...) are defined so that their positions (centers of the facets) as well as facet lengths can be tracked over time. 

Supplementary Figure S10 shows the spatial evolution of twin facets along the Y and Z axes for different loading conditions. In the case of constant strain (Figure S10(a) and (c)), interface facets move with constant velocities along the Y direction i.e. parallel to the shear direction, but slow down along the Z direction. In the case of constant stress (Figure S10(b) and (d)), all facets accelerate in both directions except TBs moving with a constant velocity along the Y direction and f-TBs slowing down along the Z direction. The specific details of such facet motion resulting from an interplay between applied stress and disconnection nucleation, propagation, and transition from one facet to another are discussed in detail in our previous works \cite{Hu2020a,Hu2020}. In such works, the twin fraction in projection on a given crystallographic direction was used to track the positions of individual facets. However, the automatic analysis and tracking of the length of individual facets was not possible by previous analysis methods but has been enabled with our method for the first time.

Figures  \hyperref[fig:supervisedresults]{4c,d} shows the evolution of facet lengths for different loading conditions. At the constant strain, the lengths of PB, BP, and f-TB interfaces consistently rise until reaching the plateau lengths around 10-15 nm each at ~60 ps. After that, such interfaces form steady-state twin tips moving with constant velocity in opposite directions (see the time evolution of the position of facet center in Supplementary Figure S10a), leading to a substantial increase in the length of TBs. This drastically decreases the fraction of PB/BP facets at twin/matrix interfaces (see Supplementary Figure S11a), making twin growth fully dominated at the later stages by TBs. However, TB motion is limited by nucleation and pinning of individual disconnections with the step of two atomic planes and not one atomic plane as in the case of PB/BP interfaces \cite{Hu2020a}, making such TB interfaces less mobile at increasing length and decreasing load (see Supplementary Figure S10c). At the constant stress, however, the situation is quite opposite. As one can see in Figure \hyperref[fig:supervisedresults]{4d}, the twin growth gets increasingly dominated by PB/BP facets, accommodating the steady-state mechanical load. However, the fraction of PB/BP interfaces does not keep increasing but instead converges to the asymptotic value, fluctuating around 70\% (see Supplementary Figure S11b). Thus, we can conclude that the twin growth at constant strain and constant stress conditions representing the extreme cases of low and high strain rates, respectively, is driven by different facet types with the increasing role of high-energy PB/BP interfaces in accommodating the load at higher strain rates.

\begin{figure}[h!]
\centering
\includegraphics[width=0.8\textwidth]{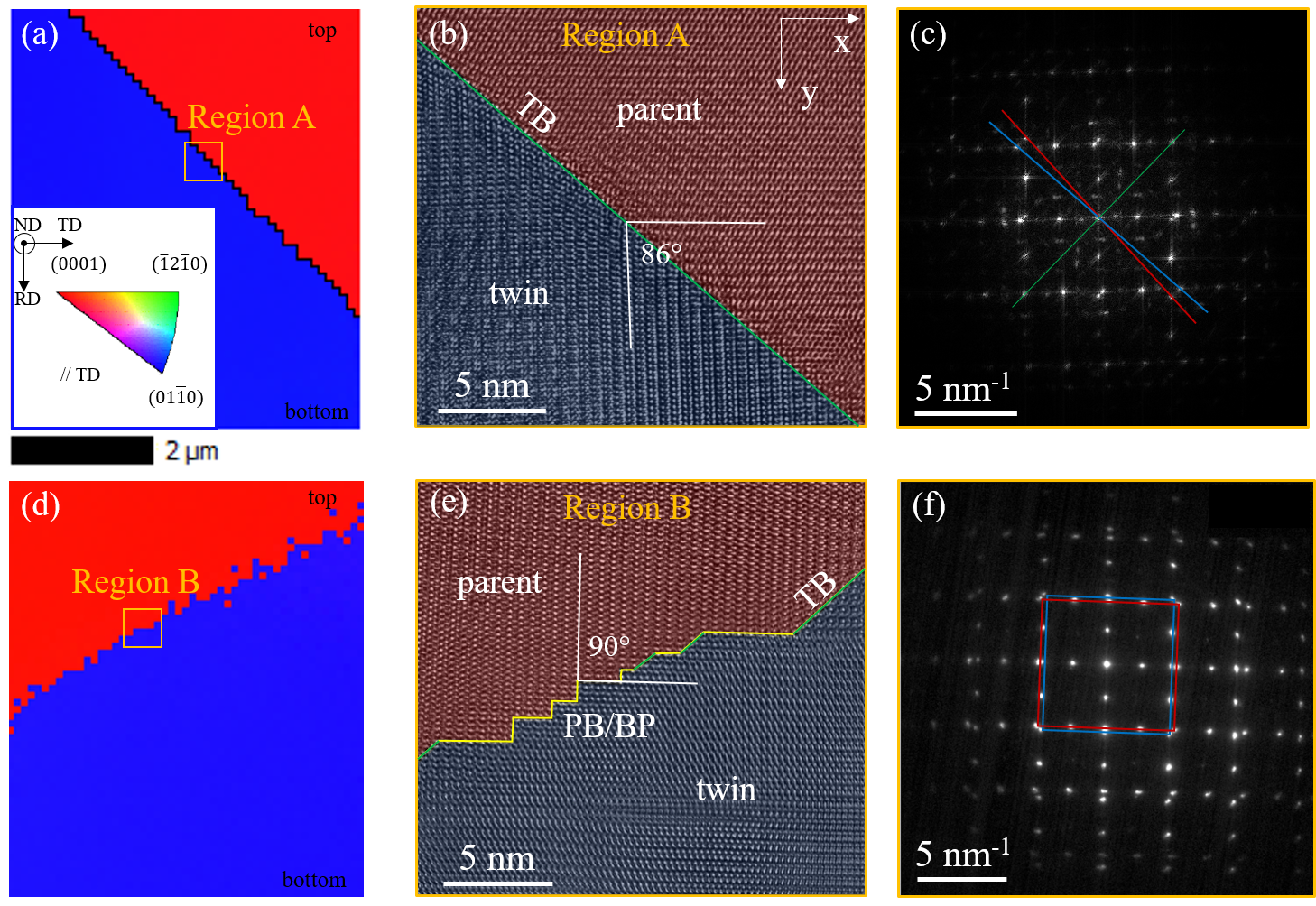}
\caption{\label{fig:expt}  (a,d) EBSD maps of two micropillars deformed under (a) quasi-static ($10^{-4}$ $s^{-1}$) and (b) shock (500 $s^{-1}$) conditions. The inverse pole figure (IPF) color code in all maps indicates the crystal orientation in the out-of-plane direction (normal direction, ND). TD: Transversal Direction, RD: Rolling Direction. Parent grain in red. Twin grain in blu. ND corresponds to the negative z-axis ([2$\bar{1}\bar{1}$0] crystal direction) in (b), TD to the x-axis ([0001] crystal direction) and RD accordingly. (b,e) HR-TEM images of regions of the twin boundaries highlighted in (a) and (d). Zone axis: [2$\bar{1}\bar{1}$0]. (b) TB constituted by a  conventional \{10$\bar{1}$2\} twin boundary. (e) Parent-twin interface characterized by the alternation of PB/BP and conventional TB interfaces (in yellow and green, respectively). The Selected Area Diffraction Pattern (SADP) in (c), taken across the TB in (b), shows 86.3° misoriented parent and twin crystals. The diffraction spot associated to the dark-green (0$\bar{1}$12) plane overlap, denoting the presence of a  crystallographic mirror plane. (f) SADP taken across the interface in (e) revealing a global 90° lattice misorientation between the basal and prismatic plane of the twin and parent grains. The (0$\bar{1}$12) diffraction spots do not overlap, showing the absence of the conventional \{10$\bar{1}$2\} twin plane.}
\end{figure}

It is important to note that the results illustrated in Fig. \hyperref[fig:supervisedresults]{4c,d} find confirmation in experimental observations. Specifically, Fig. \hyperref[fig:expt]{5} report the analyses of two parent-twin interfaces of twins nucleated at high and low strain rate conditions, respectively (more details can be found in the Methods section). BP facets constituted by a 90° parent/twin angular misorientation have been observed to characterize the twin interfaces developed at high strain rates, whilst the growth of TBs, across which a 86.3° misorientation is established, prevails at low strain rates. Together with other recent works \cite{Xie2021b,DellaVentura2022b,Jiang2022a,he2020direct}, the results illustrated in Fig. \hyperref[fig:expt]{5} suggest that the transition from a twin propagation mediated by the advance of TB to one mediated by the advance of BP facets occurs as the strain rate increases, further supporting what shown in Figure \hyperref[fig:supervisedresults]{4c,d}. Moreover, by analysis of the HR-TEM images, the fraction of PB/BP interfaces formed at high strain rate is around 67\%, which again is in line with what was predicted computationally, see Supplementary Figure S11b. It is worth noting that in Fig. \hyperref[fig:expt]{5d} the twin boundary globally deviates from the classical morphology where parent-twin interface coincides with the trace of the twin plane. This deviation is the direct consequence of a parent-twin transformation mainly accomplished through BP and PB plane conversions (Fig. \hyperref[fig:expt]{5d-f}), as a strictly defined crystallographic plane on which a shear deformation occurs cannot be identified. Nevertheless, segments of the interface correspondent to conventional TB (parent-twin misorientation of 86.3°) formed. Indeed, large coherence stresses caused by the 6.5\% lattice mismatch across BP interfaces disfavor the growth of the twin grain by lengthening these interfaces, as this would induce a significant increase in the elastic strain energy in the matrix \cite{Jiang2022a}. Consequently, misfit dislocations on PB and BP boundaries form, relaxing stresses and allowing these boundaries to extend promoting the first twin's growth. The detailed deformation steps evidencing nucleation of \{10$\bar{1}$2\} facets by relaxation process where disconnection pile-ups and coherency strains are transformed into a equivalent disclination dipoles that start to locally compensate for the difference in the 86.3° and 90° misorientations, are described by Barrett and Kadiri \cite{barrett2014roles}. In support of this scenario, the surface energy of PB interfaces is reported to be $\sim$170 mJ $m^{-2}$, above but close to the energy of the \{10$\bar{1}$2\} plane, $\sim$120 mJ $m^{-2}$ \cite{Xu2013}. The combination and migration of BP/PB interfaces and conventional TB, therefore, control the kinetics of the initial twin propagation \cite{ostapovets2014characterization}. As to why BP interfaces are formed at high strain rates, the atomic rearrangement leading to the 90° reorientation of the hcp crystal \cite{liu2014twinning} is naturally compatible with both the externally imposed deformation (compression along the [0$\bar{1}$10] direction) and the required twin growth speed at HSRs. In particular, the prismatic plane transforms into the basal by direct impinging of the flat punch on the pillar top surface, inducing a transversal basal to prismatic plane transformation caused by plastic incompressibility. In other words, and to summarize, a gradual deformation gradient in the structure favors the small angular correction required for the establishment of a minimal energetic configuration (invariant plane strain condition), whilst the fast migration of PB and BP serrations accomplished via collective atomic repositioning can accommodate a more uniform high plastic flow imposed to the entire structure at higher deformation rates.

\section{Discussion}

In this work, we proved that individual twin/matrix interfaces in Mg alloys have distinct stress signatures that can be captured by classical machine learning techniques, so the next step would be to discover more advanced fingerprints of individual interfaces to make interface analysis fully automated, more reliable, and independent of reference orientation. Examples of such fingerprints that can be used to cluster grain boundary structures based on the similarity in their properties are the atomic density \cite{Zhu2018}, the local grain boundary energy \cite{Frolov2012a}, and the local stress field, as demonstrated here. Instead of analyzing individual bins/pixels and their surroundings, more advanced deep-learning methods based on convolutional neural networks commonly applied in image and speech recognition\cite{Valueva2020, Alsobhani2021} can be used to analyze the system as a whole to automatically perform segmentation of twin/matrix interfaces into individual facets. Our naive analysis is done in the assumption of flat/linear interfaces as shown in Fig. \hyperref[fig:supervisedresults]{4b}, which can be improved by also estimating the curvature of individual facets that was well profound in the samples deformed at constant stress.

While our analysis represents a promising proof-of-concept technique for the identification and tracking of primary and forward twin boundaries (TBs and f-TBs), basal-prismatic and prismatic-basal interfaces (BP/PBs), all of which appear on the bright side of the twin, the ability of our method should be tested in more complex 3D cases. For example, at the dark side of 3D \{10$\bar{1}$2\} twin embryos in Mg, the newly reported twist-pyramidal-pyramidal (Twist-pypy1), twist-prismatic-prismatic (Twist-prpr2), and tilt-pyramidal-pyramidal interfaces (Tilt-pypy1) are usually hard to be observed in experiments but investigated in large-scale atomistic simulations \cite{Liu2019b,liu2016characterizing,wang2020characteristic,Gong2021}. In addition, a variety of twin types could be found in deformed Mg, such as tensile twins \{10$\bar{1}$2\}〈$\bar{1}$011〉 and \{11$\bar{2}$1\}〈$\bar{11}$26〉 and compressive twins \{10$\bar{1}$1\}〈10$\bar{12}$〉 and \{11$\bar{2}$2\}〈11$\bar{23}$〉, all of which are bound by distinct interfaces \cite{christian1995deformation}.

In its current form, our analysis fully relies on the local distribution of the stress field, which can be affected by twin interaction with other twins, free surfaces, lattice dislocations, precipitates, and inclusions. In literature, two kinds of such interactions are commonly considered such as direct and indirect. In the case of direct interaction, one defect gets in contact with the other one. Examples would be a twin boundary penetrated by incoming twins \cite{gong2018structural,kumar2019role,gong2020atomic} or by moving dislocations \cite{chen2019dislocation}, or a grain boundary penetrated by the growing twins \cite{ArulKumar2022,Dang2021}. Indirect interactions usually occur during the process of one twin approaching the other twin, or one dislocation approaching a twin facet, due to the long-range stress fields. Such interactions in turn affect the migration of dislocations, twins, etc. For example, in the work of Kumar et al. \cite{kumar2019role}, a “reach-out” phenomenon has been observed as one twin tip moving towards a \{10$\bar{1}$2\} twin boundary, i.e., the twin boundary in the vicinity of the impinging twin tip moves faster and forms a protrusion to reach out to the twin tip. Even in this work, the decreasing accuracy for the identification of twin/matrix interfaces during the twin growth (see Supplementary Figures S8 and S9) can be the result of indirect interactions of the twin with itself through periodic boundaries. 
In addition, interfacial dislocations and disconnections that are formed on each twin facet cause stress and strain fields that follow some analytical expressions, which can act as the governing equations for the ML model with penalizing huge deviations from such equations. With prior knowledge of possible disconnections on different facets \cite{hirth2016disconnections,Zu2017,mackain2017atomic}, various governing equations can be tested.
 
In summary, the usage of local macroscopic property grids derived from atomistic simulations can substantially reduce the noise introduced by alloying inhomogeneities and thermal fluctuations. This allowed us to confirm distinct local stress signatures/fingerprints of twin/matrix interfaces in the heavily-doped hcp Mg-Al alloy that can be captured with standard machine learning techniques. Moreover, out-of-box unsupervised learning techniques enabled identifying and tracking of individual facets for the first time. The established simple workflow of using supervised learning to identify the boundary atoms and using unsupervised learning to identify the facet type can be improved to cover more complex situation and applied to a wide range of materials. The variation of facet lengths with time suggests that twin growth is increasingly dominated by TBs at low strain rates and PB/BP interfaces at high strain rates, which was experimentally confirmed in this work. Moreover, the asymptotic fraction of PB/BP interfaces predicted by our atomistic simulations and machine learning analysis for high strain rate was found to be close to the one sampled experimentally. The consequences of our approach and its limits of applicability are discussed in detail, also highlighting further research avenues for deriving new physics from large-scale atomistic simulations.

\section{Acknowledgements}
This research was supported by the NCCR MARVEL, a National Centre of Competence in Research, funded by the Swiss National Science Foundation (grant number 205602). We thank Swiss National Supercomputing Centre (CSCS) for providing computing resources as part of s1130 production project.

\section{Data Availability}
The data supporting the findings of this study are available from the corresponding author upon reasonable request.

\section{Code Availability}
The codes used to calculate the results of this study are available from the corresponding author upon reasonable request.

\section{Author Contributions}
All authors had the original idea of the concept presented in this paper; Y.H. generated atomistic simulation data; J.F.T. analyzed the data and applied Machine Learning algorithms to characterize and identify interfaces; N.M.V performed experimental tensile testing of Mg pillars and HR-EBSD analysis; A.S. performed HR-TEM analysis of twin-matrix interfaces; X. M. supervised the experimental part and V.T. supervised the modeling part, as well as the overall project execution from the original idea to the final manuscript. All authors participated in the writing and editing of this article.

\section{Competing Interests}
The Authors declare no Competing Financial or Non-Financial Interests.

{

\bibliographystyle{elsarticle-num} 
\bibliography{Mg-ml}

\begin{thebibliography}{10}
\expandafter\ifx\csname url\endcsname\relax
  \def\url#1{\texttt{#1}}\fi
\expandafter\ifx\csname urlprefix\endcsname\relax\def\urlprefix{URL }\fi
\expandafter\ifx\csname href\endcsname\relax
  \def\href#1#2{#2} \def\path#1{#1}\fi

\bibitem{Hutchinson2010}
W.~B. Hutchinson, M.~R. Barnett,
  \href{http://dx.doi.org/10.1016/j.scriptamat.2010.05.047}{{Effective values
  of critical resolved shear stress for slip in polycrystalline magnesium and
  other hcp metals}}, Scr. Mater. 63~(7) (2010) 737--740.
\newblock \href {https://doi.org/10.1016/j.scriptamat.2010.05.047}
  {\path{doi:10.1016/j.scriptamat.2010.05.047}}.
\newline\urlprefix\url{http://dx.doi.org/10.1016/j.scriptamat.2010.05.047}

\bibitem{Sanchez-Martin2014}
R.~S{\'{a}}nchez-Mart{\'{i}}n, M.~T. P{\'{e}}rez-Prado, J.~Segurado, J.~Bohlen,
  I.~Guti{\'{e}}rrez-Urrutia, J.~Llorca, J.~M. Molina-Aldareguia, {Measuring
  the critical resolved shear stresses in Mg alloys by instrumented
  nanoindentation}, Acta Mater. 71 (2014) 283--292.
\newblock \href {https://doi.org/10.1016/j.actamat.2014.03.014}
  {\path{doi:10.1016/j.actamat.2014.03.014}}.

\bibitem{Stricker2020c}
M.~Stricker, W.~A. Curtin, {Prismatic Slip in Magnesium}, J. Phys. Chem. C
  124~(49) (2020) 27230--27240.
\newblock \href {https://doi.org/10.1021/acs.jpcc.0c09665}
  {\path{doi:10.1021/acs.jpcc.0c09665}}.

\bibitem{Sandlobes2013b}
S.~Sandl{\"{o}}bes, M.~Fri{\'{a}}k, J.~Neugebauer, D.~Raabe,
  \href{http://dx.doi.org/10.1016/j.msea.2013.03.006}{{Basal and non-basal
  dislocation slip in Mg–Y}}, Mater. Sci. Eng. A 576 (2013) 61--68.
\newblock \href {https://doi.org/10.1016/j.msea.2013.03.006}
  {\path{doi:10.1016/j.msea.2013.03.006}}.
\newline\urlprefix\url{http://dx.doi.org/10.1016/j.msea.2013.03.006}

\bibitem{Ahmad2020c}
R.~Ahmad, Z.~Wu, W.~A. Curtin,
  \href{https://doi.org/10.1016/j.actamat.2019.10.053}{{Analysis of double
  cross-slip of pyramidal I 〈c+a〉 screw dislocations and implications for
  ductility in Mg alloys}}, Acta Mater. 183 (2020) 228--241.
\newblock \href {https://doi.org/10.1016/j.actamat.2019.10.053}
  {\path{doi:10.1016/j.actamat.2019.10.053}}.
\newline\urlprefix\url{https://doi.org/10.1016/j.actamat.2019.10.053}

\bibitem{Wu2015}
Z.~Wu, W.~A. Curtin, {The origins of high hardening and low ductility in
  magnesium}, Nature 526~(7571) (2015) 62--67.
\newblock \href {https://doi.org/10.1038/nature15364}
  {\path{doi:10.1038/nature15364}}.

\bibitem{Barnett2007}
M.~R. Barnett, {Twinning and the ductility of magnesium alloys. Part I:
  "Tension" twins}, Mater. Sci. Eng. A 464~(1-2) (2007) 1--7.
\newblock \href {https://doi.org/10.1016/j.msea.2006.12.037}
  {\path{doi:10.1016/j.msea.2006.12.037}}.

\bibitem{Barnett2007a}
M.~R. Barnett, {Twinning and the ductility of magnesium alloys. Part II.
  "Contraction" twins}, Mater. Sci. Eng. A 464~(1-2) (2007) 8--16.
\newblock \href {https://doi.org/10.1016/j.msea.2007.02.109}
  {\path{doi:10.1016/j.msea.2007.02.109}}.

\bibitem{Wang2020b}
X.~Wang, L.~Jiang, C.~Cooper, K.~Yu, D.~Zhang, T.~J. Rupert, S.~Mahajan, I.~J.
  Beyerlein, E.~J. Lavernia, J.~M. Schoenung,
  \href{https://doi.org/10.1016/j.actamat.2020.05.021}{{Toughening magnesium
  with gradient twin meshes}}, Acta Mater. 195 (2020) 468--481.
\newblock \href {https://doi.org/10.1016/j.actamat.2020.05.021}
  {\path{doi:10.1016/j.actamat.2020.05.021}}.
\newline\urlprefix\url{https://doi.org/10.1016/j.actamat.2020.05.021}

\bibitem{ElKadiri2015}
H.~{El Kadiri}, C.~D. Barrett, J.~Wang, C.~N. Tom{\'{e}}, {Why are {\{}1012{\}}
  twins profuse in magnesium?}, Acta Mater. 85 (2015) 354--361.
\newblock \href {https://doi.org/10.1016/j.actamat.2014.11.033}
  {\path{doi:10.1016/j.actamat.2014.11.033}}.

\bibitem{Wang2010}
J.~Wang, I.~J. Beyerlein, C.~N. Tom{\'{e}}, {An atomic and probabilistic
  perspective on twin nucleation in Mg}, Scr. Mater. 63 (2010) 741--746.
\newblock \href {https://doi.org/10.1016/j.scriptamat.2010.01.047}
  {\path{doi:10.1016/j.scriptamat.2010.01.047}}.

\bibitem{DellaVentura2021}
N.~M. {Della Ventura}, S.~Kal{\'{a}}cska, D.~Casari, T.~E. Edwards, A.~Sharma,
  J.~Michler, R.~Log{\'{e}}, X.~Maeder,
  \href{https://doi.org/10.1016/j.matdes.2020.109206}{{{\{}10-12{\}} twinning
  mechanism during in situ micro-tensile loading of pure Mg: Role of basal slip
  and twin-twin interactions}}, Mater. Des. 197~(109206) (2021).
\newblock \href {https://doi.org/10.1016/j.matdes.2020.109206}
  {\path{doi:10.1016/j.matdes.2020.109206}}.
\newline\urlprefix\url{https://doi.org/10.1016/j.matdes.2020.109206}

\bibitem{ElKadiri2013b}
H.~{El Kadiri}, J.~Kapil, A.~L. Oppedal, L.~G. Hector, S.~R. Agnew,
  M.~Cherkaoui, S.~C. Vogel,
  \href{http://dx.doi.org/10.1016/j.actamat.2013.02.030}{{The effect of
  twin-twin interactions on the nucleation and propagation of {\{}10-12{\}}
  twinning in magnesium}}, Acta Mater. 61 (2013) 3549--3563.
\newblock \href {https://doi.org/10.1016/j.actamat.2013.02.030}
  {\path{doi:10.1016/j.actamat.2013.02.030}}.
\newline\urlprefix\url{http://dx.doi.org/10.1016/j.actamat.2013.02.030}

\bibitem{Yu2010}
Q.~Yu, Z.~W. Shan, J.~Li, X.~Huang, L.~Xiao, J.~Sun, E.~Ma, {Strong crystal
  size effect on deformation twinning}, Nature 463~(7279) (2010) 335--338.
\newblock \href {https://doi.org/10.1038/nature08692}
  {\path{doi:10.1038/nature08692}}.

\bibitem{DellaVentura2022b}
N.~M. della Ventura, A.~Sharma, S.~Kal{\'{a}}cska, M.~Jain, T.~E. Edwards,
  C.~Cayron, R.~Log{\'{e}}, J.~Michler, X.~Maeder, {Evolution of deformation
  twinning mechanisms in magnesium from low to high strain rates}, Mater. Des.
  217 (2022).
\newblock \href {https://doi.org/10.1016/j.matdes.2022.110646}
  {\path{doi:10.1016/j.matdes.2022.110646}}.

\bibitem{ArulKumar2015}
M.~{Arul Kumar}, A.~K. Kanjarla, S.~R. Niezgoda, R.~A. Lebensohn, C.~N.
  Tom{\'{e}}, {Numerical study of the stress state of a deformation twin in
  magnesium}, Acta Mater. 84 (2015) 349--358.
\newblock \href {https://doi.org/10.1016/j.actamat.2014.10.048}
  {\path{doi:10.1016/j.actamat.2014.10.048}}.

\bibitem{Spearot2020}
D.~E. Spearot, V.~Taupin, K.~Dang, L.~Capolungo,
  \href{https://doi.org/10.1016/j.mechmat.2020.103314}{{Structure and kinetics
  of three-dimensional defects on the {\{}10-12{\}} twin boundary in magnesium:
  Atomistic and phase-field simulations}}, Mech. Mater. 143 (2020) 103314.
\newblock \href {https://doi.org/10.1016/j.mechmat.2020.103314}
  {\path{doi:10.1016/j.mechmat.2020.103314}}.
\newline\urlprefix\url{https://doi.org/10.1016/j.mechmat.2020.103314}

\bibitem{Capolungo2008}
L.~Capolungo, I.~J. Beyerlein, {Nucleation and stability of twins in hcp
  metals}, Phys. Rev. B 78~(024117) (2008).
\newblock \href {https://doi.org/10.1103/PhysRevB.78.024117}
  {\path{doi:10.1103/PhysRevB.78.024117}}.

\bibitem{Xu2013}
B.~Xu, L.~Capolungo, D.~Rodney, {On the importance of prismatic/basal
  interfaces in the growth of (-1012) twins in hexagonal close packed
  crystals}, Scr. Mater. 68 (2013) 901--904.
\newblock \href {https://doi.org/10.1016/j.scriptamat.2013.02.023}
  {\path{doi:10.1016/j.scriptamat.2013.02.023}}.

\bibitem{Hu2020a}
Y.~Hu, V.~Turlo, I.~J. Beyerlein, S.~Mahajan, E.~J. Lavernia, J.~M. Schoenung,
  T.~J. Rupert,
  \href{https://doi.org/10.1016/j.actamat.2020.04.010}{{Disconnection-mediated
  twin embryo growth in Mg}}, Acta Mater. 194 (2020) 437--451.
\newblock \href {https://doi.org/10.1016/j.actamat.2020.04.010}
  {\path{doi:10.1016/j.actamat.2020.04.010}}.
\newline\urlprefix\url{https://doi.org/10.1016/j.actamat.2020.04.010}

\bibitem{Hu2020}
Y.~Hu, V.~Turlo, I.~J. Beyerlein, S.~Mahajan, E.~J. Lavernia, J.~M. Schoenung,
  T.~J. Rupert,
  \href{https://doi.org/10.1103/PhysRevLett.125.205503}{{Embracing the Chaos:
  Alloying Adds Stochasticity to Twin Embryo Growth}}, Phys. Rev. Lett.
  125~(205503) (2020).
\newblock \href {https://doi.org/10.1103/PhysRevLett.125.205503}
  {\path{doi:10.1103/PhysRevLett.125.205503}}.
\newline\urlprefix\url{https://doi.org/10.1103/PhysRevLett.125.205503}

\bibitem{Gong2021}
M.~Gong, J.~Graham, V.~Taupin, L.~Capolungo,
  \href{https://doi.org/10.1016/j.actamat.2020.116603}{{The effects of stress,
  temperature and facet structure on growth of {\{}101¯2{\}} twins in Mg: A
  molecular dynamics and phase field study}}, Acta Mater. 208 (2021) 116603.
\newblock \href {https://doi.org/10.1016/j.actamat.2020.116603}
  {\path{doi:10.1016/j.actamat.2020.116603}}.
\newline\urlprefix\url{https://doi.org/10.1016/j.actamat.2020.116603}

\bibitem{Huang2021}
Z.~Huang, V.~Turlo, X.~Wang, F.~Chen, Q.~Shen, L.~Zhang, I.~J. Beyerlein, T.~J.
  Rupert,
  \href{https://doi.org/10.1016/j.commatsci.2020.110241}{{Dislocation-induced Y
  segregation at basal-prismatic interfaces in Mg}}, Comput. Mater. Sci.
  188~(110241) (2021).
\newblock \href {https://doi.org/10.1016/j.commatsci.2020.110241}
  {\path{doi:10.1016/j.commatsci.2020.110241}}.
\newline\urlprefix\url{https://doi.org/10.1016/j.commatsci.2020.110241}

\bibitem{Dang2021}
K.~Dang, J.~Graham, R.~J. McCabe, V.~Taupin, C.~N. Tom{\'{e}}, L.~Capolungo,
  \href{https://doi.org/10.1016/j.mtla.2021.101247}{{Atomistic and phase field
  simulations of three dimensional interactions of {\{}101¯2{\}} twins with
  grain boundaries in Mg: Twin transmission and dislocation emission}},
  Materialia 20~(August) (2021) 101247.
\newblock \href {https://doi.org/10.1016/j.mtla.2021.101247}
  {\path{doi:10.1016/j.mtla.2021.101247}}.
\newline\urlprefix\url{https://doi.org/10.1016/j.mtla.2021.101247}

\bibitem{ArulKumar2022}
M.~{Arul Kumar}, K.~Dang, V.~Taupin, R.~J. McCabe, C.~N. Tom{\'{e}},
  L.~Capolungo, \href{https://doi.org/10.1016/j.mtla.2022.101437}{{Numerical
  and experimental characterization of twin transmission across grain
  boundaries along the forward and lateral directions}}, Materialia 23~(April)
  (2022) 101437.
\newblock \href {https://doi.org/10.1016/j.mtla.2022.101437}
  {\path{doi:10.1016/j.mtla.2022.101437}}.
\newline\urlprefix\url{https://doi.org/10.1016/j.mtla.2022.101437}

\bibitem{Howe2009}
J.~M. Howe, R.~C. Pond, J.~P. Hirth,
  \href{http://dx.doi.org/10.1016/j.pmatsci.2009.04.001}{{The role of
  disconnections in phase transformations}}, Prog. Mater. Sci. 54~(6) (2009)
  792--838.
\newblock \href {https://doi.org/10.1016/j.pmatsci.2009.04.001}
  {\path{doi:10.1016/j.pmatsci.2009.04.001}}.
\newline\urlprefix\url{http://dx.doi.org/10.1016/j.pmatsci.2009.04.001}

\bibitem{Pond2016}
R.~C. Pond, J.~P. Hirth, A.~Serra, D.~J. Bacon,
  \href{https://doi.org/10.1080/21663831.2016.1165298}{{Atomic displacements
  accompanying deformation twinning: shears and shuffles}}, Mater. Res. Lett.
  4~(4) (2016) 185--190.
\newblock \href {https://doi.org/10.1080/21663831.2016.1165298}
  {\path{doi:10.1080/21663831.2016.1165298}}.
\newline\urlprefix\url{https://doi.org/10.1080/21663831.2016.1165298}

\bibitem{Zu2017}
Q.~Zu, X.~Z. Tang, S.~Xu, Y.~F. Guo,
  \href{http://dx.doi.org/10.1016/j.actamat.2017.03.035}{{Atomistic study of
  nucleation and migration of the basal/prismatic interfaces in Mg single
  crystals}}, Acta Mater. 130 (2017) 310--318.
\newblock \href {https://doi.org/10.1016/j.actamat.2017.03.035}
  {\path{doi:10.1016/j.actamat.2017.03.035}}.
\newline\urlprefix\url{http://dx.doi.org/10.1016/j.actamat.2017.03.035}

\bibitem{Han2022}
J.~Han, D.~J. Srolovitz, M.~Salvalaglio,
  \href{https://doi.org/10.1016/j.actamat.2021.117178}{{Disconnection-mediated
  migration of interfaces in microstructures: I. continuum model}}, Acta Mater.
  227 (2022) 117178.
\newblock \href {http://arxiv.org/abs/2103.09688} {\path{arXiv:2103.09688}},
  \href {https://doi.org/10.1016/j.actamat.2021.117178}
  {\path{doi:10.1016/j.actamat.2021.117178}}.
\newline\urlprefix\url{https://doi.org/10.1016/j.actamat.2021.117178}

\bibitem{Salvalaglio2022}
M.~Salvalaglio, D.~J. Srolovitz, J.~Han,
  \href{https://doi.org/10.1016/j.actamat.2021.117463}{{Disconnection-Mediated
  migration of interfaces in microstructures: II. diffuse interface
  simulations}}, Acta Mater. 227 (2022) 117463.
\newblock \href {http://arxiv.org/abs/2103.09689} {\path{arXiv:2103.09689}},
  \href {https://doi.org/10.1016/j.actamat.2021.117463}
  {\path{doi:10.1016/j.actamat.2021.117463}}.
\newline\urlprefix\url{https://doi.org/10.1016/j.actamat.2021.117463}

\bibitem{Drozdenko2019}
D.~Drozdenko, J.~{\v{C}}apek, B.~Clausen, A.~Vinogradov, K.~M{\'{a}}this,
  {Influence of the solute concentration on the anelasticity in Mg-Al alloys: A
  multiple-approach study}, J. Alloys Compd. 786 (2019) 779--790.
\newblock \href {https://doi.org/10.1016/j.jallcom.2019.01.358}
  {\path{doi:10.1016/j.jallcom.2019.01.358}}.

\bibitem{Honeycutt1987}
J.~D. Honeycutt, H.~C. Andersen, {Molecular dynamics study of melting and
  freezing of small Lennard-Jones clusters}, J. Phys. Chem. 91 (1987)
  4950--4963.
\newblock \href {https://doi.org/10.1021/j100303a014}
  {\path{doi:10.1021/j100303a014}}.

\bibitem{Ackland2006}
G.~J. Ackland, A.~P. Jones, {Applications of local crystal structure measures
  in experiment and simulation}, Phys. Rev. B - Condens. Matter Mater. Phys.
  73~(054104) (2006).
\newblock \href {https://doi.org/10.1103/PhysRevB.73.054104}
  {\path{doi:10.1103/PhysRevB.73.054104}}.

\bibitem{Larsen2016a}
P.~M. Larsen, S.~Schmidt, J.~Schi{\O}tz, {Robust structural identification via
  polyhedral template matching}, Model. Simul. Mater. Sci. Eng. 24~(055007)
  (2016).
\newblock \href {http://arxiv.org/abs/1603.05143} {\path{arXiv:1603.05143}},
  \href {https://doi.org/10.1088/0965-0393/24/5/055007}
  {\path{doi:10.1088/0965-0393/24/5/055007}}.

\bibitem{Voronoi1908}
G.~Voronoi, {Nouvelles applications des param{\`{e}}tres continus {\`{a}} la
  th{\'{e}}orie des formes quadratiques. Deuxi{\`{e}}me m{\'{e}}moire.
  Recherches sur les parall{\'{e}}llo{\`{e}}dres primitifs}, J. f{\"{u}}r die
  reine und Angew. Math. 134 (1908) 198--287.

\bibitem{Carleo2019}
G.~Carleo, I.~Cirac, K.~Cranmer, L.~Daudet, M.~Schuld, N.~Tishby,
  L.~Vogt-Maranto, L.~Zdeborov{\'{a}},
  \href{https://doi.org/10.1103/RevModPhys.91.045002}{{Machine learning and the
  physical sciences}}, Rev. Mod. Phys. 91~(4) (2019) 45002.
\newblock \href {http://arxiv.org/abs/1903.10563} {\path{arXiv:1903.10563}},
  \href {https://doi.org/10.1103/RevModPhys.91.045002}
  {\path{doi:10.1103/RevModPhys.91.045002}}.
\newline\urlprefix\url{https://doi.org/10.1103/RevModPhys.91.045002}

\bibitem{Behler2016}
J.~Behler, \href{http://dx.doi.org/10.1063/1.4966192}{{Perspective: Machine
  learning potentials for atomistic simulations}}, J. Chem. Phys. 145~(170901)
  (2016).
\newblock \href {https://doi.org/10.1063/1.4966192}
  {\path{doi:10.1063/1.4966192}}.
\newline\urlprefix\url{http://dx.doi.org/10.1063/1.4966192}

\bibitem{Rasmussen2006}
C.~E. Rasmussen, C.~K.~I. Williams, {Gaussian Processes for Machine Learning},
  the MIT Press, Cambridge, Massachusetts, 2006.

\bibitem{Cristianini2000}
N.~Cristianini, J.~Shawe-Taylor, {An Introduction to Support Vector Machines
  and other Kernel-Based Learning Methods Nello}, Cambridge University Press,
  Cambridge, 2000.

\bibitem{Rabunal2005}
J.~R. Rabu{\~{n}}al, J.~Dorado, {Artificial neural networks in real-life
  applications}, Idea Group Publishing, London, 2005.
\newblock \href {https://doi.org/10.4018/978-1-59140-902-1}
  {\path{doi:10.4018/978-1-59140-902-1}}.

\bibitem{Mishin2021a}
Y.~Mishin,
  \href{https://doi.org/10.1016/j.actamat.2021.116980}{{Machine-learning
  interatomic potentials for materials science}}, Acta Mater. 214 (2021)
  116980.
\newblock \href {http://arxiv.org/abs/2102.06163} {\path{arXiv:2102.06163}},
  \href {https://doi.org/10.1016/j.actamat.2021.116980}
  {\path{doi:10.1016/j.actamat.2021.116980}}.
\newline\urlprefix\url{https://doi.org/10.1016/j.actamat.2021.116980}

\bibitem{Rupp2012}
M.~Rupp, A.~Tkatchenko, K.~R. M{\"{u}}ller, O.~A. {Von Lilienfeld}, {Fast and
  accurate modeling of molecular atomization energies with machine learning},
  Phys. Rev. Lett. 108~(058301) (2012).
\newblock \href {http://arxiv.org/abs/1109.2618} {\path{arXiv:1109.2618}},
  \href {https://doi.org/10.1103/PhysRevLett.108.058301}
  {\path{doi:10.1103/PhysRevLett.108.058301}}.

\bibitem{Faber2015}
F.~Faber, A.~Lindmaa, O.~A. {Von Lilienfeld}, R.~Armiento, {Crystal structure
  representations for machine learning models of formation energies}, Int. J.
  Quantum Chem. 115~(16) (2015) 1094--1101.
\newblock \href {http://arxiv.org/abs/1503.07406} {\path{arXiv:1503.07406}},
  \href {https://doi.org/10.1002/qua.24917} {\path{doi:10.1002/qua.24917}}.

\bibitem{Behler2011}
J.~Behler, {Atom-centered symmetry functions for constructing high-dimensional
  neural network potentials}, J. Chem. Phys. 134~(074106) (2011).
\newblock \href {https://doi.org/10.1063/1.3553717}
  {\path{doi:10.1063/1.3553717}}.

\bibitem{Bartok2013}
A.~P. Bart{\'{o}}k, R.~Kondor, G.~Cs{\'{a}}nyi, {On representing chemical
  environments}, Phys. Rev. B 87~(184115) (2013).
\newblock \href {http://arxiv.org/abs/1209.3140} {\path{arXiv:1209.3140}},
  \href {https://doi.org/10.1103/PhysRevB.87.184115}
  {\path{doi:10.1103/PhysRevB.87.184115}}.

\bibitem{Rosenbrock2017}
C.~W. Rosenbrock, E.~R. Homer, G.~Cs{\'{a}}nyi, G.~L. Hart, {Discovering the
  building blocks of atomic systems using machine learning: Application to
  grain boundaries}, npj Comput. Mater. 3~(1) (2017).
\newblock \href {http://arxiv.org/abs/arXiv:1703.06236v1}
  {\path{arXiv:arXiv:1703.06236v1}}, \href
  {https://doi.org/10.1038/s41524-017-0027-x}
  {\path{doi:10.1038/s41524-017-0027-x}}.

\bibitem{Homer2019}
E.~R. Homer, D.~M. Hensley, C.~W. Rosenbrock, A.~H. Nguyen, G.~L. Hart,
  {Machine-learning informed representations for grain boundary structures},
  Front. Mater. 6~(July) (2019) 1--11.
\newblock \href {https://doi.org/10.3389/fmats.2019.00168}
  {\path{doi:10.3389/fmats.2019.00168}}.

\bibitem{Troncoso2021}
J.~{F. Troncoso},
  \href{https://doi.org/10.1016/j.commatsci.2020.110167}{{ClasSOMfier: A neural
  network for cluster analysis and detection of lattice defects}}, Comput.
  Mater. Sci. 188~(110167) (2021).
\newblock \href {https://doi.org/10.1016/j.commatsci.2020.110167}
  {\path{doi:10.1016/j.commatsci.2020.110167}}.
\newline\urlprefix\url{https://doi.org/10.1016/j.commatsci.2020.110167}

\bibitem{lammpsWeb}
http://lammps.sandia.gov.

\bibitem{Plimpton1995}
S.~Plimpton, \href{http://www.cs.sandia.gov/∼sjplimp/main.html}{{Short-Range
  Molecular Dynamics}}, J. Comput. Phys. 117~(6) (1995) 1--19.
\newblock \href {http://arxiv.org/abs/nag.2347} {\path{arXiv:nag.2347}}.
\newline\urlprefix\url{http://www.cs.sandia.gov/∼sjplimp/main.html}

\bibitem{Liu1998}
X.~Y. Liu, J.~B. Adams, {Grain-boundary segregation in Al-10{\%}Mg alloys at
  hot working temperatures}, Acta Mater. 46~(10) (1998) 3467--3476.
\newblock \href {https://doi.org/10.1016/S1359-6454(98)00038-X}
  {\path{doi:10.1016/S1359-6454(98)00038-X}}.

\bibitem{Thompson2009}
A.~P. Thompson, S.~J. Plimpton, W.~Mattson, {General formulation of pressure
  and stress tensor for arbitrary many-body interaction potentials under
  periodic boundary conditions}, J. Chem. Phys. 131~(154107) (2009).
\newblock \href {https://doi.org/10.1063/1.3245303}
  {\path{doi:10.1063/1.3245303}}.

\bibitem{Stukowski2010}
A.~Stukowski, {Visualization and analysis of atomistic simulation data with
  OVITO-the Open Visualization Tool}, Model. Simul. Mater. Sci. Eng.
  18~(015012) (2010).
\newblock \href {https://doi.org/10.1088/0965-0393/18/1/015012}
  {\path{doi:10.1088/0965-0393/18/1/015012}}.

\bibitem{FausettLaurene1994}
F.~Laurene, {Fundamentals of Neural Network, Architectures, Algorithms And
  Applications}, Prentice-Hall, Inc., 1994.
\newblock \href {https://doi.org/10.1109/T-C.1969.222718}
  {\path{doi:10.1109/T-C.1969.222718}}.

\bibitem{Larose2004}
D.~T. Larose, {Discovering knowledge in data : an introduction to data mining},
  Wiley, New Jersey, 2004.

\bibitem{Angus2006}
A.~J. Wilkinson, G.~Meaden, D.~J. Dingley,
  \href{https://www.sciencedirect.com/science/article/pii/S0304399105002251}{High-resolution
  elastic strain measurement from electron backscatter diffraction patterns:
  New levels of sensitivity}, Ultramicroscopy 106~(4) (2006) 307--313.
\newblock \href
  {https://doi.org/https://doi.org/10.1016/j.ultramic.2005.10.001}
  {\path{doi:https://doi.org/10.1016/j.ultramic.2005.10.001}}.
\newline\urlprefix\url{https://www.sciencedirect.com/science/article/pii/S0304399105002251}

\bibitem{Angus2010}
A.~J. Wilkinson, D.~Randman,
  \href{https://doi.org/10.1080/14786430903304145}{Determination of elastic
  strain fields and geometrically necessary dislocation distributions near
  nanoindents using electron back scatter diffraction}, Philosophical Magazine
  90~(9) (2010) 1159--1177.
\newblock \href
  {http://arxiv.org/abs/https://doi.org/10.1080/14786430903304145}
  {\path{arXiv:https://doi.org/10.1080/14786430903304145}}, \href
  {https://doi.org/10.1080/14786430903304145}
  {\path{doi:10.1080/14786430903304145}}.
\newline\urlprefix\url{https://doi.org/10.1080/14786430903304145}

\bibitem{Dudamell2013}
N.~V. Dudamell, P.~Hidalgo-Manrique, A.~Chakkedath, Z.~Chen, C.~J. Boehlert,
  F.~G{\'{a}}lvez, S.~Yi, J.~Bohlen, D.~Letzig, M.~T. P{\'{e}}rez-Prado,
  {Influence of strain rate on the twin and slip activity of a magnesium alloy
  containing neodymium}, Mater. Sci. Eng. A 583 (2013) 220--231.
\newblock \href {https://doi.org/10.1016/j.msea.2013.07.003}
  {\path{doi:10.1016/j.msea.2013.07.003}}.

\bibitem{Zhou2021a}
P.~Zhou, G.~Z. Zhu, {Strain accommodations among twin variants in Ti and Mg},
  Crystals 11~(453) (2021).
\newblock \href {https://doi.org/10.3390/cryst11040453}
  {\path{doi:10.3390/cryst11040453}}.

\bibitem{Zhang2021}
M.~Zhang, H.~Zhang, A.~Ma, J.~Llorca, {Experimental and numerical analysis of
  cyclic deformation and fatigue behavior of a Mg-RE alloy}, Int. J. Plast. 139
  (2021) 1--31.
\newblock \href {http://arxiv.org/abs/2010.02774} {\path{arXiv:2010.02774}},
  \href {https://doi.org/10.1016/j.ijplas.2020.102885}
  {\path{doi:10.1016/j.ijplas.2020.102885}}.

\bibitem{Hawick2010}
K.~A. Hawick, D.~P. Playne, {Automated and Parallel Code Generation for
  Finite-Differencing Stencils with Arbitrary Data Types}, Procedia Comput.
  Sci. 00 (2010) 1--9.

\bibitem{panzarino2015quantitative}
J.~F. Panzarino, J.~J. Ramos, T.~J. Rupert, Quantitative tracking of grain
  structure evolution in a nanocrystalline metal during cyclic loading,
  Modelling and Simulation in Materials Science and Engineering 23~(2) (2015)
  025005.

\bibitem{priedeman2018quantifying}
J.~L. Priedeman, C.~W. Rosenbrock, O.~K. Johnson, E.~R. Homer, Quantifying and
  connecting atomic and crystallographic grain boundary structure using local
  environment representation and dimensionality reduction techniques, Acta
  Materialia 161 (2018) 431--443.

\bibitem{Xie2021b}
K.~Y. Xie, K.~Hazeli, N.~Dixit, L.~Ma, K.~T. Ramesh, K.~J. Hemker, {Twin
  boundary migration mechanisms in quasi-statically compressed and
  plate-impacted mg single crystals}, Sci. Adv. 7~(42) (2021) 1--7.
\newblock \href {https://doi.org/10.1126/sciadv.abg3443}
  {\path{doi:10.1126/sciadv.abg3443}}.

\bibitem{Jiang2022a}
L.~Jiang, M.~Gong, J.~Wang, Z.~Pan, X.~Wang, D.~Zhang, Y.~M. Wang, J.~Ciston,
  A.~M. Minor, M.~Xu, X.~Pan, T.~J. Rupert, S.~Mahajan, E.~J. Lavernia, I.~J.
  Beyerlein, J.~M. Schoenung, {Visualization and validation of twin nucleation
  and early-stage growth in magnesium}, Nat. Commun. 13~(1) (2022) 1--11.
\newblock \href {https://doi.org/10.1038/s41467-021-27591-z}
  {\path{doi:10.1038/s41467-021-27591-z}}.

\bibitem{he2020direct}
Y.~He, B.~Li, C.~Wang, S.~X. Mao, Direct observation of dual-step twinning
  nucleation in hexagonal close-packed crystals, Nature communications 11~(1)
  (2020) 2483.

\bibitem{barrett2014roles}
C.~D. Barrett, H.~El~Kadiri, The roles of grain boundary dislocations and
  disclinations in the nucleation of $\{$1 0 1 2$\}$ twinning, Acta materialia
  63 (2014) 1--15.

\bibitem{ostapovets2014characterization}
A.~Ostapovets, A.~Serra, Characterization of the matrix--twin interface of a
  (101-2) twin during growth, Philosophical Magazine 94~(25) (2014) 2827--2839.

\bibitem{liu2014twinning}
B.-Y. Liu, J.~Wang, B.~Li, L.~Lu, X.-Y. Zhang, Z.-W. Shan, J.~Li, C.-L. Jia,
  J.~Sun, E.~Ma, Twinning-like lattice reorientation without a crystallographic
  twinning plane, Nature communications 5~(1) (2014) 1--6.

\bibitem{Zhu2018}
Q.~Zhu, A.~Samanta, B.~Li, R.~E. Rudd, T.~Frolov,
  \href{http://dx.doi.org/10.1038/s41467-018-02937-2
  http://www.nature.com/articles/s41467-018-02937-2}{{Predicting phase behavior
  of grain boundaries with evolutionary search and machine learning}}, Nat.
  Commun. 9~(1) (2018) 467.
\newblock \href {http://arxiv.org/abs/1707.09699} {\path{arXiv:1707.09699}},
  \href {https://doi.org/10.1038/s41467-018-02937-2}
  {\path{doi:10.1038/s41467-018-02937-2}}.
\newline\urlprefix\url{http://dx.doi.org/10.1038/s41467-018-02937-2
  http://www.nature.com/articles/s41467-018-02937-2}

\bibitem{Frolov2012a}
T.~Frolov, Y.~Mishin, {Thermodynamics of coherent interfaces under mechanical
  stresses. I. Theory}, Phys. Rev. B - Condens. Matter Mater. Phys. 85~(22)
  (2012) 12--15.
\newblock \href {http://arxiv.org/abs/1304.0144} {\path{arXiv:1304.0144}},
  \href {https://doi.org/10.1103/PhysRevB.85.224106}
  {\path{doi:10.1103/PhysRevB.85.224106}}.

\bibitem{Valueva2020}
M.~V. Valueva, N.~N. Nagornov, P.~A. Lyakhov, G.~V. Valuev, N.~I. Chervyakov,
  \href{https://doi.org/10.1016/j.matcom.2020.04.031}{{Application of the
  residue number system to reduce hardware costs of the convolutional neural
  network implementation}}, Math. Comput. Simul. 177 (2020) 232--243.
\newblock \href {https://doi.org/10.1016/j.matcom.2020.04.031}
  {\path{doi:10.1016/j.matcom.2020.04.031}}.
\newline\urlprefix\url{https://doi.org/10.1016/j.matcom.2020.04.031}

\bibitem{Alsobhani2021}
A.~Alsobhani, H.~M. Alabboodi, H.~Mahdi, {Speech Recognition using Convolution
  Deep Neural Networks}, J. Phys. Conf. Ser. 1973~(1) (2021) 0--10.
\newblock \href {https://doi.org/10.1088/1742-6596/1973/1/012166}
  {\path{doi:10.1088/1742-6596/1973/1/012166}}.

\bibitem{Liu2019b}
Y.~Liu, P.~Z. Tang, M.~Y. Gong, R.~J. McCabe, J.~Wang, C.~N. Tom{\'{e}},
  \href{http://dx.doi.org/10.1038/s41467-019-10573-7}{{Three-dimensional
  character of the deformation twin in magnesium}}, Nat. Commun. 10~(1) (2019)
  1--7.
\newblock \href {https://doi.org/10.1038/s41467-019-10573-7}
  {\path{doi:10.1038/s41467-019-10573-7}}.
\newline\urlprefix\url{http://dx.doi.org/10.1038/s41467-019-10573-7}

\bibitem{liu2016characterizing}
Y.~Liu, N.~Li, S.~Shao, M.~Gong, J.~Wang, R.~McCabe, Y.~Jiang, C.~Tom{\'e},
  Characterizing the boundary lateral to the shear direction of deformation
  twins in magnesium, Nature communications 7~(1) (2016) 1--6.

\bibitem{wang2020characteristic}
S.~Wang, M.~Gong, R.~J. McCabe, L.~Capolungo, J.~Wang, C.~N. Tom{\'e},
  Characteristic boundaries associated with three-dimensional twins in
  hexagonal metals, Science advances 6~(28) (2020) eaaz2600.

\bibitem{christian1995deformation}
J.~W. Christian, S.~Mahajan, Deformation twinning, Progress in materials
  science 39~(1-2) (1995) 1--157.

\bibitem{gong2018structural}
M.~Gong, S.~Xu, Y.~Jiang, Y.~Liu, J.~Wang, Structural characteristics of
  $\{$1{\={}} 012$\}$ non-cozone twin-twin interactions in magnesium, Acta
  Materialia 159 (2018) 65--76.

\bibitem{kumar2019role}
M.~A. Kumar, M.~Gong, I.~Beyerlein, J.~Wang, C.~N. Tom{\'e}, Role of local
  stresses on co-zone twin-twin junction formation in hcp magnesium, Acta
  Materialia 168 (2019) 353--361.

\bibitem{gong2020atomic}
M.~Gong, W.~Wu, Atomic-level study of twin--twin interactions in hexagonal
  metals, Journal of Materials Research 35~(13) (2020) 1647--1659.

\bibitem{chen2019dislocation}
P.~Chen, F.~Wang, B.~Li, Dislocation absorption and transmutation at $\{$101
  2$\}$ twin boundaries in deformation of magnesium, Acta Materialia 164 (2019)
  440--453.

\bibitem{hirth2016disconnections}
J.~Hirth, J.~Wang, C.~Tom{\'e}, Disconnections and other defects associated
  with twin interfaces, Progress in Materials Science 83 (2016) 417--471.

\bibitem{mackain2017atomic}
O.~MacKain, M.~Cottura, D.~Rodney, E.~Clouet, Atomic-scale modeling of twinning
  disconnections in zirconium, Physical Review B 95~(13) (2017) 134102.

\end{thebibliography}

}
\renewcommand{\thefigure}{S\arabic{figure}}
\renewcommand{\thetable}{S\arabic{table}}  
\setcounter{figure}{0} 
\newpage
\section{Supplementary material}
\subsection{Evolution of the atomic stress}

\begin{figure}[h!]
\centering
    \includegraphics{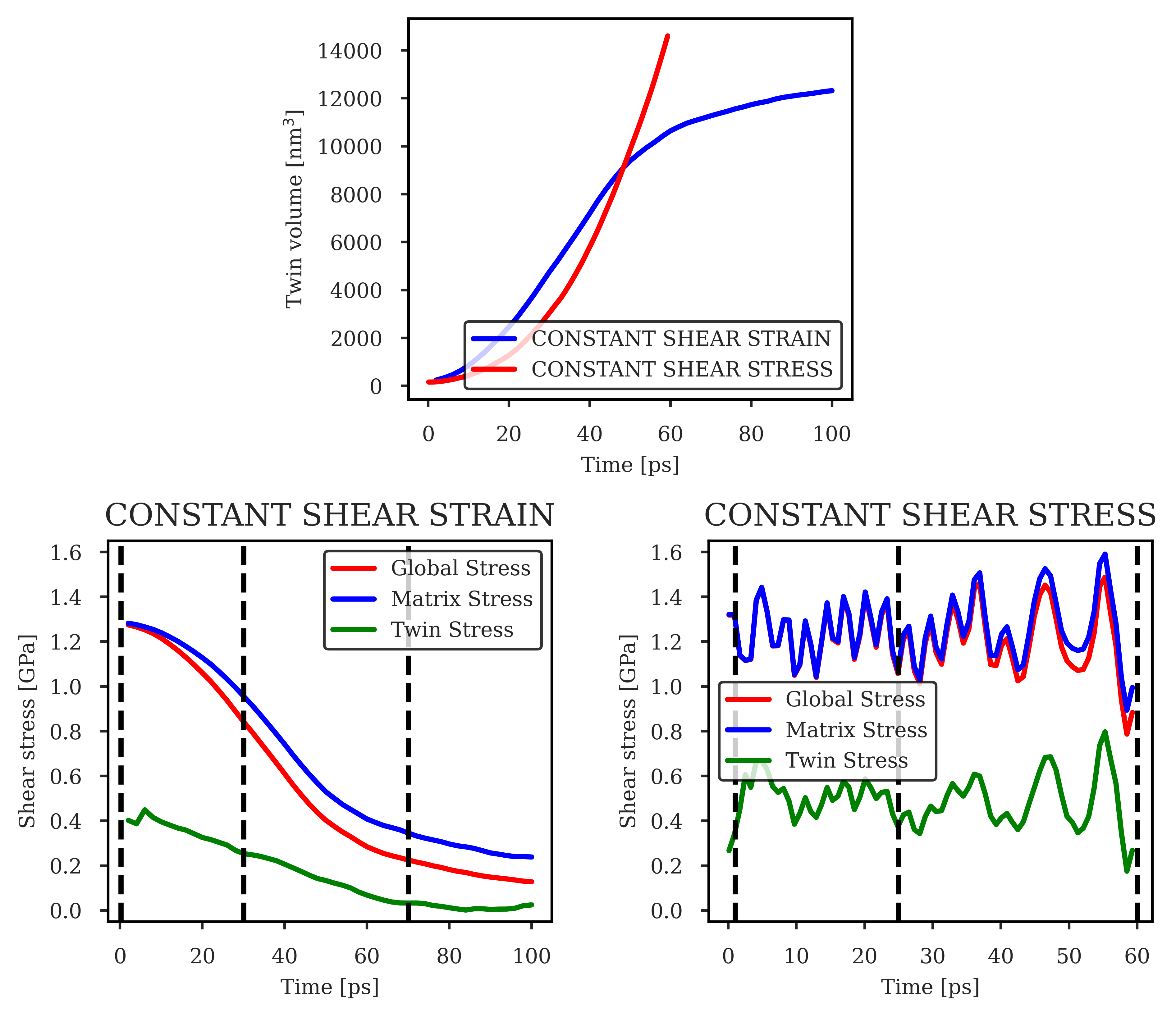}
\caption{\label{fig:simulationbox2} \label{fig:stressvstime} (a) Evolution of the twin volume in the simulation at constant shear strain (blue) and shear stress (red). Evolution of the average YZ shear stress in the Mg-10at.\%Al samples with the initial stress of 1.2 GPa under (b) constant strain and (c) constant shear stress conditions. The three lines represent the volume-averaged shear stress for all atoms (global stress), for matrix atoms (matrix stress), and for twin atoms (twin stress). During twin growth, all stresses are substantially reduced under constant strain conditions while remaining constant under constant stress conditions with a standard deviation of 0.15 GPa. The vertical dashed lines correspond to the snapshots shown in Figure 1 in the main text. }
\end{figure}

\newpage

\subsection{Atomic stress field, $S$, as given by LAMMPS}

\begin{figure}[h!]
\centering
  \vspace{-0.4cm}
  \begin{center}
      \textbf{Constant Strain:}
  \end{center}
  \vspace{-0.4cm}
    \includegraphics{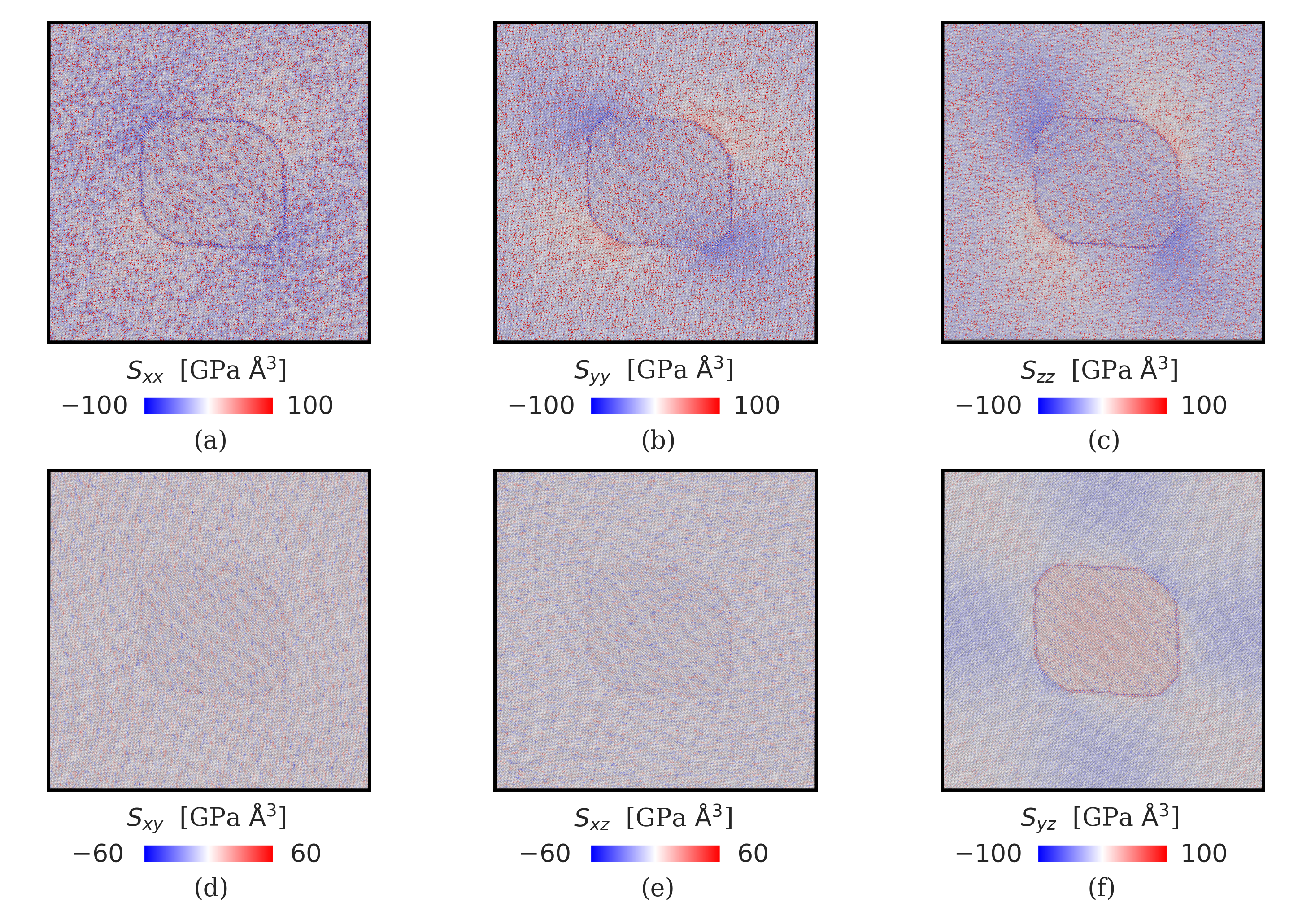}
  \begin{center}
  \vspace{-0.6cm}
      \textbf{Constant Stress:}
  \end{center}
  \vspace{-0.33cm}
    \includegraphics{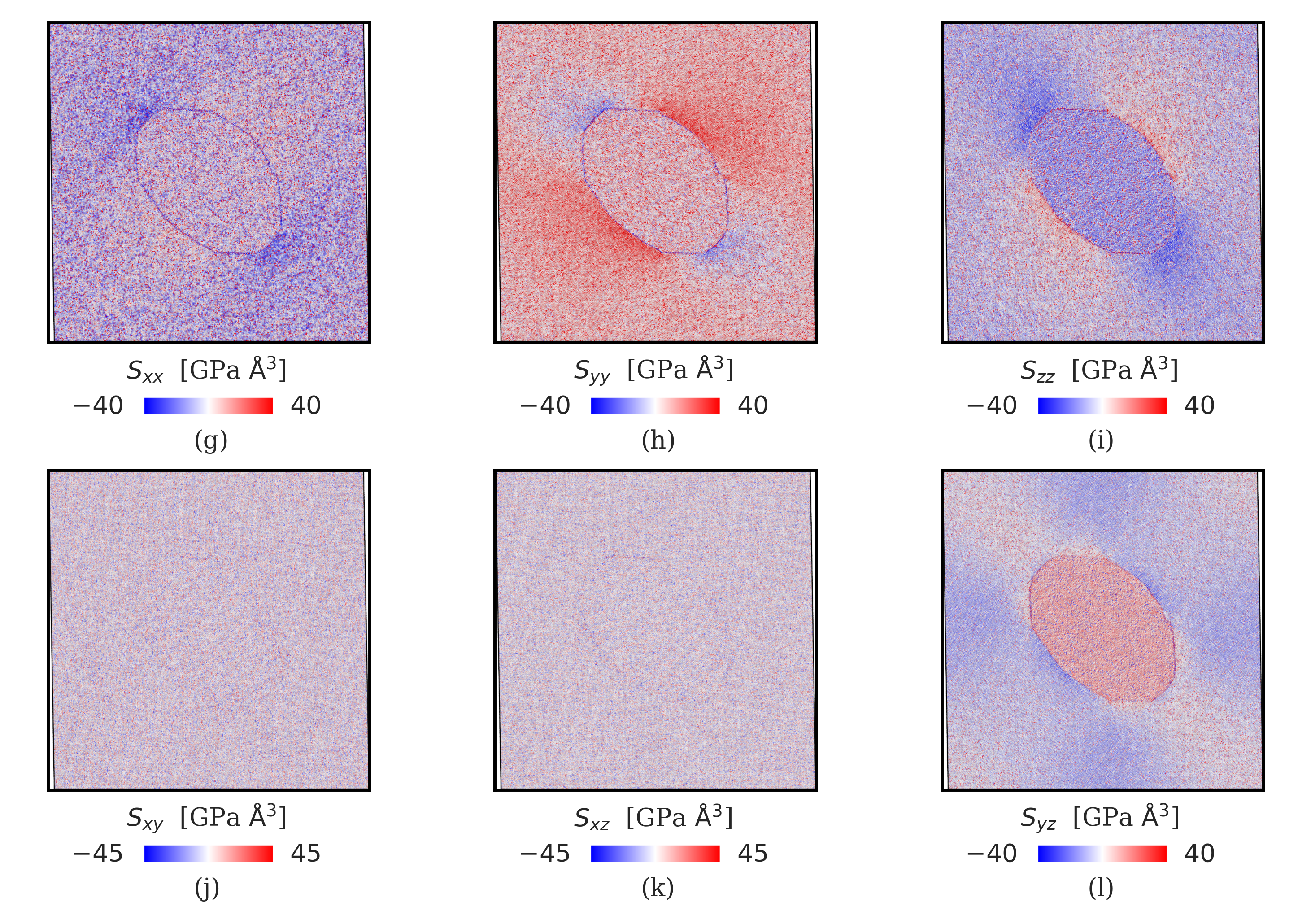}
\caption{\label{fig:simulationbox2} Distribution of the raw stress $S$ as given by LAMMPS i.e., not volume-averaged. }
\end{figure}

\newpage
\subsection{Dependence on the grid bin size}

\begin{figure}[h!]
\centering
  
\includegraphics{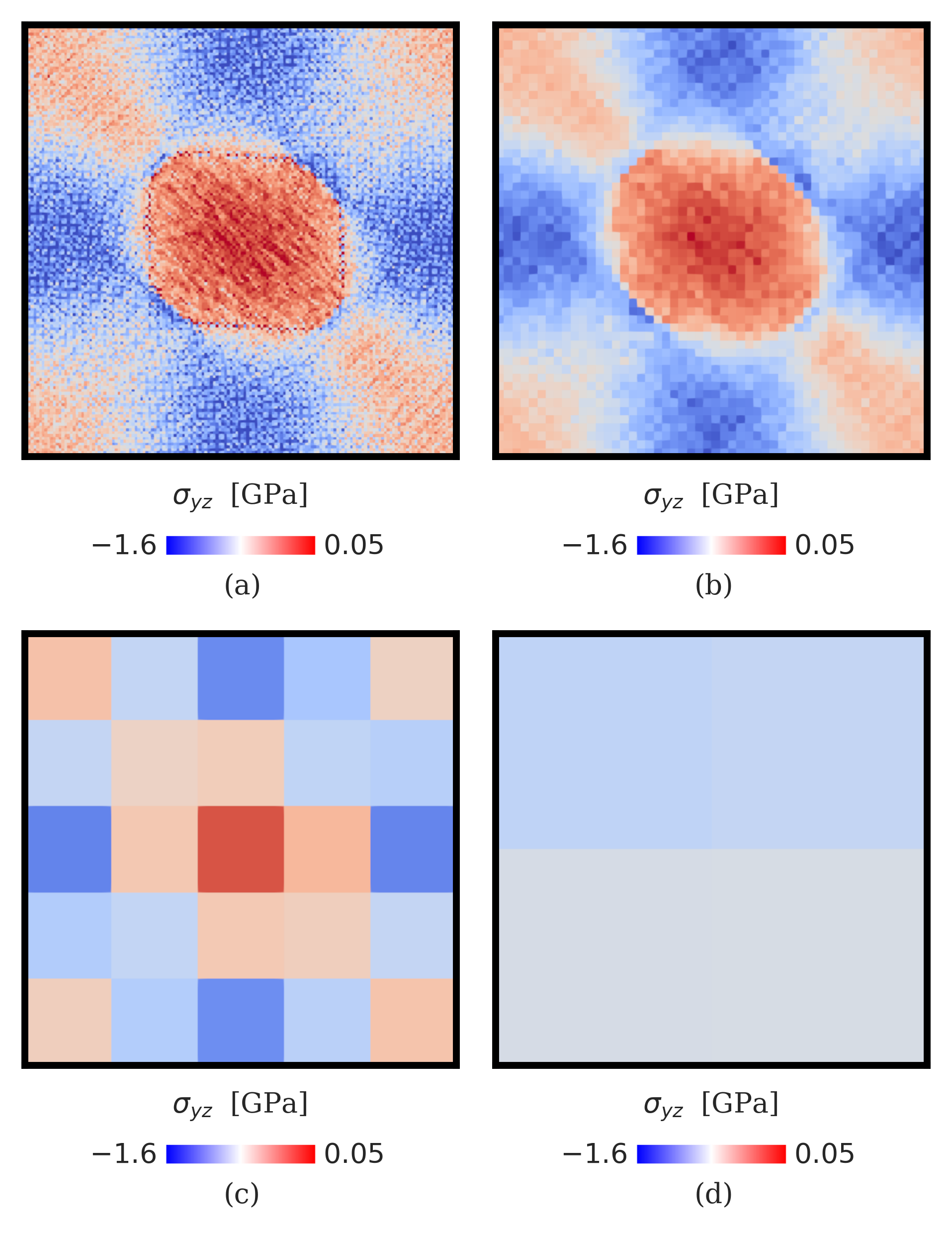}
\caption{\label{fig:stderrorstress} Dependence of the stress field tensor  $\sigma$ with the bin size, $dh$: (a) $dh=0.5$ nm, (b) $dh=1.5$ nm, (c) $dh=25$ nm and (d) $dh=70$ nm.}
\end{figure}

\newpage

\subsection{Volume-averaged stress field $\sigma$}

\begin{figure}[h!]
\centering
  \vspace{-0.4cm}
  \begin{center}
      \textbf{Constant Strain:}
  \end{center}
  \vspace{-0.4cm}
    \includegraphics{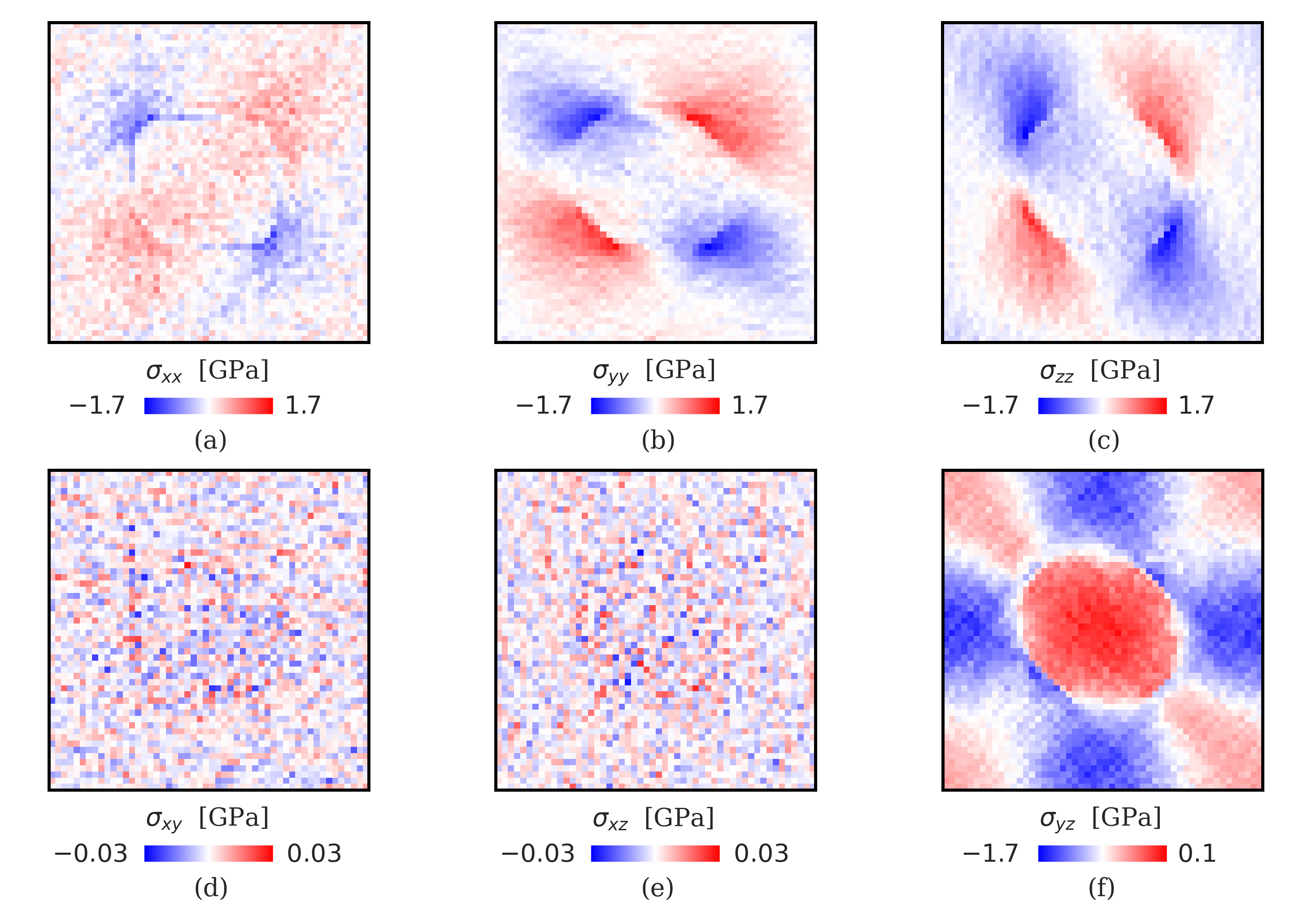}
  \begin{center}
  \vspace{-0.6cm}
      \textbf{Constant Stress:}
  \end{center}
  \vspace{-0.33cm}
    \includegraphics{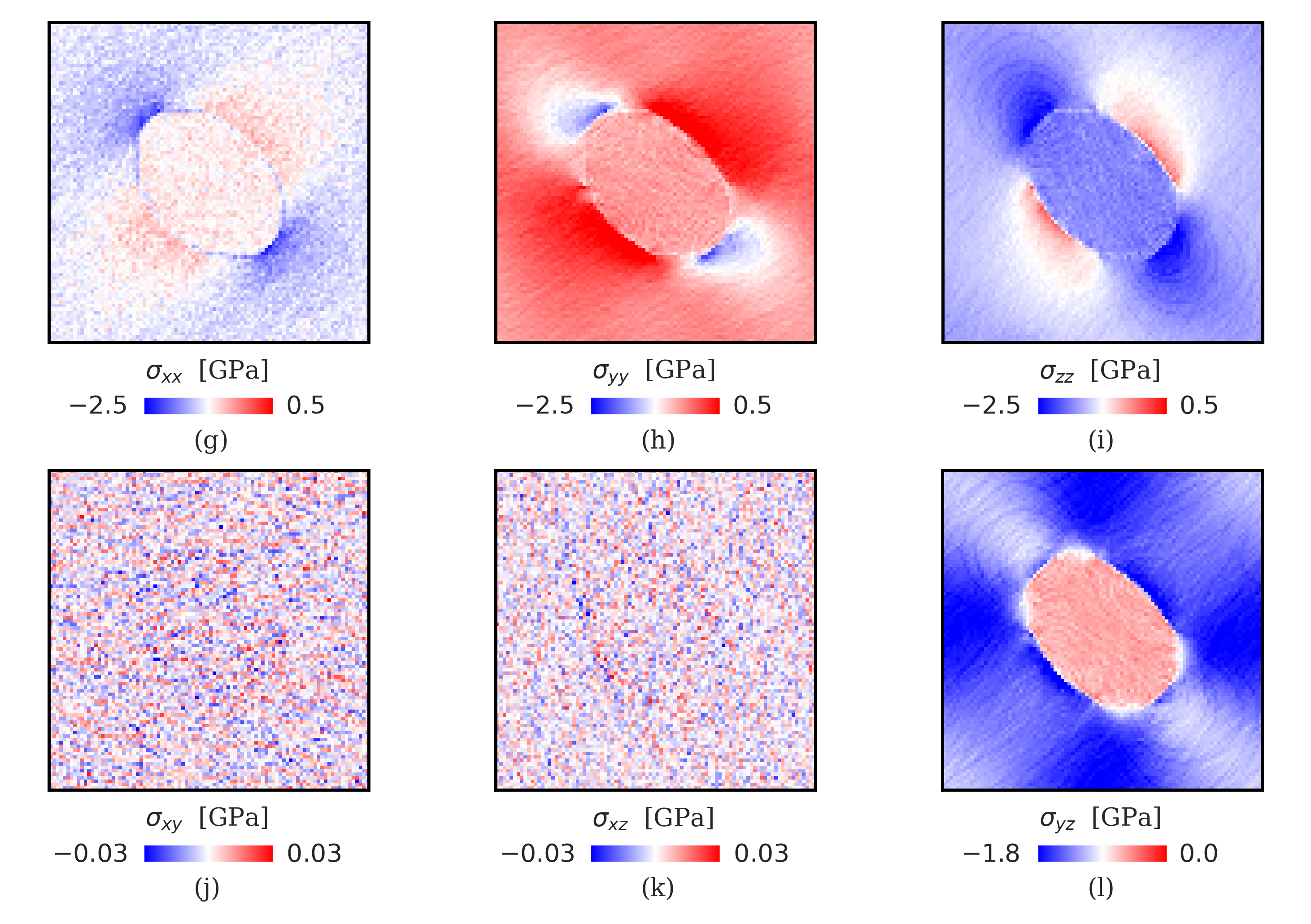}
\caption{\label{fig:simulationbox2} Values of the operator $f_0$ acting on the components of the stress field tensor  $\sigma$. }
\end{figure}

\newpage

\subsection{Operators applied to the stress field}
For the bin described by the indices $(i,j)$, the following operators on an arbitrary field $g$ are defined below.

Deviation from the mean:
\begin{equation}
f_0[g(i,j)]=g(i,j)-\langle g(i,j)\rangle,
\end{equation}

Module gradient:
\begin{equation}
f_1[g(i,j)]=\frac{\sqrt{\left(g(i+1,j)-g(i-1,j)\right)^2+\left(g(i,j+1)-g(i,j-1)\right)^2}}{2dh},
\end{equation}

Bi-harmonic operator:
\begin{equation}
  \begin{split}
f_2[g(i,j)]=[2g(i+1,j+1)+2g(i-1,j-1)+2g(i+1,j-1)\\
+2g(i-1,j+1)-8g(i+1,j)-8g(i,j+1)\\
-8g(i-1,j)-8g(i,j-1)+g(i+2,j)\\
+g(i,j+2)+g(i-2,j)+g(i,j-2)\\+20g(i,j)]/\left(120dh^2\right).
  \end{split}
  \label{eq:operator2}
\end{equation}

\newpage

\subsection{Operator $f_0$ on the stress field $\sigma$}

\begin{figure}[h!]
\centering
  \vspace{-0.4cm}
  \begin{center}
      \textbf{Constant Strain:}
  \end{center}
  \vspace{-0.4cm}
    \includegraphics{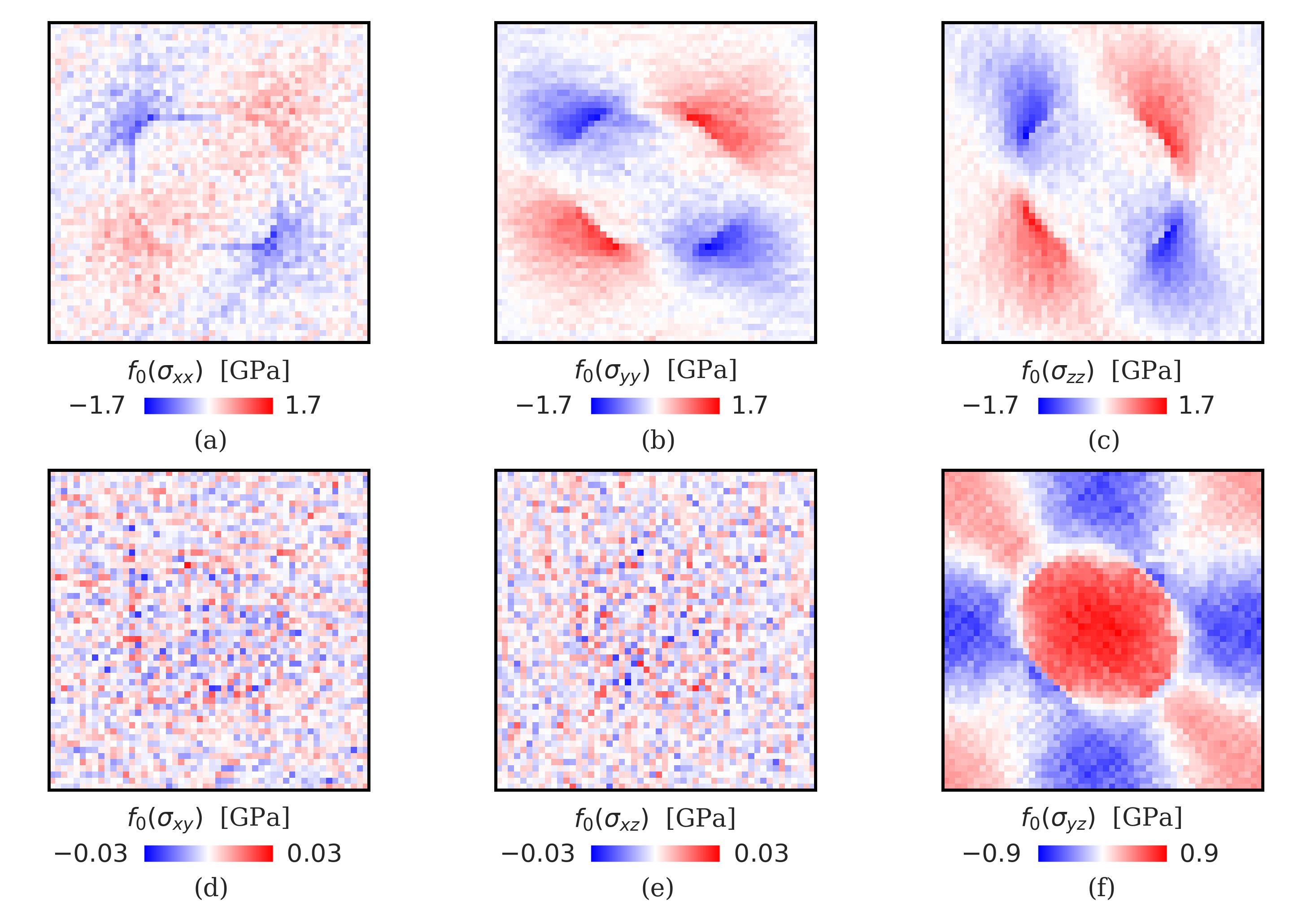}
  \begin{center}
  \vspace{-0.6cm}
      \textbf{Constant Stress:}
  \end{center}
  \vspace{-0.33cm}
    \includegraphics{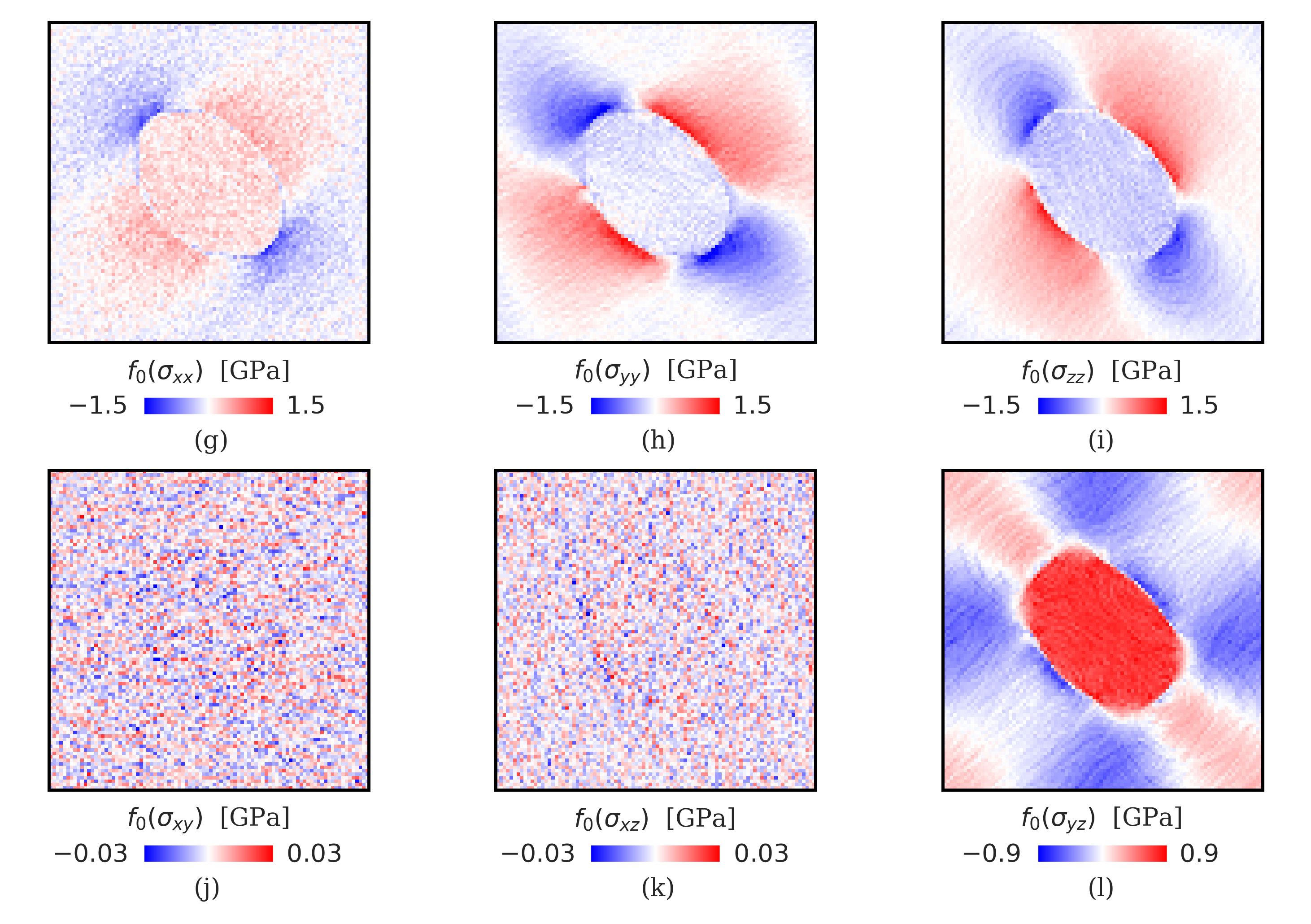}
\caption{\label{fig:simulationbox2} Values of the operator $f_0$ acting on the components of the stress field tensor  $\sigma$. }
\end{figure}

\newpage

\subsection{Operator $f_1$ on the stress field $\sigma$}

\begin{figure}[h!]
\centering
  \vspace{-0.4cm}
  \begin{center}
      \textbf{Constant Strain:}
  \end{center}
  \vspace{-0.4cm}
    \includegraphics{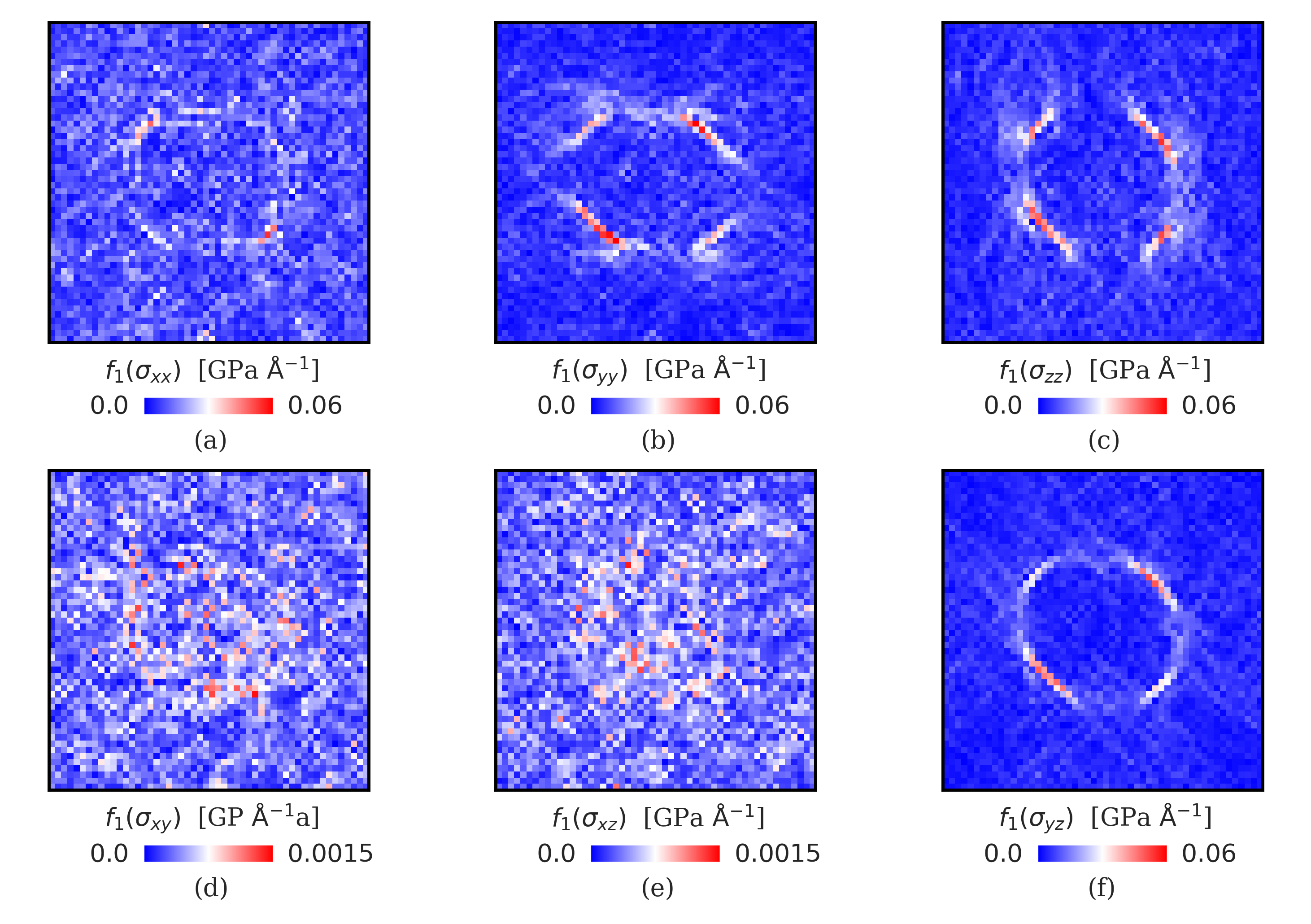}
  \begin{center}
  \vspace{-0.6cm}
      \textbf{Constant Stress:}
  \end{center}
  \vspace{-0.33cm}
    \includegraphics{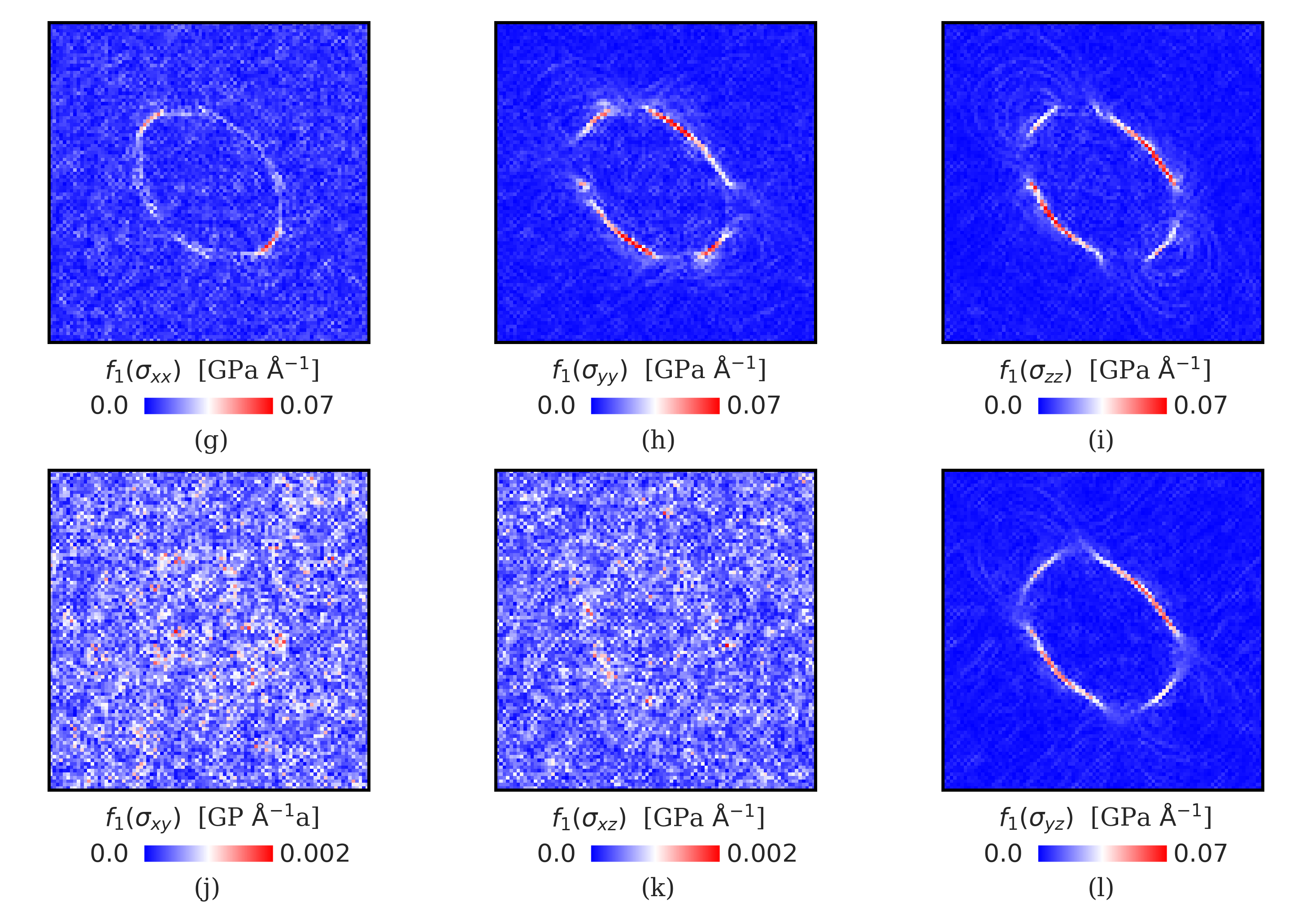}
\caption{\label{fig:simulationbox2} Values of the operator $f_1$ acting on the components of the stress field tensor  $\sigma$. }
\end{figure}

\newpage
\subsection{Operator $f_2$ on the stress field $\sigma$}

\begin{figure}[h!]
\centering
  \vspace{-0.4cm}
  \begin{center}
      \textbf{Constant Strain:}
  \end{center}
  \vspace{-0.4cm}
    \includegraphics{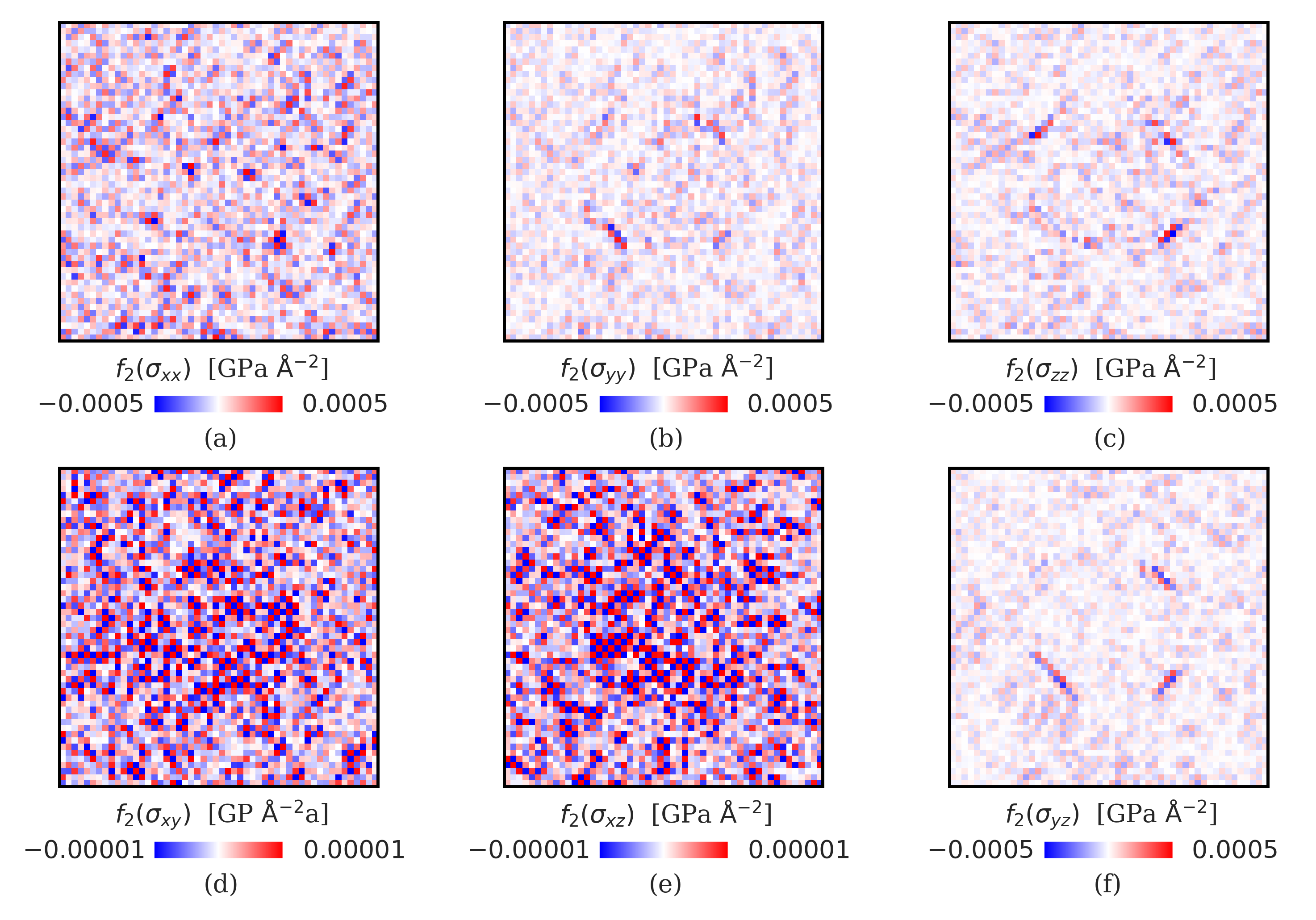}
  \begin{center}
  \vspace{-0.6cm}
      \textbf{Constant Stress:}
  \end{center}
  \vspace{-0.33cm}
    \includegraphics{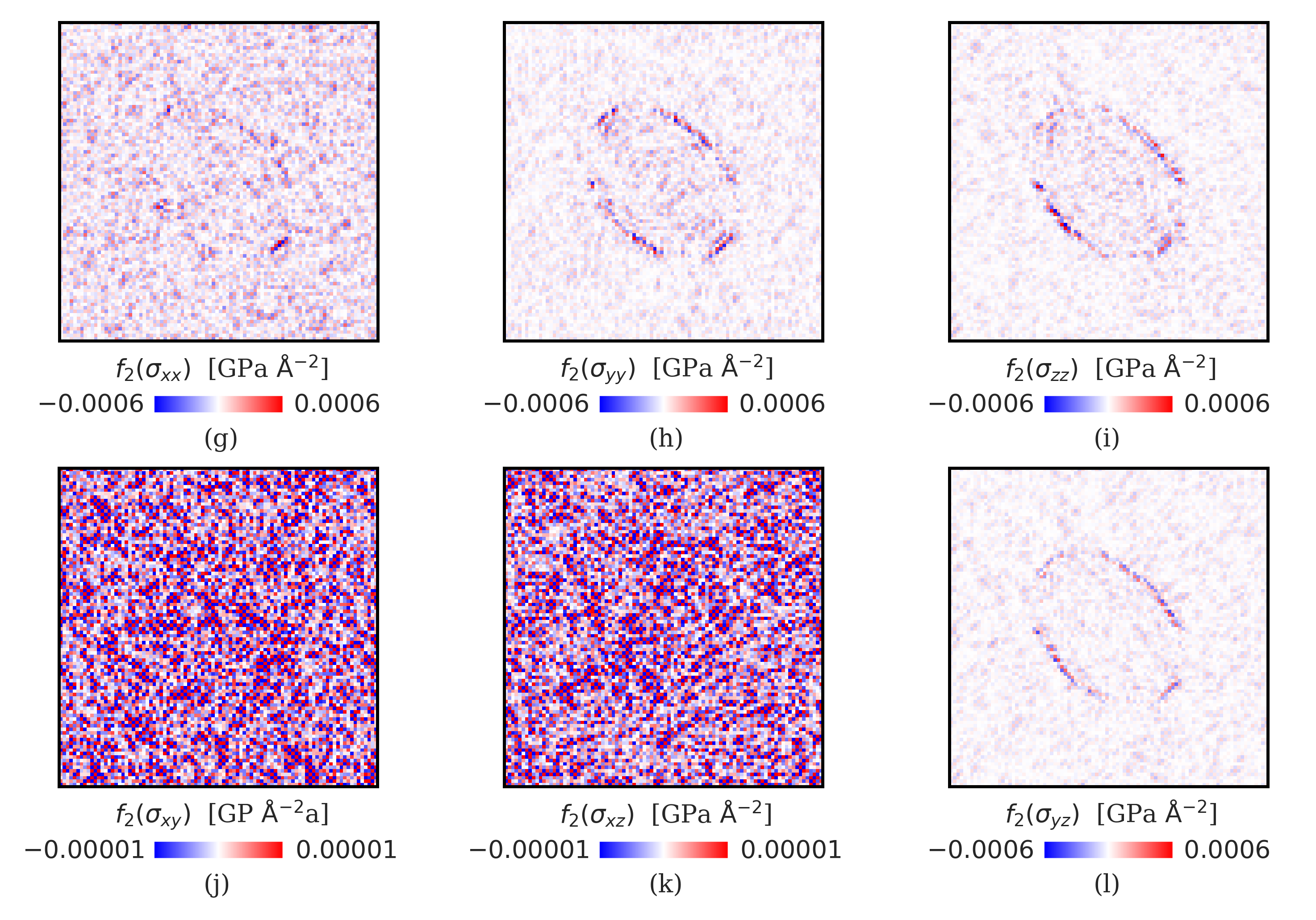}
\caption{\label{fig:simulationbox2} Values of the operator $f_2$ acting on the components of the stress field tensor  $\sigma$. }
\end{figure}

\newpage
\subsection{Different sets of input features used in the present work.}

\begin{table}[h!]
\begin{center}
\caption{Sets of input features used for training supervised NNs with the corresponding accuracy for validation set that is used to avoid overfitting. }
\label{table:inputstructures} 
\begin{tabular}{ccrc} 
\hline
Ref. & Input Features & Accuracy [\%]\\
\hline
 \#1&  $f_0[\sigma_{xx}],f_0[\sigma_{yy}],f_0[\sigma_{zz}],f_0[\sigma_{yz}]$&97.0\\
\cdashline{1-4}
 &  $f_0[\sigma_{xx}],f_0[\sigma_{yy}],f_0[\sigma_{zz}],f_0[\sigma_{yz}]$\\
 \#2&  $f_1[\sigma_{xx}],f_1[\sigma_{yy}],f_1[\sigma_{zz}],f_1[\sigma_{yz}]$ &98.0\\
\cdashline{1-4}
 &  $f_1[\sigma_{xx}],f_1[\sigma_{yy}],f_1[\sigma_{zz}],f_1[\sigma_{yz}]$\\
 \#3&$f_2[\sigma_{xx}],f_2[\sigma_{yy}],f_2[\sigma_{zz}],f_2[\sigma_{yz}]$ &97.3\\
\cdashline{1-4}
 &  $f_0[\sigma_{xx}],f_0[\sigma_{yy}],f_0[\sigma_{zz}],f_0[\sigma_{yz}]$\\
 \#4&  $f_1[\sigma_{xx}],f_1[\sigma_{yy}],f_1[\sigma_{zz}],f_1[\sigma_{yz}]$ &98.0\\
 &$f_2[\sigma_{xx}],f_2[\sigma_{yy}],f_2[\sigma_{zz}],f_2[\sigma_{yz}]$\\
\hline
 \#5&  $f_0[\sigma_{xx}],f_0[\sigma_{yy}],f_0[\sigma_{zz}],f_0[\sigma_{xy}],f_0[\sigma_{xz}],f_0[\sigma_{yz}]$&96.9\\
\cdashline{1-4}
 &  $f_0[\sigma_{xx}],f_0[\sigma_{yy}],f_0[\sigma_{zz}],f_0[\sigma_{xy}],f_0[\sigma_{xz}],f_0[\sigma_{yz}]$\\
 \#6&  $f_1[\sigma_{xx}],f_1[\sigma_{yy}],f_1[\sigma_{zz}],f_1[\sigma_{xy}],f_1[\sigma_{xz}],f_1[\sigma_{yz}]$ &97.9\\
\cdashline{1-4}
 &  $f_1[\sigma_{xx}],f_1[\sigma_{yy}],f_1[\sigma_{zz}],f_1[\sigma_{xy}],f_1[\sigma_{xz}],f_1[\sigma_{yz}]$\\
 \#7&$f_2[\sigma_{xx}],f_2[\sigma_{yy}],f_2[\sigma_{zz}],f_2[\sigma_{xy}],f_2[\sigma_{xz}],f_2[\sigma_{yz}]$ &97.3\\
\cdashline{1-4}
 &  $f_0[\sigma_{xx}],f_0[\sigma_{yy}],f_0[\sigma_{zz}],f_0[\sigma_{xy}],f_0[\sigma_{xz}],f_0[\sigma_{yz}]$\\
 \#8&  $f_1[\sigma_{xx}],f_1[\sigma_{yy}],f_1[\sigma_{zz}],f_1[\sigma_{xy}],f_1[\sigma_{xz}],f_1[\sigma_{yz}]$ &98.2\\
 &$f_2[\sigma_{xx}],f_2[\sigma_{yy}],f_2[\sigma_{zz}],f_2[\sigma_{xy}],f_2[\sigma_{xz}],f_2[\sigma_{yz}]$\\
\hline
\end{tabular}
\end{center}
\end{table}

\newpage
\subsection{Accuracy in the sample at constant strain for different input features.}

\begin{figure}[h!]

  \hspace*{\fill}   
  \begin{subfigure}{0.35\textwidth}
    \includegraphics[width=\linewidth]{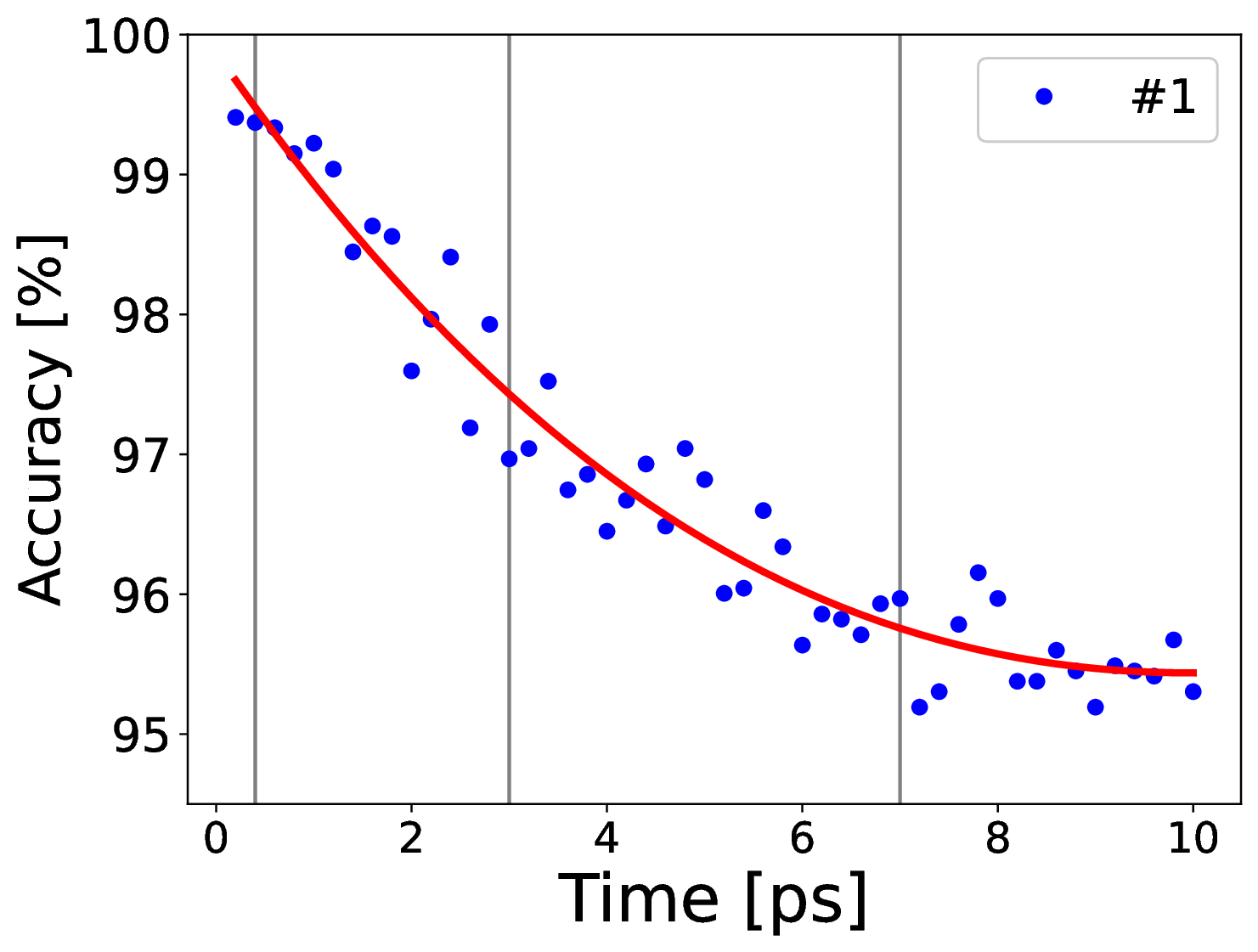}
    \caption{} 
  \end{subfigure}%
  \hspace*{\fill}   
  \begin{subfigure}{0.35\textwidth}
    \includegraphics[width=\linewidth]{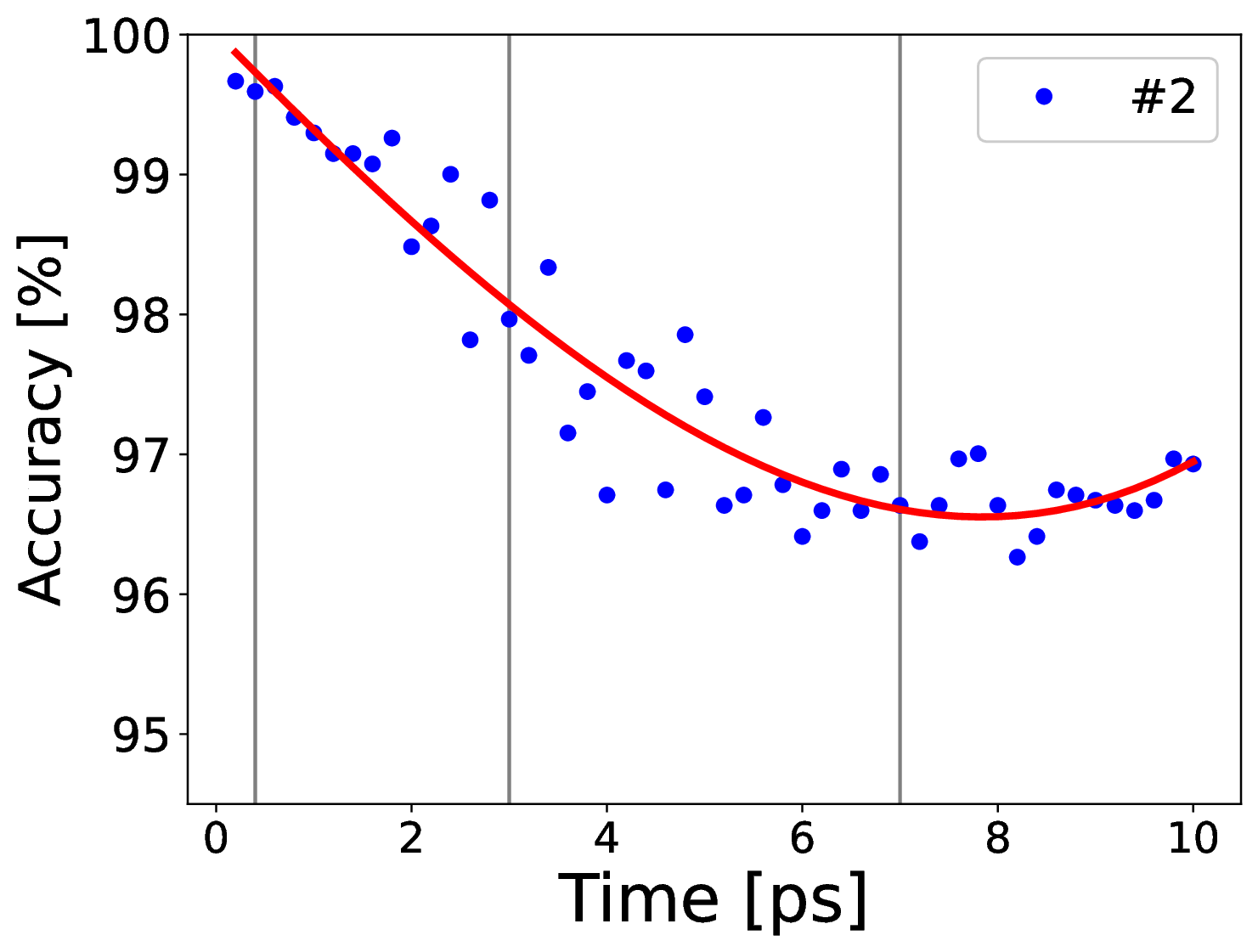}
    \caption{} 
  \end{subfigure}%
  \hspace*{\fill}   
  
  \hspace*{\fill}   
  \begin{subfigure}{0.35\textwidth}
    \includegraphics[width=\linewidth]{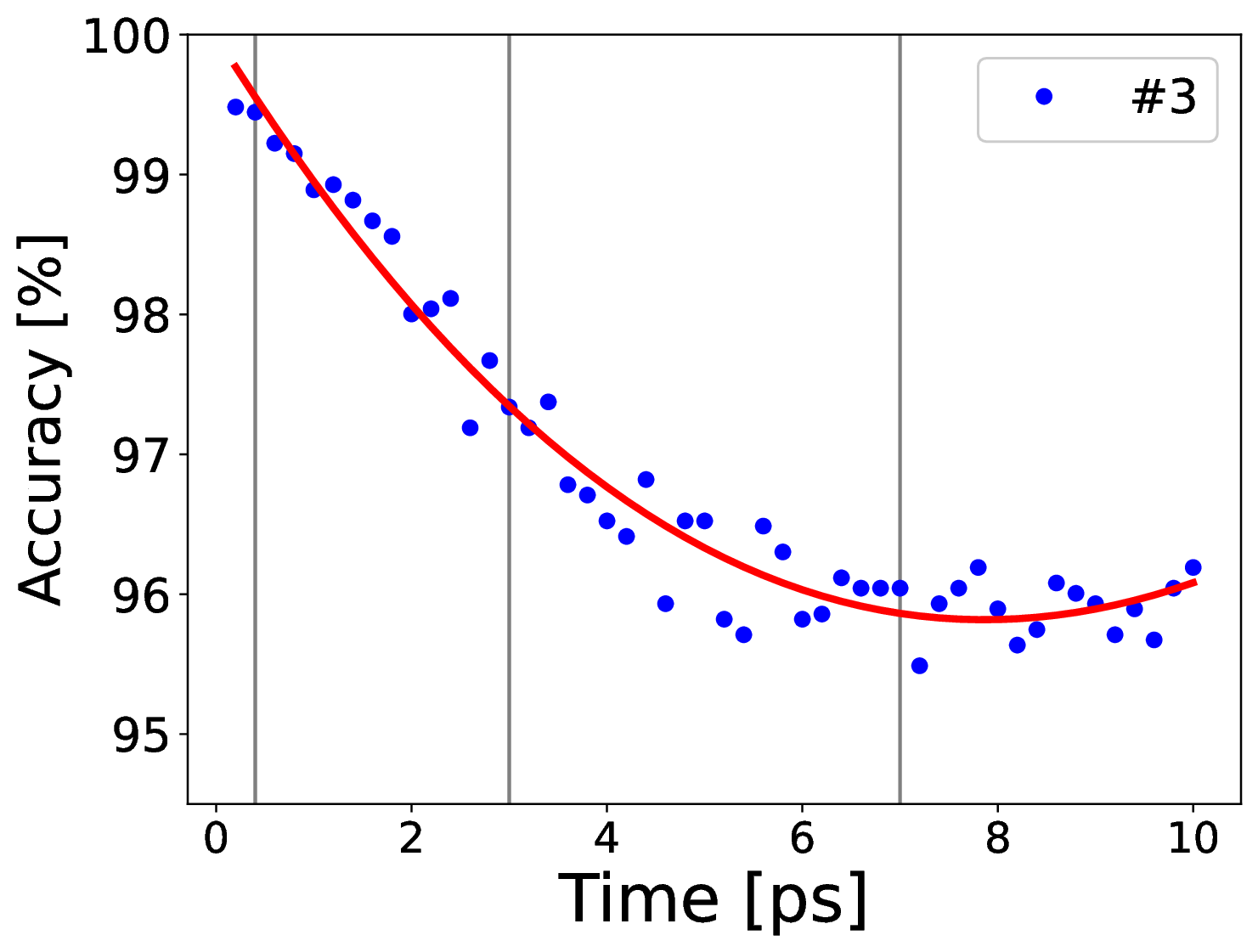}
    \caption{}
  \end{subfigure}%
  \hspace*{\fill}   
  \begin{subfigure}{0.35\textwidth}
    \includegraphics[width=\linewidth]{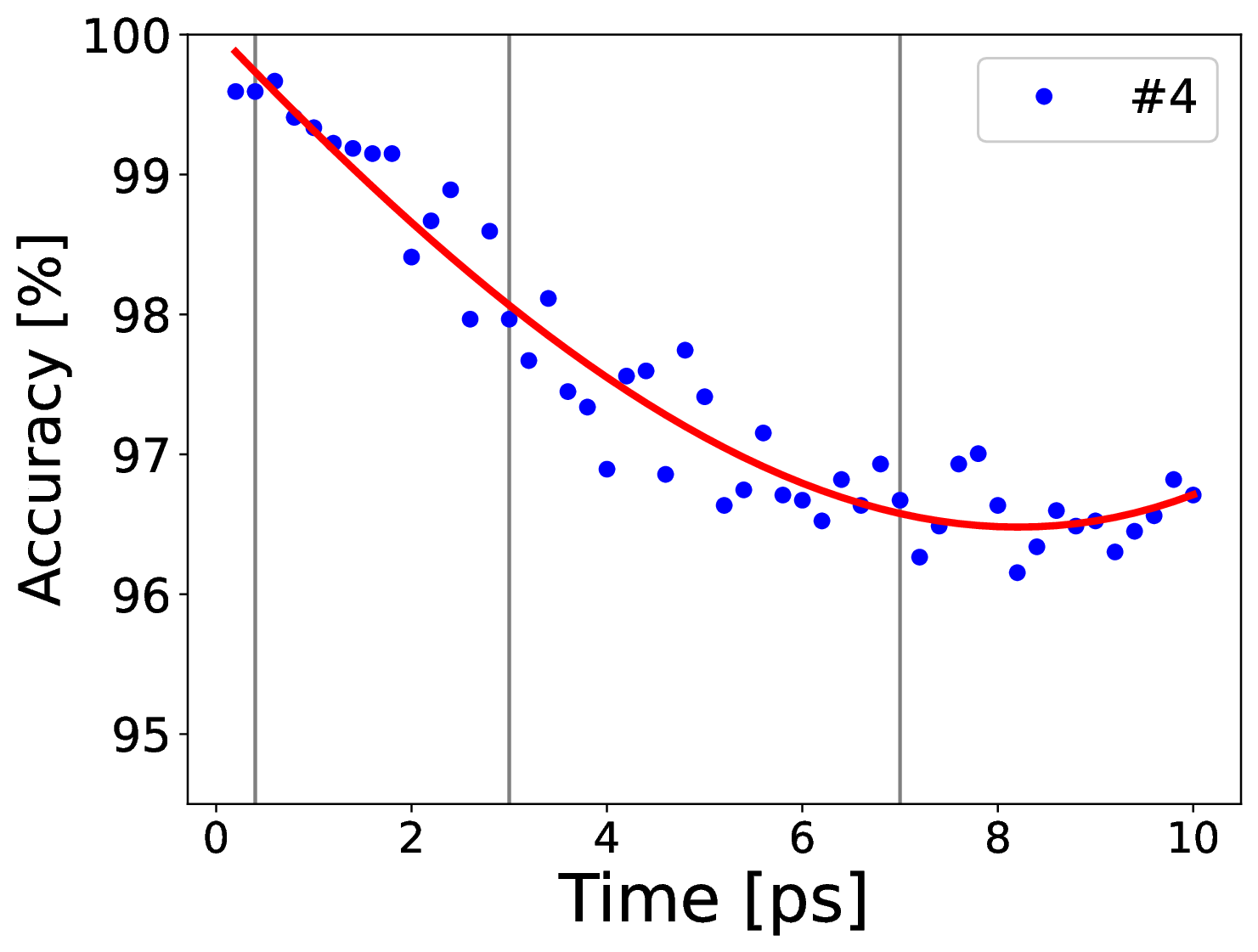}
    \caption{} 
  \end{subfigure}%
  \hspace*{\fill}   
  
  \hspace*{\fill}   
  \begin{subfigure}{0.35\textwidth}
    \includegraphics[width=\linewidth]{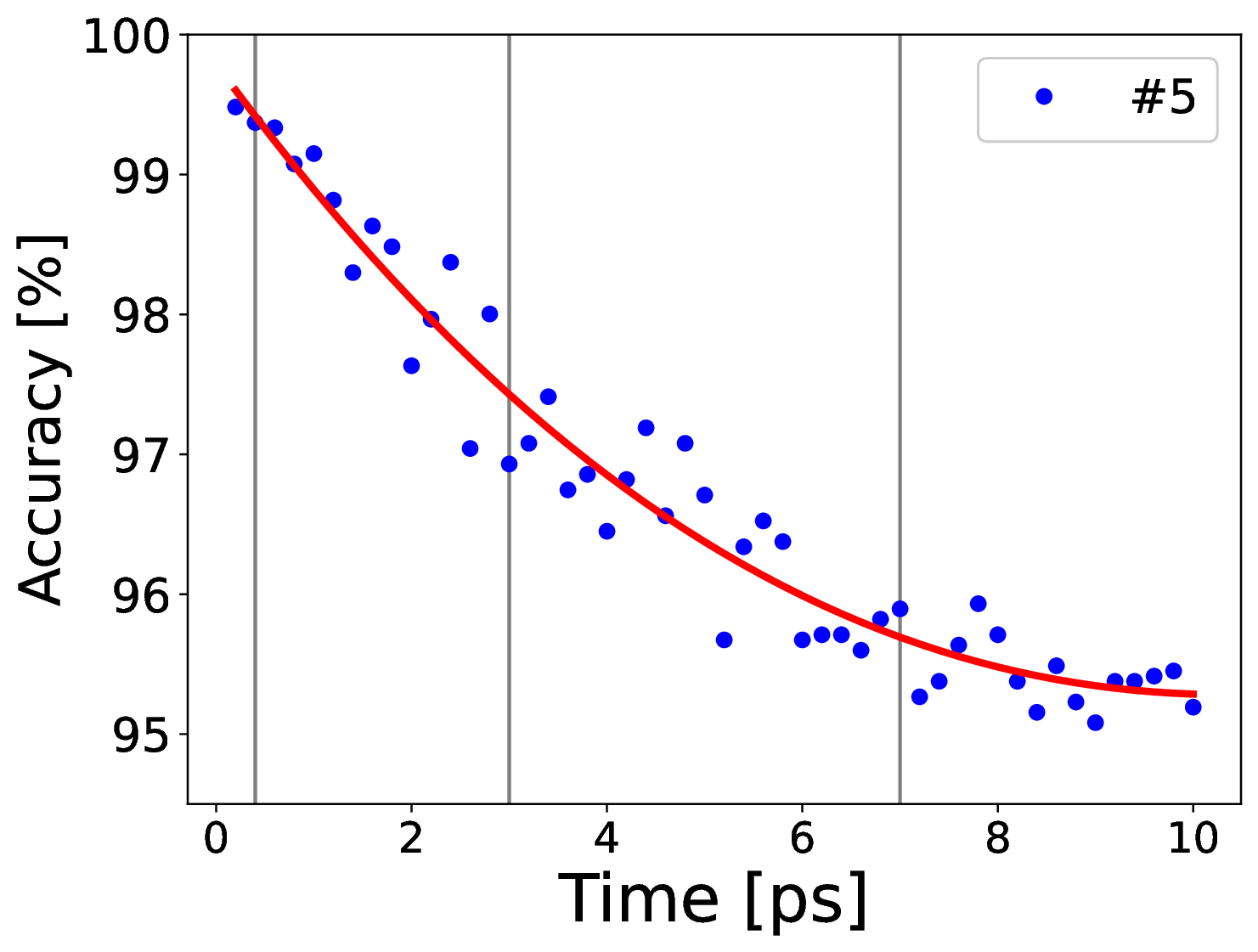}
    \caption{} 
  \end{subfigure}%
  \hspace*{\fill}   
  \begin{subfigure}{0.35\textwidth}
    \includegraphics[width=\linewidth]{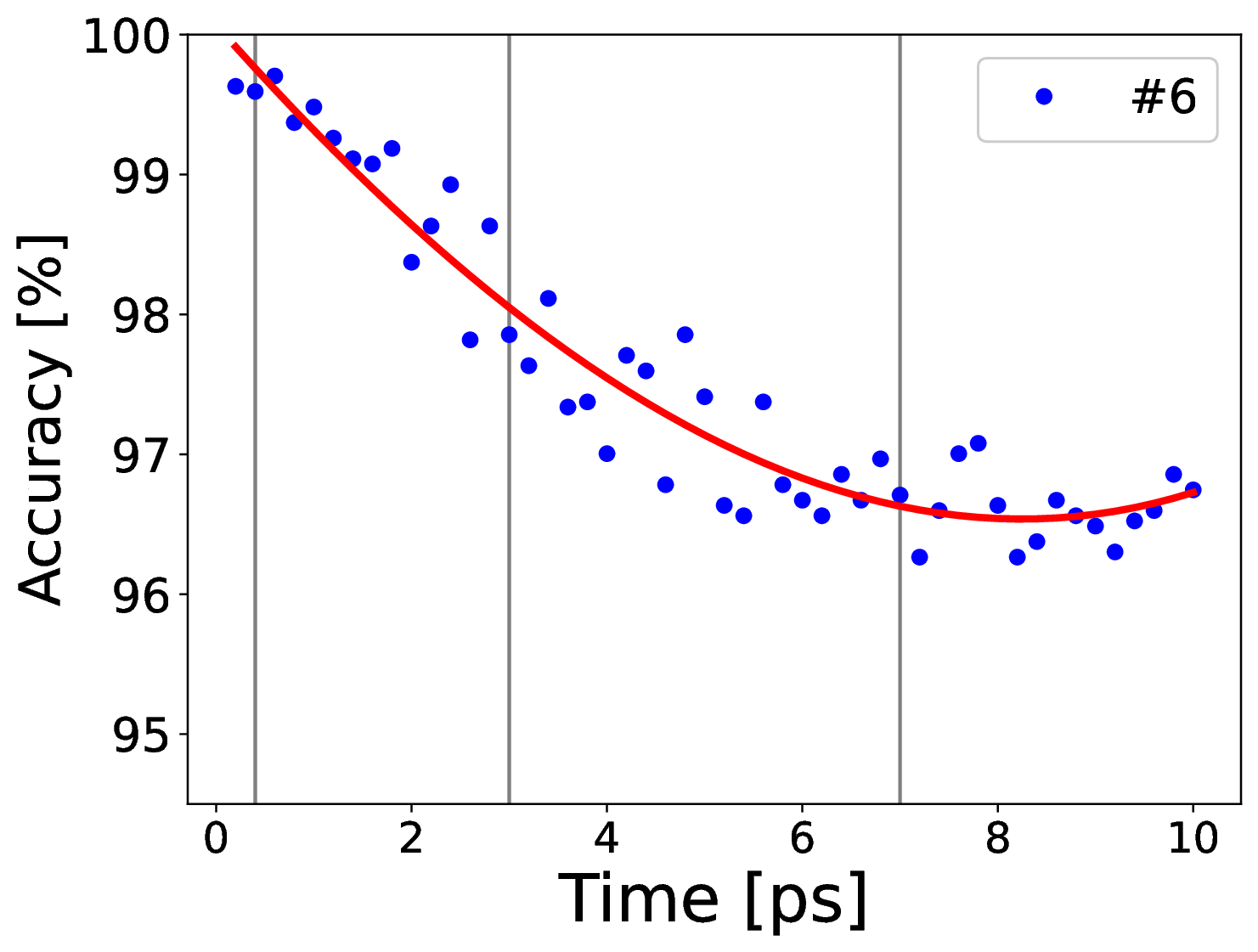}
    \caption{}
  \end{subfigure}%
  \hspace*{\fill}   

  \hspace*{\fill}   
  \begin{subfigure}{0.35\textwidth}
    \includegraphics[width=\linewidth]{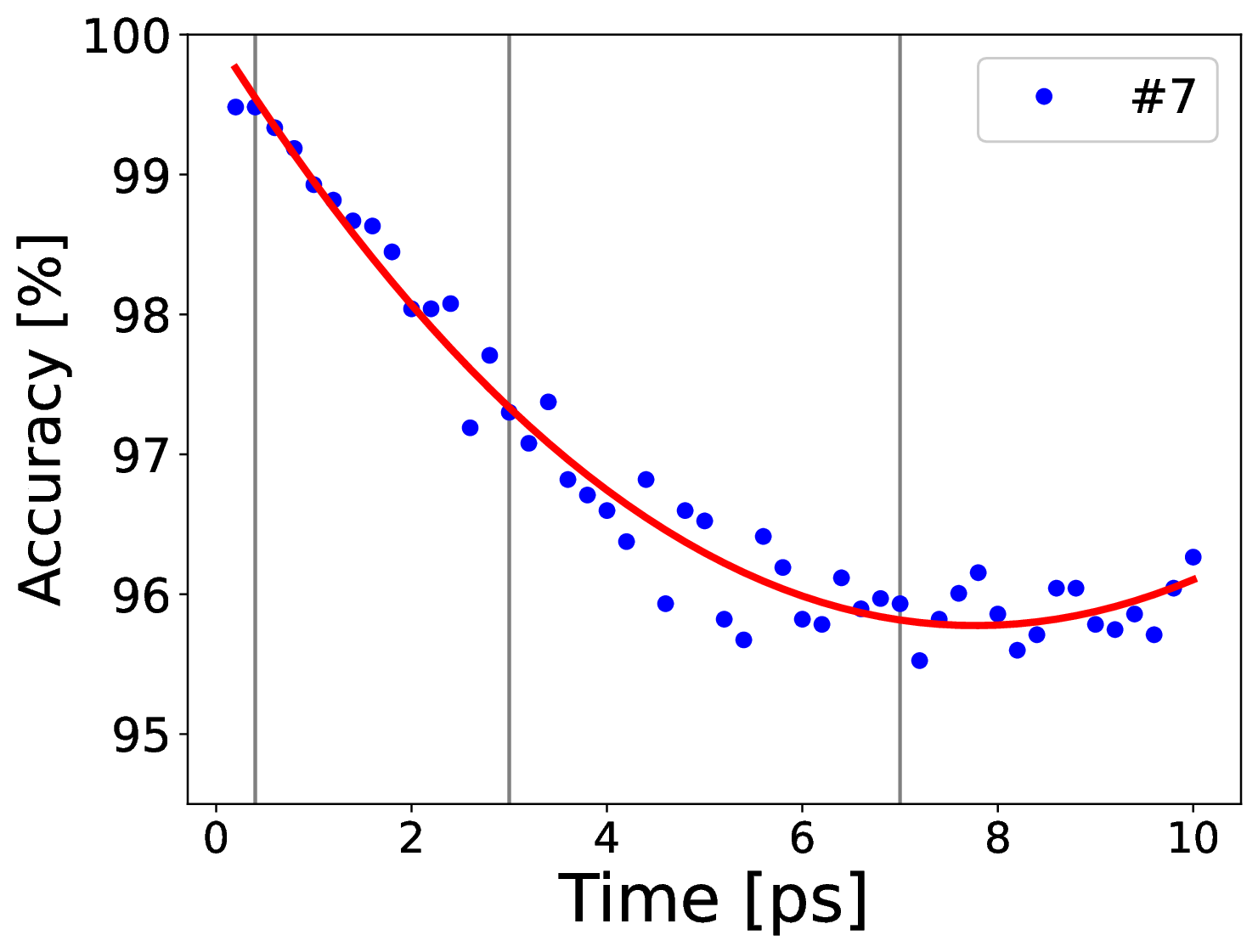}
    \caption{} 
  \end{subfigure}%
  \hspace*{\fill}   
  \begin{subfigure}{0.35\textwidth}
    \includegraphics[width=\linewidth]{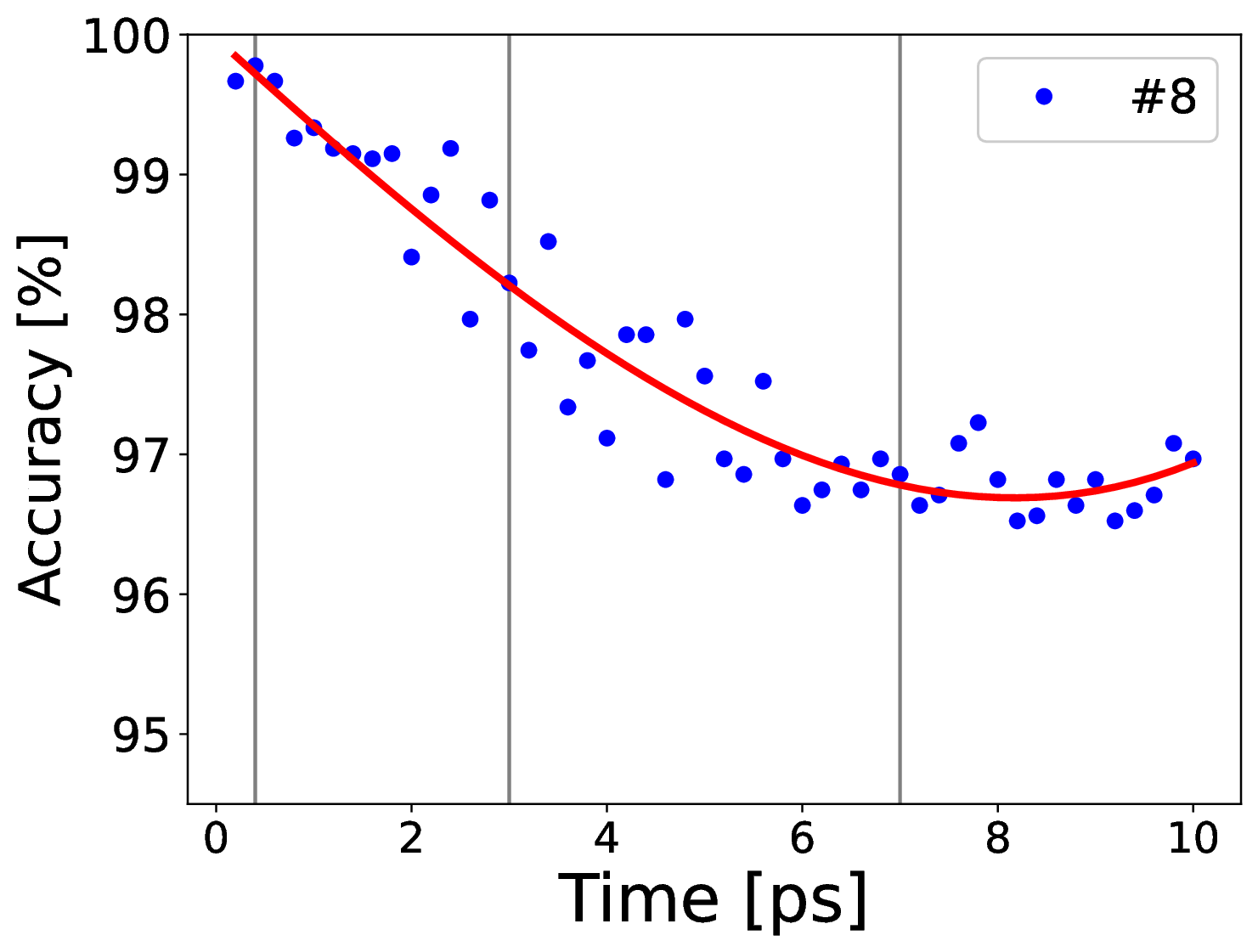}
    \caption{} 
  \end{subfigure}%
  \hspace*{\fill}   

\centering
\caption{\label{fig:accuracyvstime} Accuracy of the NNs trained with the input datasets listed in Table 1 in the main manuscript  in the detection of the grain boundary at different time steps. Vertical grey lines correspond to the time steps used in the training: initial, with small twin, intermediate and final step.  }
\end{figure}

\newpage
\subsection{Accuracy in the sample at constant stress for different input features.}

\begin{figure}[h!]

  \hspace*{\fill}   
  \begin{subfigure}{0.35\textwidth}
    \includegraphics[width=\linewidth]{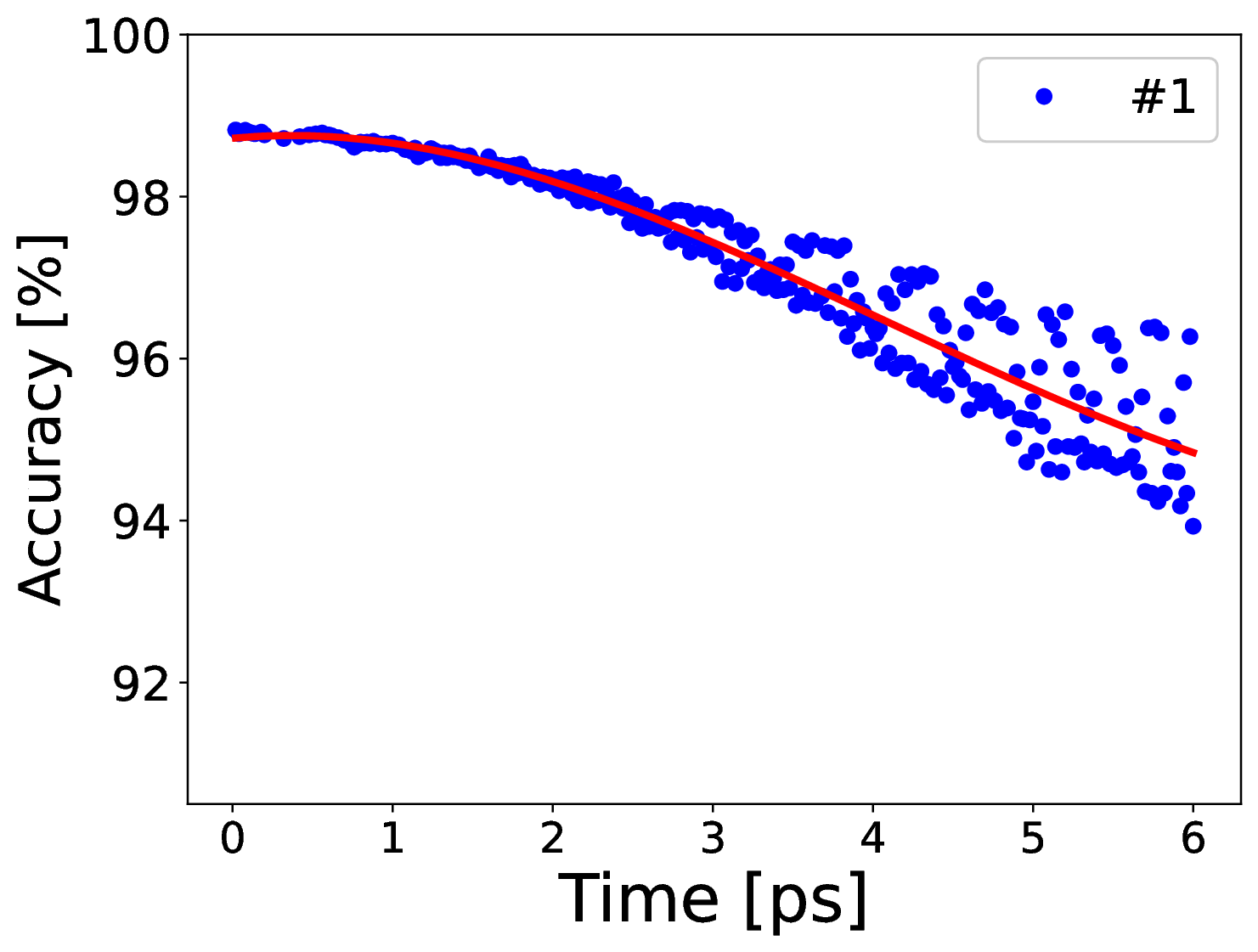}
    \caption{} 
  \end{subfigure}%
  \hspace*{\fill}   
  \begin{subfigure}{0.35\textwidth}
    \includegraphics[width=\linewidth]{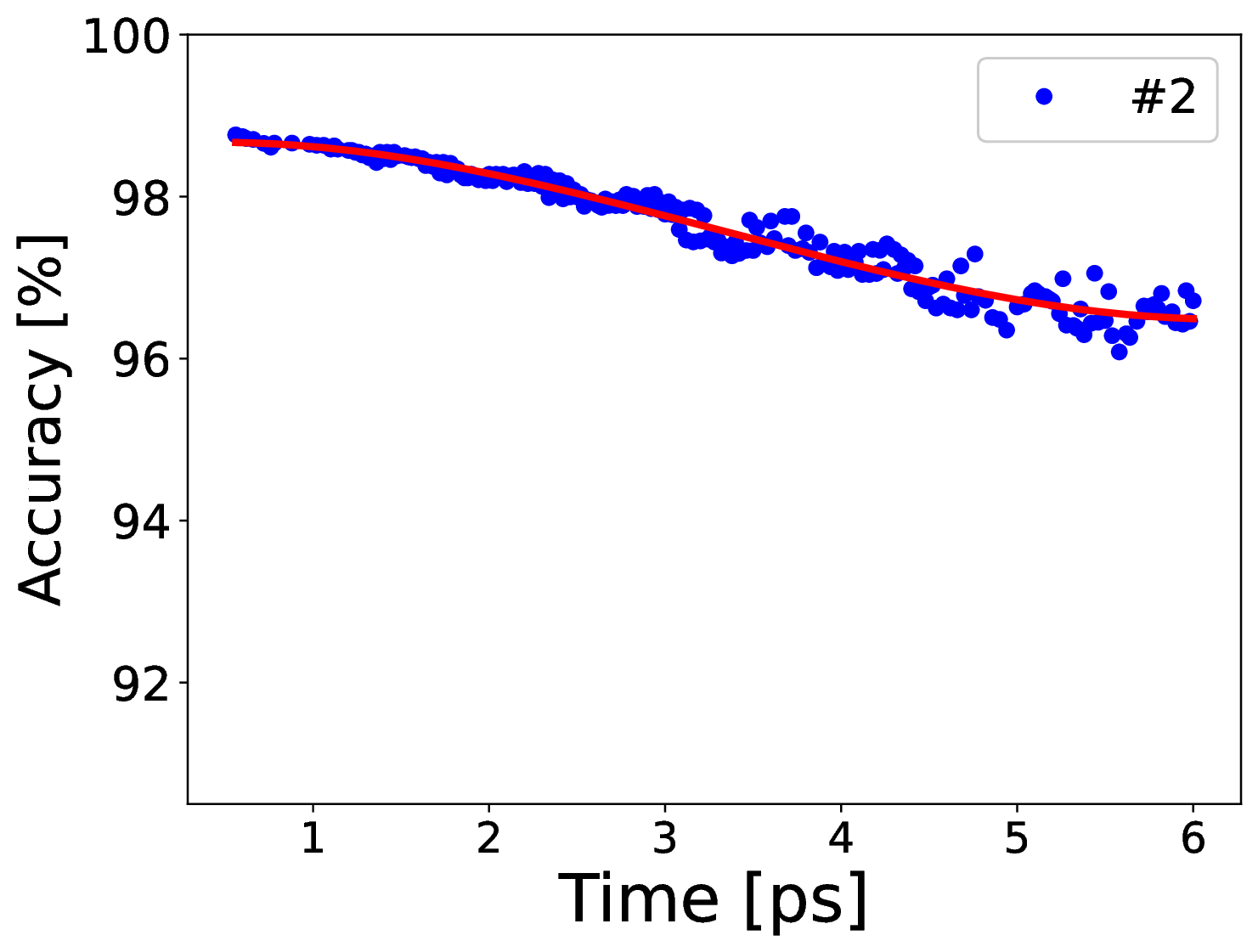}
    \caption{} 
  \end{subfigure}%
  \hspace*{\fill}   
  
  \hspace*{\fill}   
  \begin{subfigure}{0.35\textwidth}
    \includegraphics[width=\linewidth]{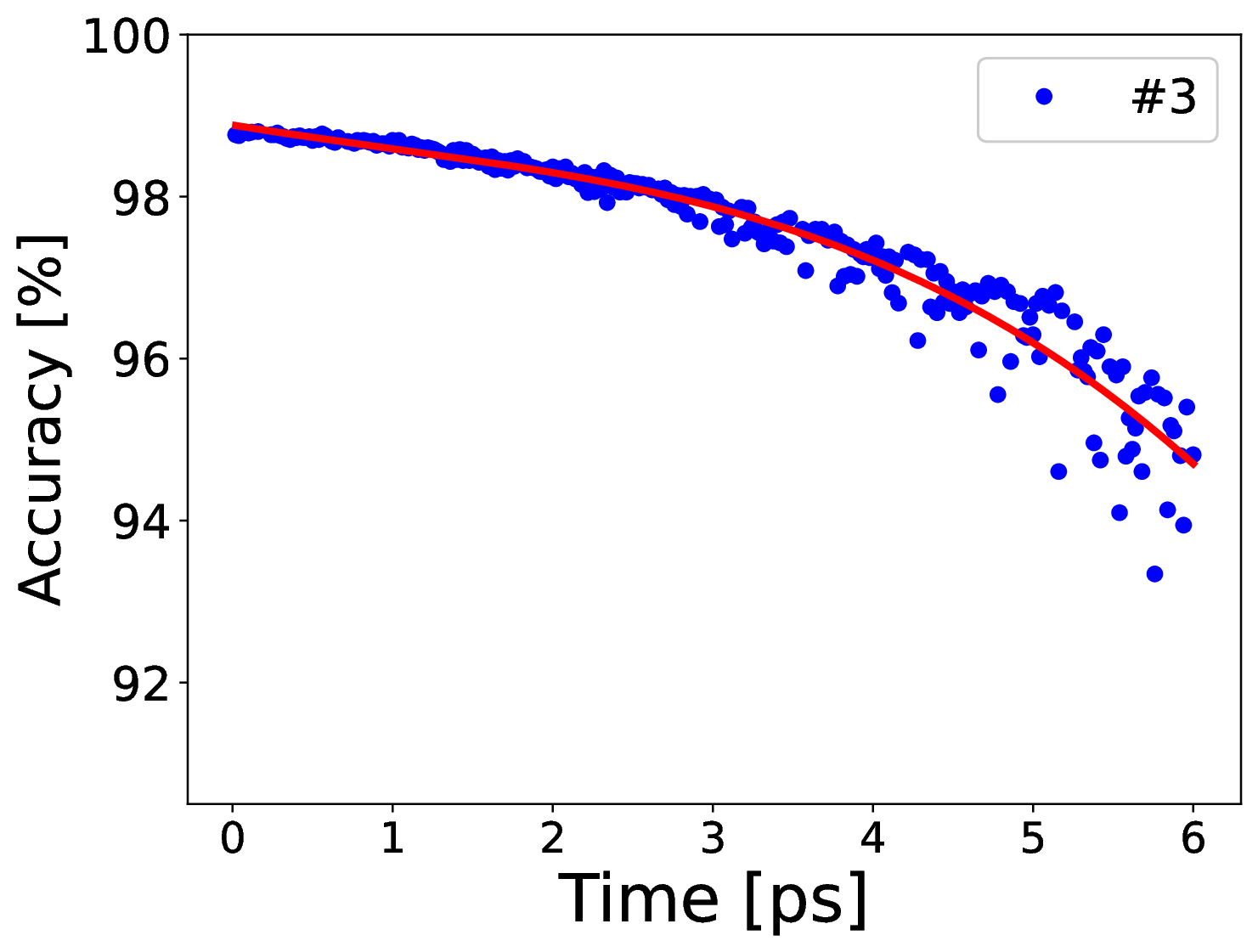}
    \caption{}
  \end{subfigure}%
  \hspace*{\fill}   
  \begin{subfigure}{0.35\textwidth}
    \includegraphics[width=\linewidth]{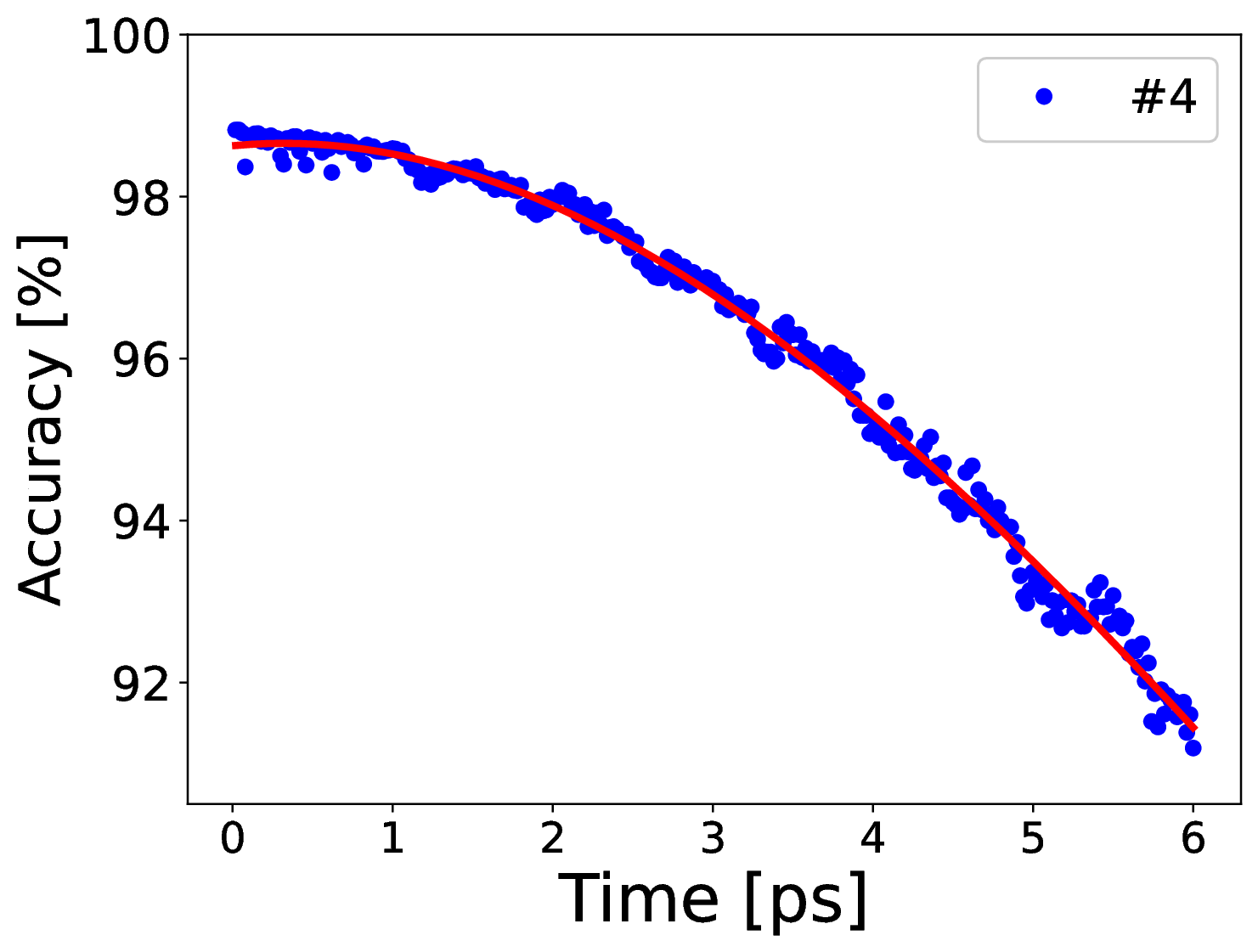}
    \caption{} 
  \end{subfigure}%
  \hspace*{\fill}   
  
  \hspace*{\fill}   
  \begin{subfigure}{0.35\textwidth}
    \includegraphics[width=\linewidth]{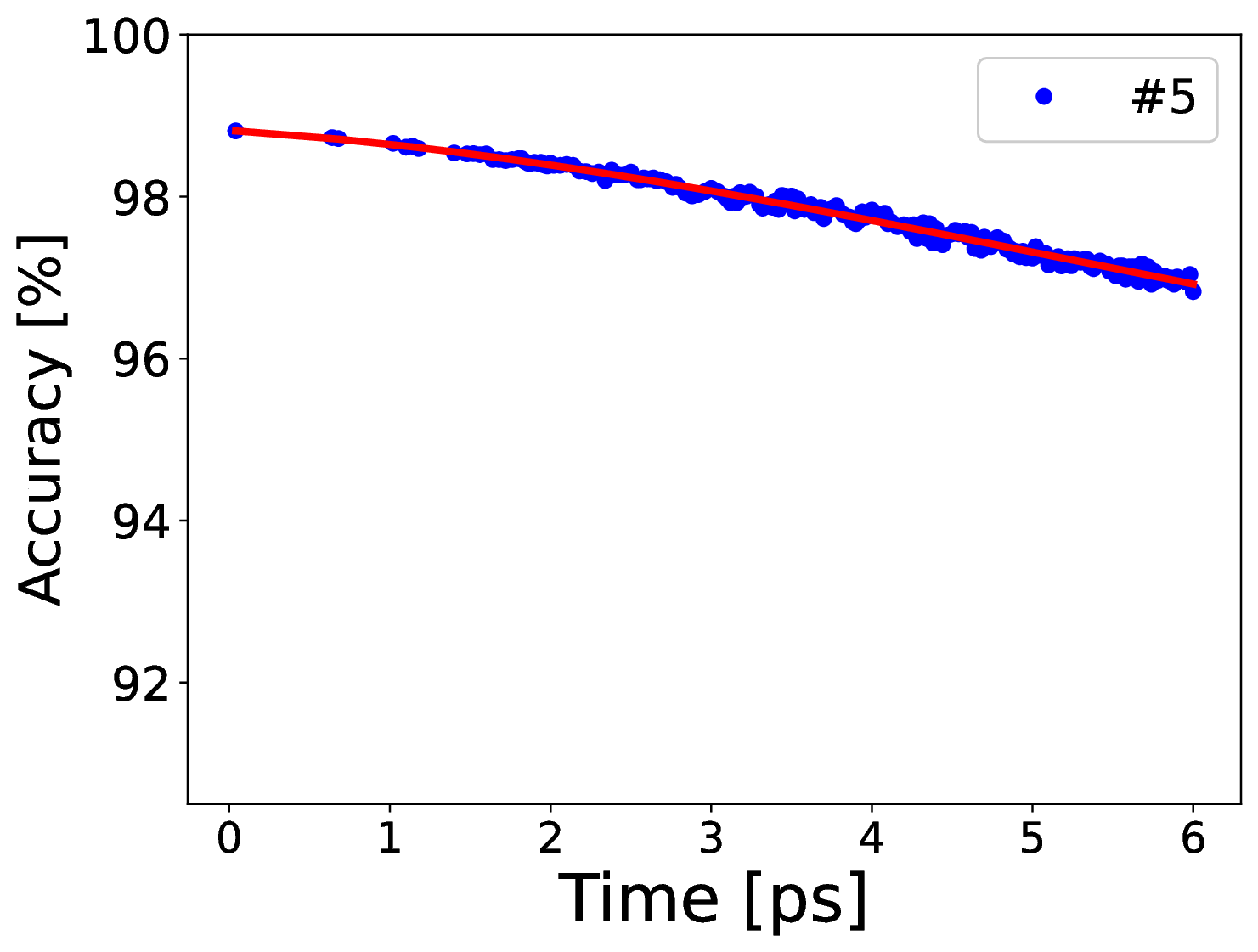}
    \caption{} 
  \end{subfigure}%
  \hspace*{\fill}   
  \begin{subfigure}{0.35\textwidth}
    \includegraphics[width=\linewidth]{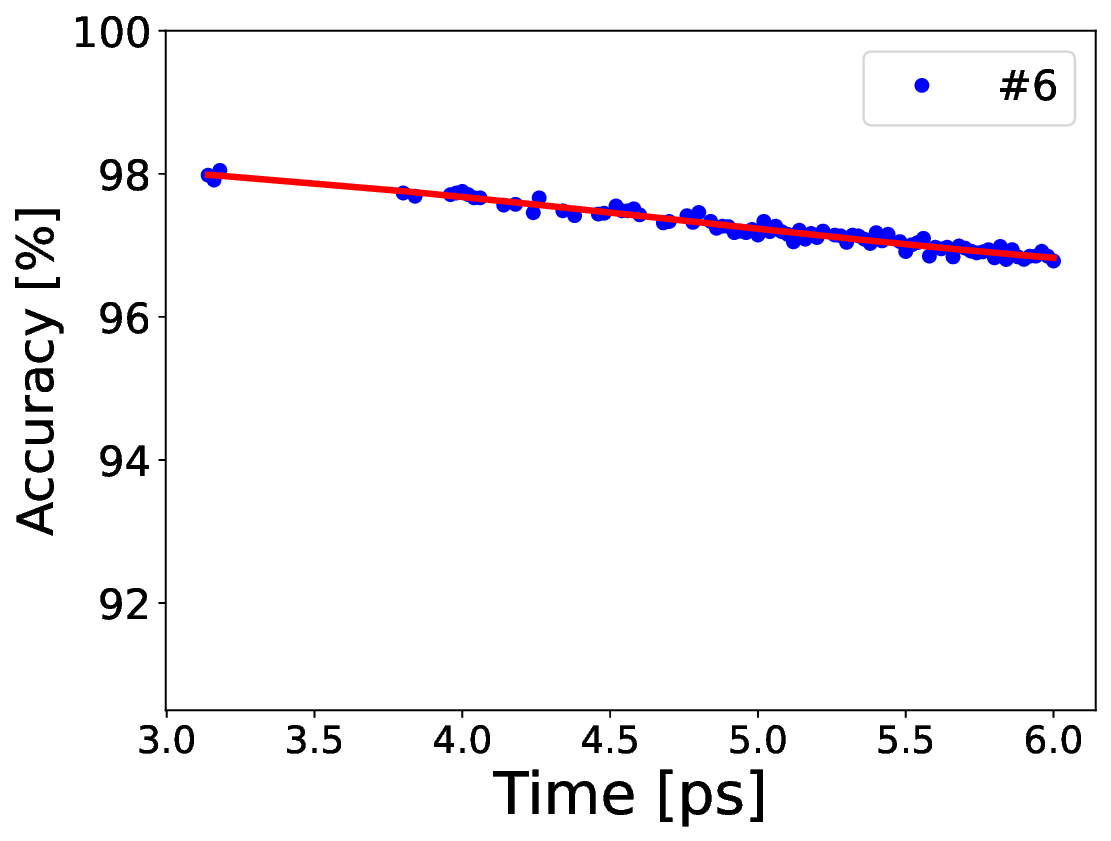}
    \caption{}
  \end{subfigure}%
  \hspace*{\fill}   

  \hspace*{\fill}   
  \begin{subfigure}{0.35\textwidth}
    \includegraphics[width=\linewidth]{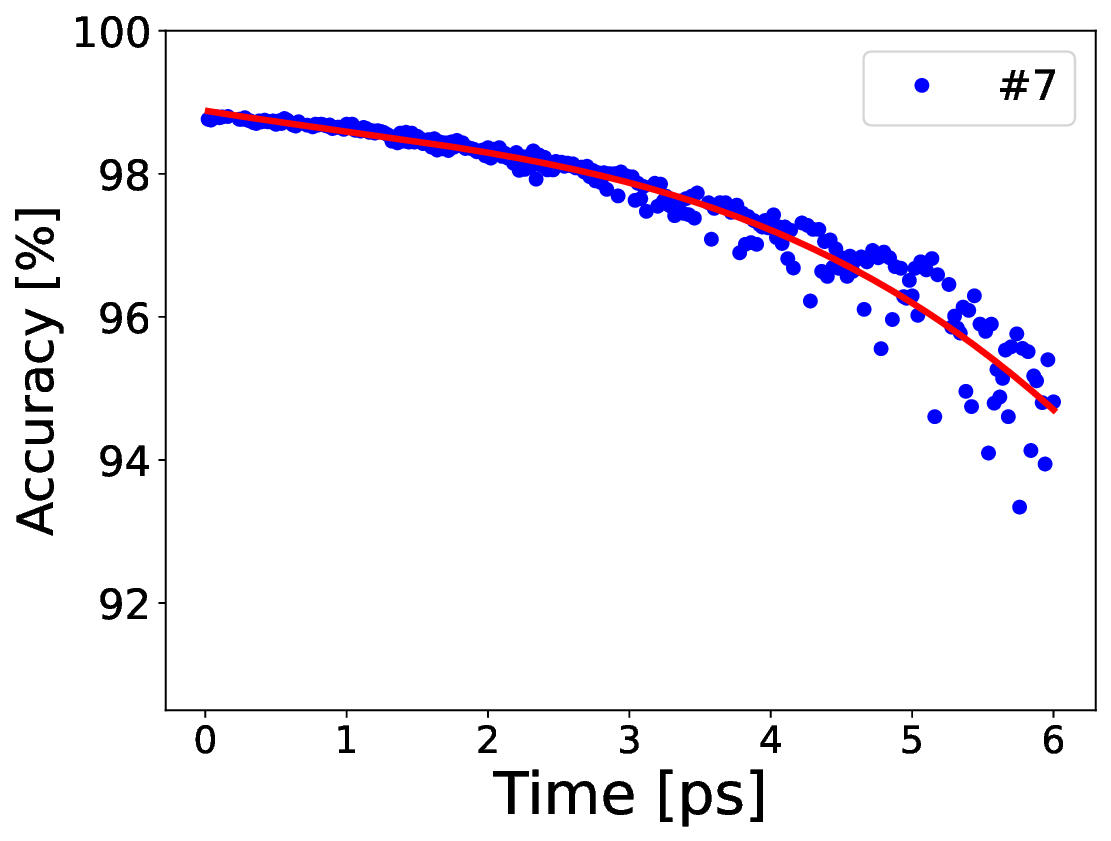}
    \caption{} 
  \end{subfigure}%
  \hspace*{\fill}   
  \begin{subfigure}{0.35\textwidth}
    \includegraphics[width=\linewidth]{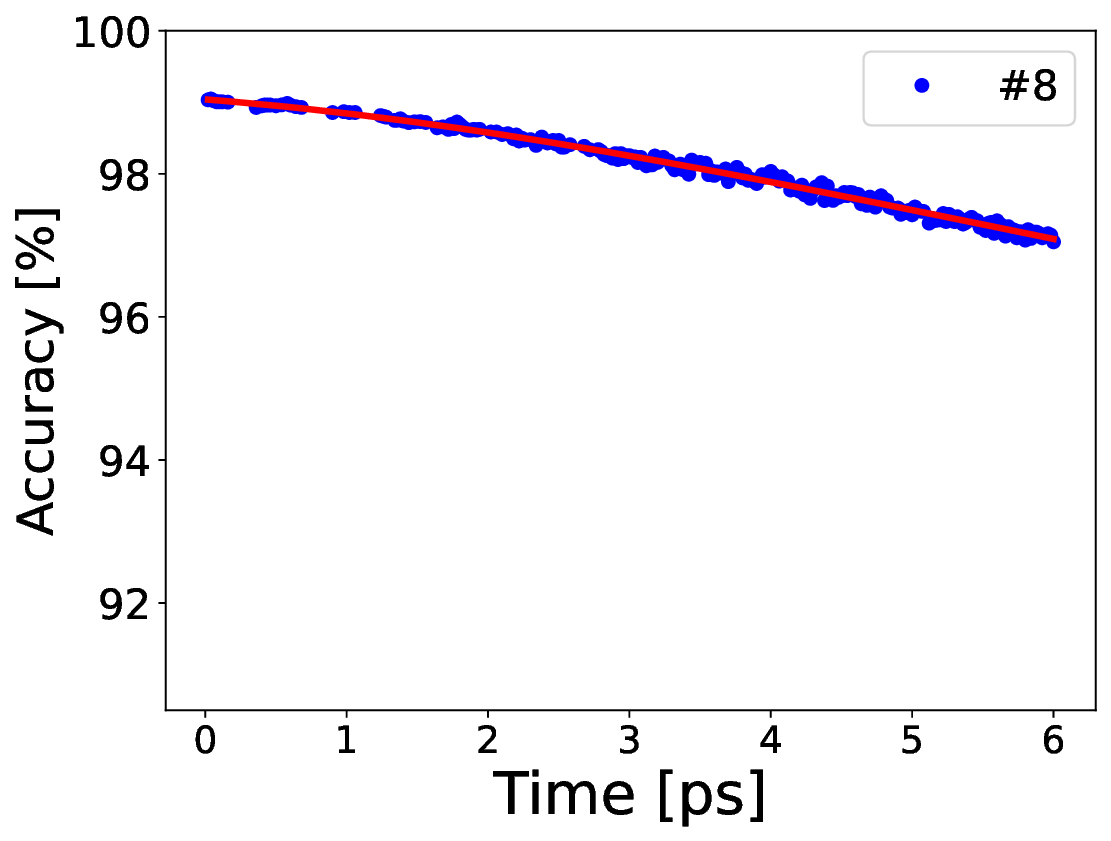}
    \caption{} 
  \end{subfigure}%
  \hspace*{\fill}   

\centering
\caption{\label{fig:accuracyvstime} Accuracy of the NNs trained with the input datasets listed in  Table 1 in the main manuscript in their application to the detection of the grain boundary at different time steps in the simulations at constant stress.  }
\end{figure}

\newpage
\subsection{Evolution of the mean positions of each facet}

\begin{figure}[h!]
\centering
    \includegraphics{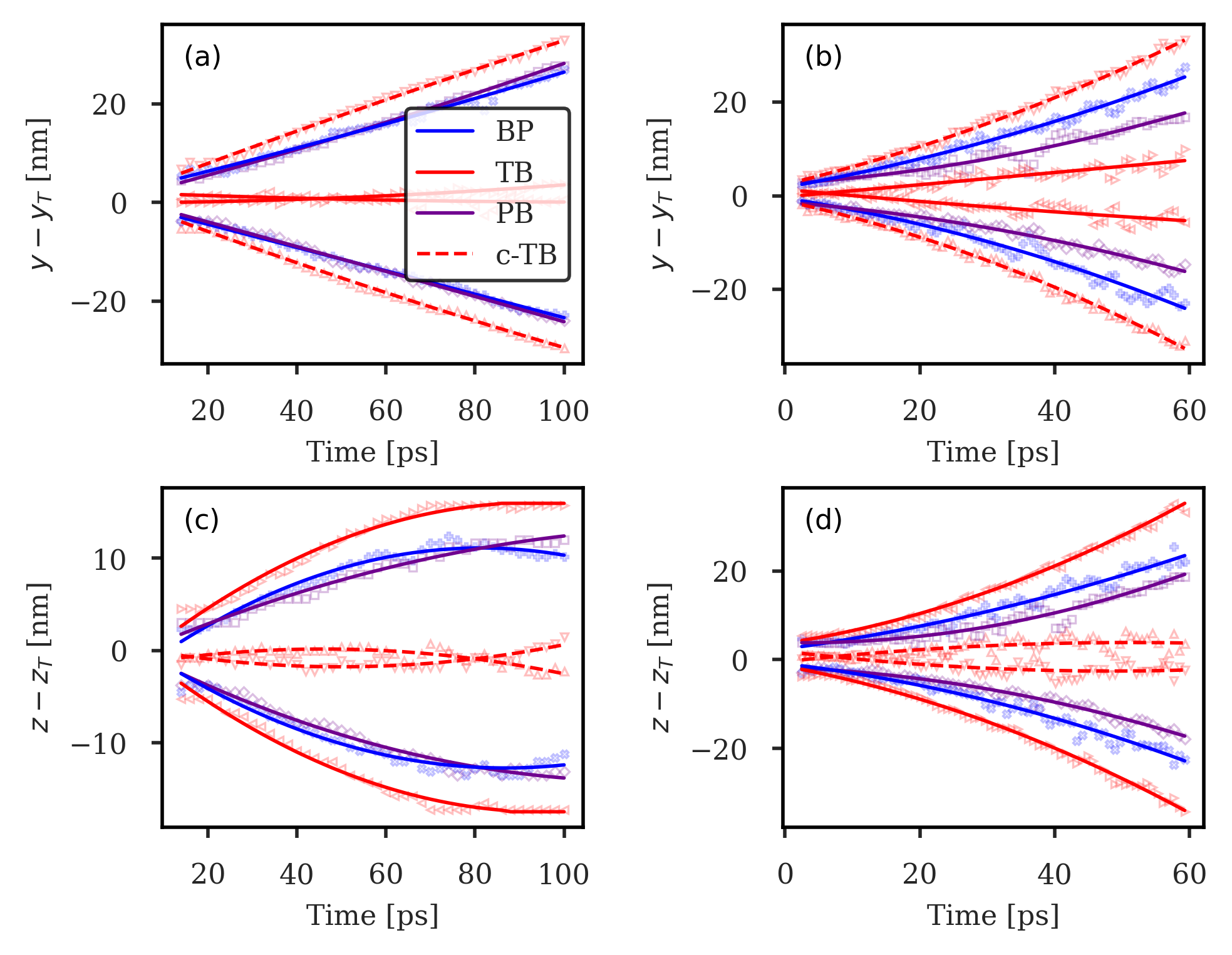}
\caption{\label{fig:analysis} Evolution of the   mean positions of each facet with respect to the twin centre. Simulation at constant strain on the LHS and constant stress on the RHS. "c-TB" refers to "f-TB" in the main text, both are used in previous literature. The equivalence between symbols and facets can be found in Fig. S9 in the Supplementary Material.}
\end{figure}

\newpage
\subsection{Evolution of $\phi$}
The fraction of BP/PB facets at twin/matrix interfaces is characterized by the following ratio:
\begin{equation}
    \phi=\frac{\overline{AB}+\overline{CD}+\overline{EF}+\overline{GH}}{\overline{AB}+\overline{BC}+\overline{CD}+\overline{DE}+\overline{EF}+\overline{FG}+\overline{GH}+\overline{HA}}.
\end{equation}
where AB, BC, etc. are individual twin facets determined as shown in Figure 4a,b in the main text.

\begin{figure}[h!]
\centering
    \includegraphics{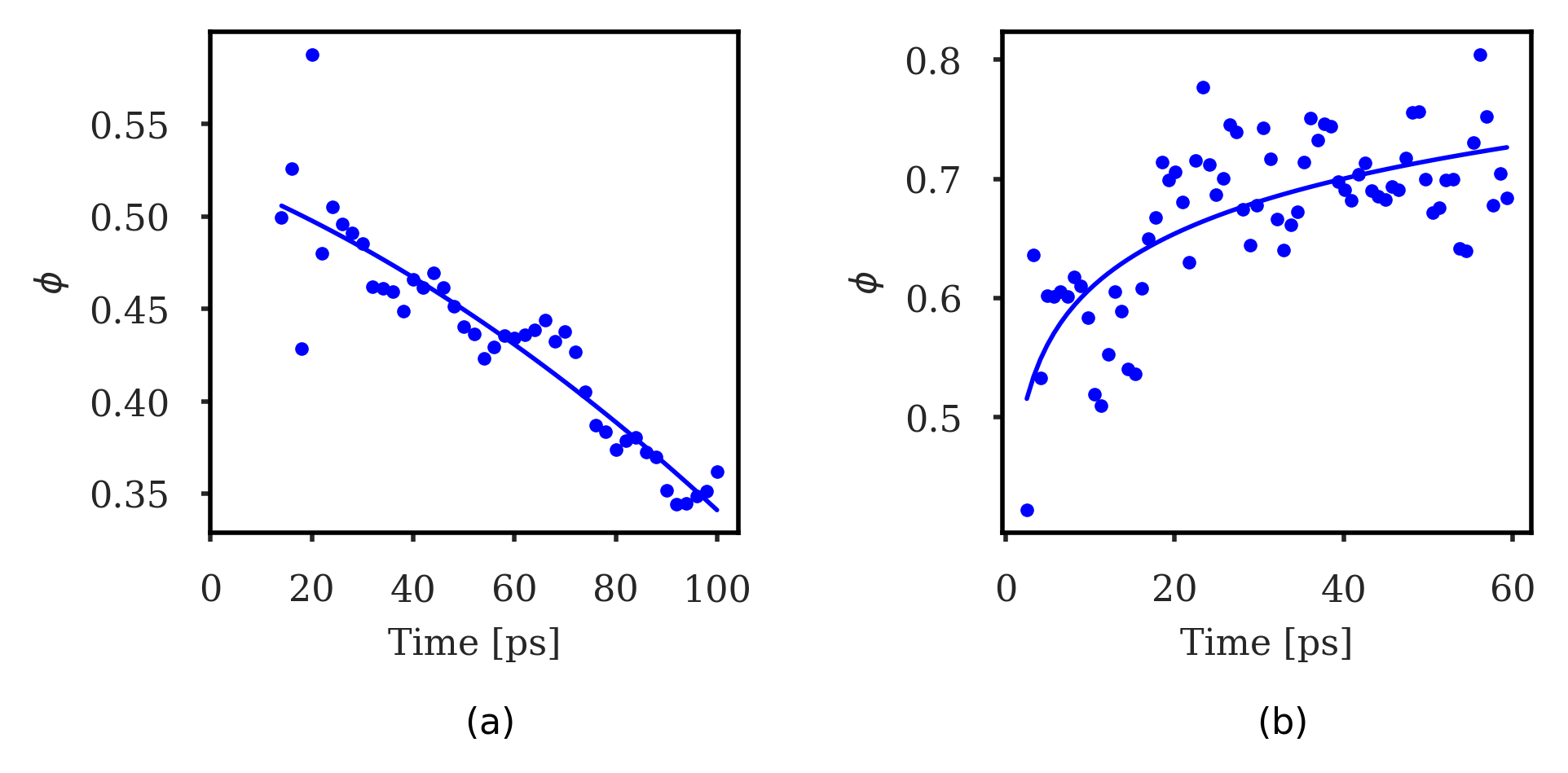}
\caption{\label{fig:analysis} Ratio $\phi$ in MD simulations at (a) constant strain and (b) constant stress.}
\end{figure}

\end{document}